\documentclass{aa}
\usepackage{txfonts}
\usepackage{natbib}
\usepackage{graphicx}
\usepackage{multirow}

\bibpunct[, ]{(}{)}{;}{a}{}{,}

\def\width9{{\sc WIDTH9}}

\def\logg{\log(g)}
\def\teff{T_{\rm eff}}

\def\kms{km/s}
\def\bs{\langle B_{\mathrm{s}} \rangle}

\def\synth3{{\sc Synth3}}

\def\synthmag{{\sc Synthmag}}
\def\synmast{{\sc Synmast}}

\def\vsini{\upsilon\sin i}

\def\hunda{Hund's case~(a)}
\def\hundb{Hund's case~(b)}
\def\br{B_{\rm r}}
\def\bm{B_{\rm m}}
\def\ba{B_{\rm a}}
\def\b{|\mathrm{\mathbf{B}}|}

\def\btimesf{(|\mathrm{\mathbf{B}}|f)}
\def\btimesfi{\sum|\mathbf{B}_i|f_i}

\def\gammaw{\gamma_\mathrm{Waals}}

\def\ll{\lambda\lambda}

\newcommand{\abn}[1]{\alpha(\mathrm{#1})}
\def\mum{$\mu$m}

\begin{document}

\title{
Exploring the magnetic field complexity in M dwarfs at the boundary to full convection
\thanks{Based on observations collected at the European Southern Observatory, Paranal, Chile (program 385.D-0273)}
}
\titlerunning{Magnetic field complexity in M dwarfs}

\author{D. Shulyak\inst{1}  \and A. Reiners\inst{1} \and U. Seemann\inst{1} \and O. Kochukhov\inst{2} \and N. Piskunov\inst{2}}
\authorrunning{D. Shulyak et al.}

\offprints{D. Shulyak, \\
\email{denis.shulyak@gmail.com}}
\institute{
Institute of Astrophysics, Georg-August University, Friedrich-Hund-Platz 1, D-37077 G\"ottingen, Germany \and
Department of Physics and Astronomy, Uppsala University, Box 515, 751 20, Uppsala, Sweden
}

\date{Received / Accepted}

\abstract 
{Magnetic {fields} have a pivotal role in the formation and evolution of low-mass
  stars, but dynamo mechanisms generating these fields are poorly
  understood. The measurement of cool star magnetism is a complicated task
  because of the complexity of cool star spectra and the subtle signatures of
  magnetic fields.}
{Based on detailed spectral synthesis we carry out quantitative measurements
  of the strength and complexity of surface magnetic fields in the four
  well-known M-dwarfs \object{GJ~388}, \object{GJ~729}, \object{GJ~285}, and
  \object{GJ~406} populating the mass regime around the boundary between
  partially and fully convective stars.  Very high resolution ($R=100\,000$),
  high signal-to-noise (up to 400) near-infrared Stokes~I spectra were obtained
  with CRIRES at ESO's Very Large Telescope covering regions of the FeH
  Wing-Ford transitions at $1$~\mum\ and \ion{Na}{i} lines at $2.2$~\mum.}
{A modified version of the Molecular Zeeman Library (MZL) was used to compute
  Land\'e g-factors for FeH lines. We determine the distribution of magnetic
  fields by magnetic spectral synthesis performed with the \synmast\ code. We
  test two different magnetic geometries to probe the influence of field orientation
  effects.}
{Our analysis confirms that FeH lines are {excellent} indicators of surface magnetic
  fields in low-mass stars of type~M, particularly in comparison to profiles
  of \ion{Na}{i} lines that are heavily affected by water lines and suffer
  problems with continuum normalization. The field distributions in all four
  stars are characterized by three distinct groups of field components, the
  data are neither consistent with a smooth distribution of different field
  strengths, nor with one average field strength covering the full star. We
  find evidence of a subtle difference in the field distribution of GJ~285
  compared to the other three targets. GJ~285 also has the highest average
  field of 3.5\,kG and the strongest maximum field component of 7--7.5\,kG. The maximum
  local field strengths in our sample seem to be correlated with rotation
  rate. While the average field strength is saturated, the maximum local field
  strengths in our sample show no evidence for saturation.}
{We find no difference between the field distributions of
  partially and fully convective stars. The one star with evidence for a field
  distribution different to the other three is the most active star 
  {(i.e. with largest x-ray luminosity and mean surface magnetic field)}
  rotating relatively fast. A possible explanation is that rotation determines the
  distribution of surface magnetic fields, and that local field strengths grow
  with rotation even in stars in which the average field is already
  saturated. }


\maketitle

\section{Introduction}

Low-mass stars of spectral type M are subjects of intensive studies today because of several attractive characteristics.
Most important are the high level of activity accompanied by strong X-ray fluxes, appearance of emission lines, and
global magnetic fields of the order of a few kilogauss detected in many stars. The later signature is different from
what we know about the Sun as such strong magnetic fields can only be found in small localized areas known as solar
spots. In case of M-dwarfs we observe magnetic regions that have scales comparable to the size of the stars themselves
\citep[see, e.g.,][]{2010MNRAS.407.2269M}.
{Stellar} evolution predicts that stars of spectral types later than M$3.5$ become fully convective and do not host
an interface layer of strong differential rotation. Both partially
and fully convective stars can host magnetic fields of similar intensities but likely with different dynamo mechanisms
operating in their interiors. All this provides a unique testing ground and challenges for theory of stellar evolution
and magnetism. In addition, due to their low mass, M-dwarfs
are attractive stars to search for Earth-size planets in the habitable zone
since such planets will be {relatively} close to their central star and will have
correspondingly shorter orbital periods and will induce relatively large
radial velocity signatures.

The first direct measurements of magnetic fields from Zeeman broadened line profiles was done by \citet{1985ApJ...299L..47S}.
However, atomic lines are often blended by rich molecular absorption which complicated the analysis. 
\citet{1996ApJ...459L..95J} applied a relative analysis based on a few atomic lines including the strong
\ion{Fe}{i}~$\lambda8468$ line to measure magnetic fields in a number of M-dwarfs \citep{2000ASPC..198..371J,2009AIPC.1094..124K}.
For dwarfs cooler than mid~-~M, atomic line {intensities}
decay rapidly and molecular lines of FeH Wing-Ford $F^4\,\Delta-X^4\,\Delta$ transitions around
$0.99$~\mum\ were proposed as alternative magnetic field indicators \citep{2001ASPC..223.1579V,2006ApJ...644..497R}.
Some of these lines do show strong magnetic sensitivity, as seen, for instance, in the
sunspot spectra \citep{1999asus.book.....W}.

Unfortunately,  theoretical attempts to compute Zeeman patterns of FeH lines
have not achieved much success. This is because the Born-Oppenheimer approximation,
which is usually used in theoretical descriptions of level splitting, fails for the FeH molecule.
We refer to works of \citet{2006ApJ...636..548A} and \citet{2002A&A...385..701B} for more details.

A promising solution was then suggested by
\citet{2006ApJ...644..497R} and \citet{2007ApJ...656.1121R} (hereafter RB07), who estimated the magnetic fields
in a number of M-dwarfs by simple linear interpolation between
the spectral features of two reference stars with known magnetic
fields. The reference stars were \object{GJ~873} (EV~Lac) with $\btimesfi = 3.9$\,kG estimated
by \cite{2000ASPC..198..371J} (here $f$ is a filling factor), 
and the non-magnetic GJ~1002.
See \citet{2012LRSP....9....1R} for a review.

The later attempts tried to combine theoretical and empirical approaches to obtain better estimates
of magnetic fields. \citet{2008A&A...482..387A} made use of a semi-empirical approach to
estimate the Land\'e g-factor of FeH lines in the sunspot spectra.
\citet{2008ApJ...686.1426H} presented
the empirical g-factors for a number of FeH lines originating from
different levels but limited to low magnetic $J$-numbers. 

Finally, \citet{2010A&A...523A..37S} presented a slightly different semi-empirical
approach where Land\'e g-factors are computed according to the level's $\Omega$ and $J$ quantum numbers.
Using sunspot spectra as a reference, these authors showed a good consistency between observed 
and predicted splitting patterns for numerous FeH lines and applied their method to measure magnetic fields in selected M-dwarfs. 
It was found that the magnetic fields needed to fit characteristic magnetic features are by $\sim15-30$\% smaller
than those reported in previous studies using a FeH molecule. However, throughout the analysis it was assumed that the magnetic field
has only one single component, i.e. the corresponding filling factor $f=1$.

The fact that surface magnetic fields of M-dwarfs can be localized in areas similar to Sun spots and therefore have
rather complex geometry that correspond to particular sets of filling factors 
(i.e. set of surface areas of the star covered by a magnetic field of particular intensity) 
had already been noted in previous works \citep[see, e.g.,][]{2000ASPC..198..371J,2009AIPC.1094..124K}. These complex geometries
reveal themselves as an impossibility to simultaneously fit narrow cores and wide wings of magnetically sensitive lines with
a single magnetic field component.

Spectropolarimetry and Doppler Imaging have a potential to reconstruct the true magnetic field geometry in M-stars.
However, these techniques are very challenging to apply for such faint objects. Nevertheless, it is still possible to
observe the brightest objects and use a Least Square Deconvolution (LSD) method \citep[see][]{1997MNRAS.291..658D}
to construct an average line
profile based on selected individual atomic lines for which Land\'e factors are accurately known, as was
done in a series of papers by \citet{2006Sci...311..633D,2008MNRAS.390..545D,2008MNRAS.390..567M}. 
The authors found the presence of regions with 
strong magnetic fields that occupy large surface areas in selected M-dwarfs, but the field geometry of the majority of them 
are dominated by the poloidal component, which indicates rather homogeneous magnetic fields. Interestingly, these authors also
find a signature of different magnetic field geometry observed in partly and fully convective stars with the former
hosting mostly non-axisymmetric toroidal fields while axisymmetric poloidal fields are common for the later.
{However}, new observations show that there are exceptions in both groups \citep{2010MNRAS.407.2269M}. Because of complicated spectra
of M-dwarfs and the use of only Stokes $V$ spectra for Doppler mapping
(which is sensitive only to large scale magnetic fields)  the results of {these} studies should be re-confirmed
with more precise analysis involving at least Stokes $I$ and $V$ spectra as well as direct synthesis of both atomic and molecular
features.

Since the direct detection of strong surface magnetic fields in M-dwarfs from magnetically split lines 
in mid eighties, there is still no clear understanding about the true geometry
of these fields. Because of flaring events that affect fluxes of stars in a dramatic and irregular way, 
it remains very challenging to confirm the presence of stellar spots via light curve analysis and 
answer the question whether the magnetic fields originate from such active regions or they cover
the whole stellar surface. What is known from spectroscopic and spectropolarimetric studies
is that the magnetic fields can have complex structures covering a large ($\approx70-80$\%) fraction of
the stellar surface. 

In this work we make use of very high resolution infrared spectra of four benchmark M-dwarfs GJ~388, GJ~285, GJ~729, and GJ~406
obtained with CRIRES@VLT to measure the complexity of their surface magnetic fields by studying individual line profiles.
We attempt to derive distributions of filling factors that provide a best agreement between observed and theoretical spectra and to
address the question whether there are any differences between fully and partly convective stars solely from spectroscopic
analysis.

\section{Observations}
Our study is based on high-resolution near-infrared spectra obtained with CRIRES, \citep[the CRyogenic high-resolution InfraRed Echelle Spectrograph;][]{2004SPIE.5492.1218K}, 
mounted on UT1 at ESO's VLT. The data were collected between June 2010 and January 2011 under PID 385.D-0273. 
The CRIRES instrument is the only spectrograph capable of providing a resolving power of $R=100\,000$ in the near-infrared ($1-5\mu$m).
We observed each of our four program stars in two wavelength regions around $1.0\mu$m and $2.2\mu$m, to cover the FeH molecular band and the \ion{Na}{I} line. 
In each wavelength region, two adjacent wavelength settings were observed to bridge the three small gaps in wavelength coverage introduced by the spacing between 
the four CRIRES detectors. 
For all observations, we used an entrance slit width of 0\farcs2, resulting in a nominal resolving power of $R=10^5$. The adaptive optics system was also utilized 
to improve throughput. Table~\ref{tab:obs} contains basic information about observations obtained.

\begin{table}
\caption{\textbf{Observations used in the paper.}}
\label{tab:obs}
\begin{scriptsize}
\begin{center}
\begin{tabular}{cccc}
\hline
\hline
Target & MJD started-finished &  $\lambda$, nm & $<$SNR$>$\\
\hline
GJ~1002 & 55372.36459129~--~55372.41802987 & 991.0-1000.6  & 50\\
        &                               & 2232.4-2239.1 & 100\\
        & 55373.38971144~--~55373.42115227 & 991.0-1000.6  & 30\\
        &                               & 2232.4-2239.1 & 100\\
\hline
GJ~406  & 55290.13828172~--~55290.17037657 & 991.0-1000.6  & 120\\
        &                               & 2232.4-2239.1 & 190\\
\hline
GJ~729  & 55362.38028065~--~55362.41393174 & 991.0-1000.6  & 130\\
        &                               & 2232.4-2239.1 & 190\\
\hline
GJ~285  & 55551.15463425~--~55551.18593180 & 991.0-1000.6  & 200\\
        &                               & 2232.4-2239.1 & 200\\
\hline
GJ~388  & 55580.25172658~--~55580.28725773 & 991.0-1000.6  & 500\\
        &                               & 2232.4-2239.1 & 350\\
\hline
\end{tabular}                             
\end{center}
\end{scriptsize}
\end{table}

In addition to the four program stars, we obtained CRIRES data for another target, \object{GJ~1002}, observed over several nights in October 2008 (PID 079.D-0357). These data cover some strong 
\ion{Ti}{i} lines in the $1-1.1\mu$m interval.

Data reduction of the nodded observations followed mostly standard procedures, where each detector chip is treated separately. 
Averaged dark frames are created for  all the raw frames except for the nodded science frames, from appropriate exposure times. 
This is an important step to treat 
the odd-even effect and the amplifier glow inherent to the CRIRES detectors
\footnote{http://www.eso.org/observing/dfo/quality/CRIRES/pipeline/ \\ recipe\_science.html}.
To correct for the non-linear detector response, we measure the linearity deviation as a function of flux 
per pixel as obtained from a series of flat field images 
with varying illumination levels, recorded close in time to our actual observations. These data also serve to identify bad pixels, which are then flagged. 
The bad pixel mask constructed in this way greatly helps to mask out detector defects and cosmetics. 
The corrections are applied to all science frames, and to their corresponding flat field images. 
The so-corrected flat fields are averaged, and subsequently utilized to calibrate spatial inhomogeneities of the science exposures. 
Finally, the science frames obtained in two nodding positions along the slit are subtracted, 
thereby removing the sky background and dark and bias signal in the 2D spectra. 
We usually obtained eleven pairs of nodded observations per setting, which are then aligned and combined to boost the SNR. 
Lastly, an optimal extraction is performed to retrieve 1D spectra, which have a typical SNR of 50 to 400 depending on the object faintness. 
A wavelength solution is assigned per chip, based on the comparison with the synthetic spectra (see below).

\section{Methods}
\label{sect:methods}

\begin{figure*}
\centerline{
\includegraphics[width=0.25\hsize]{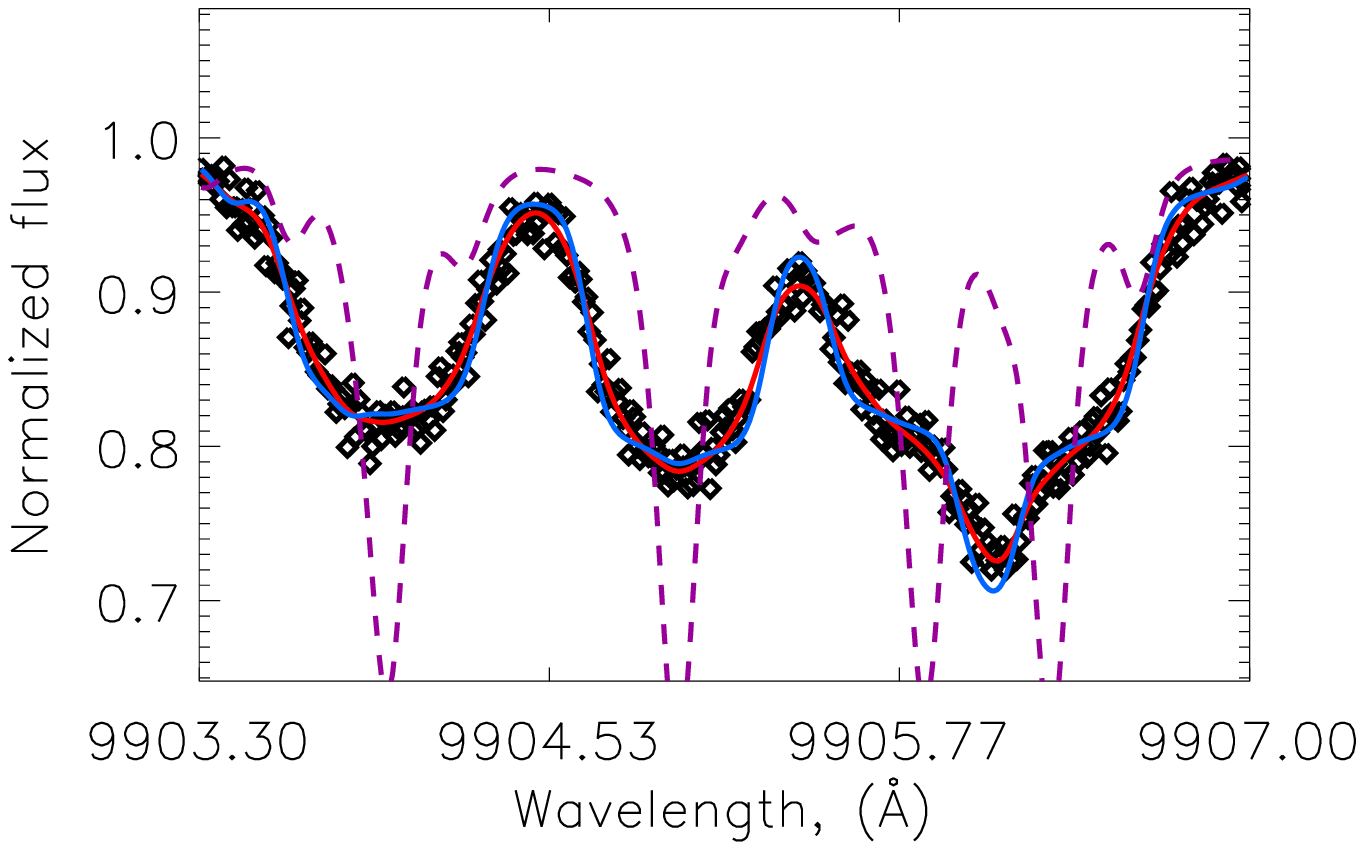}
\includegraphics[width=0.25\hsize]{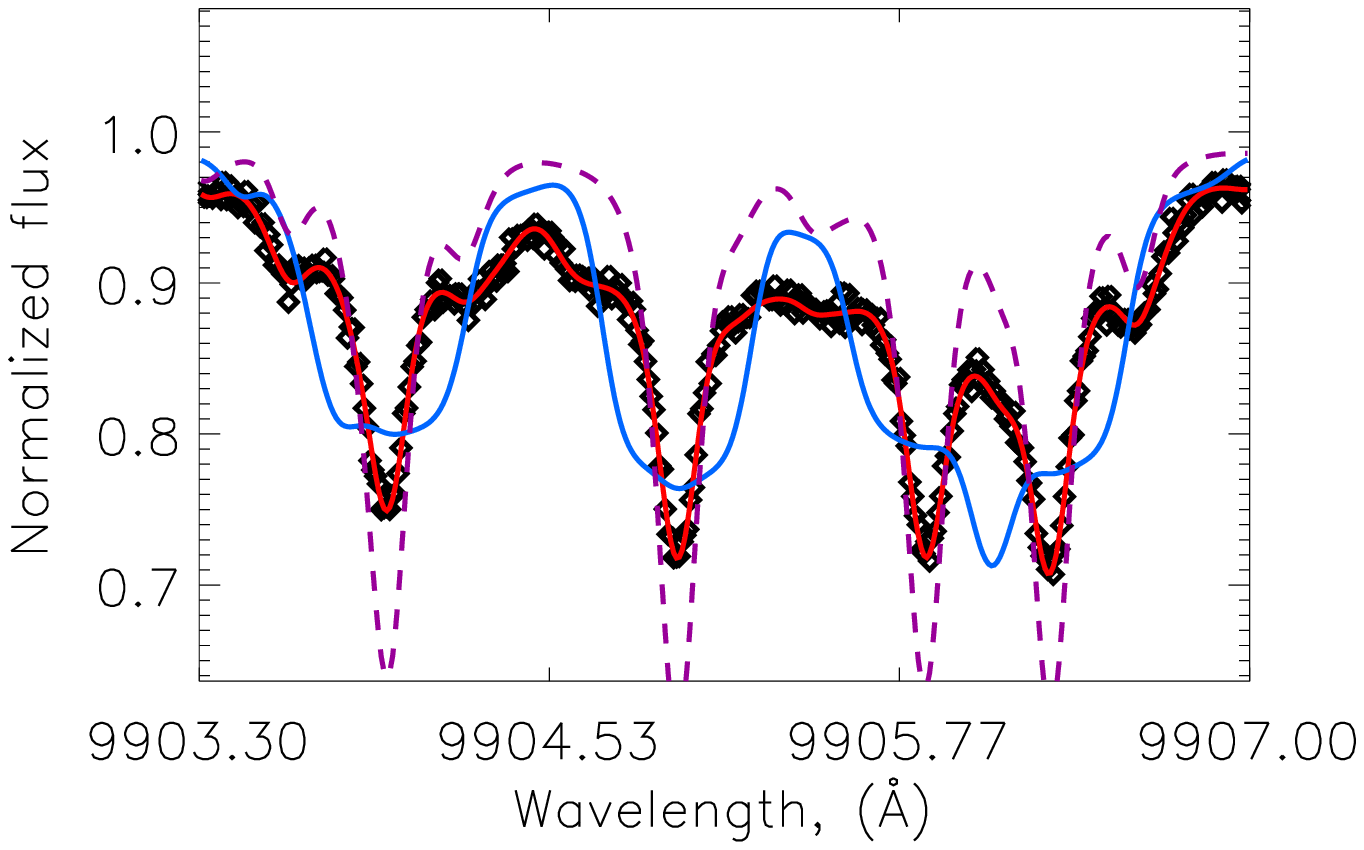}
\includegraphics[width=0.25\hsize]{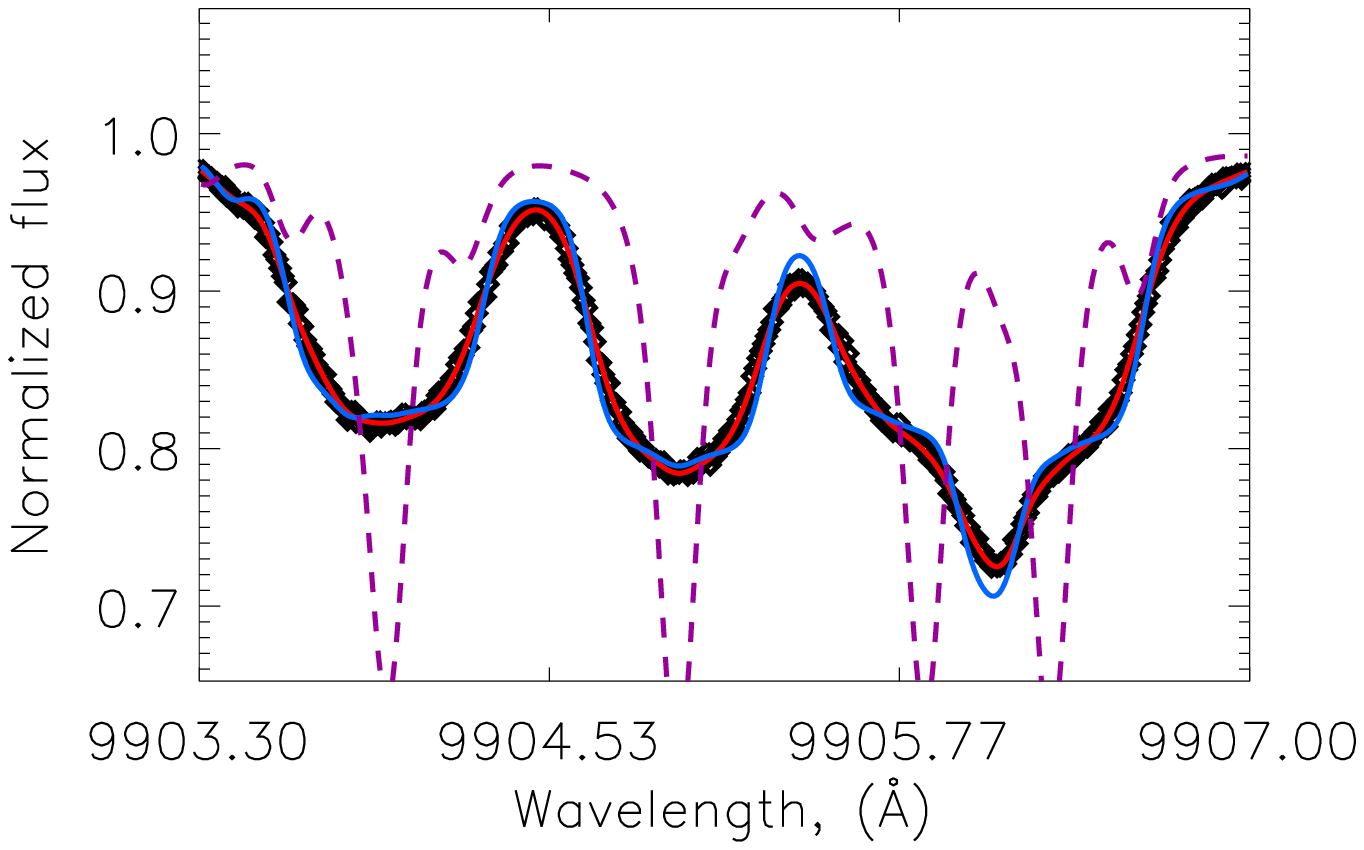}
\includegraphics[width=0.25\hsize]{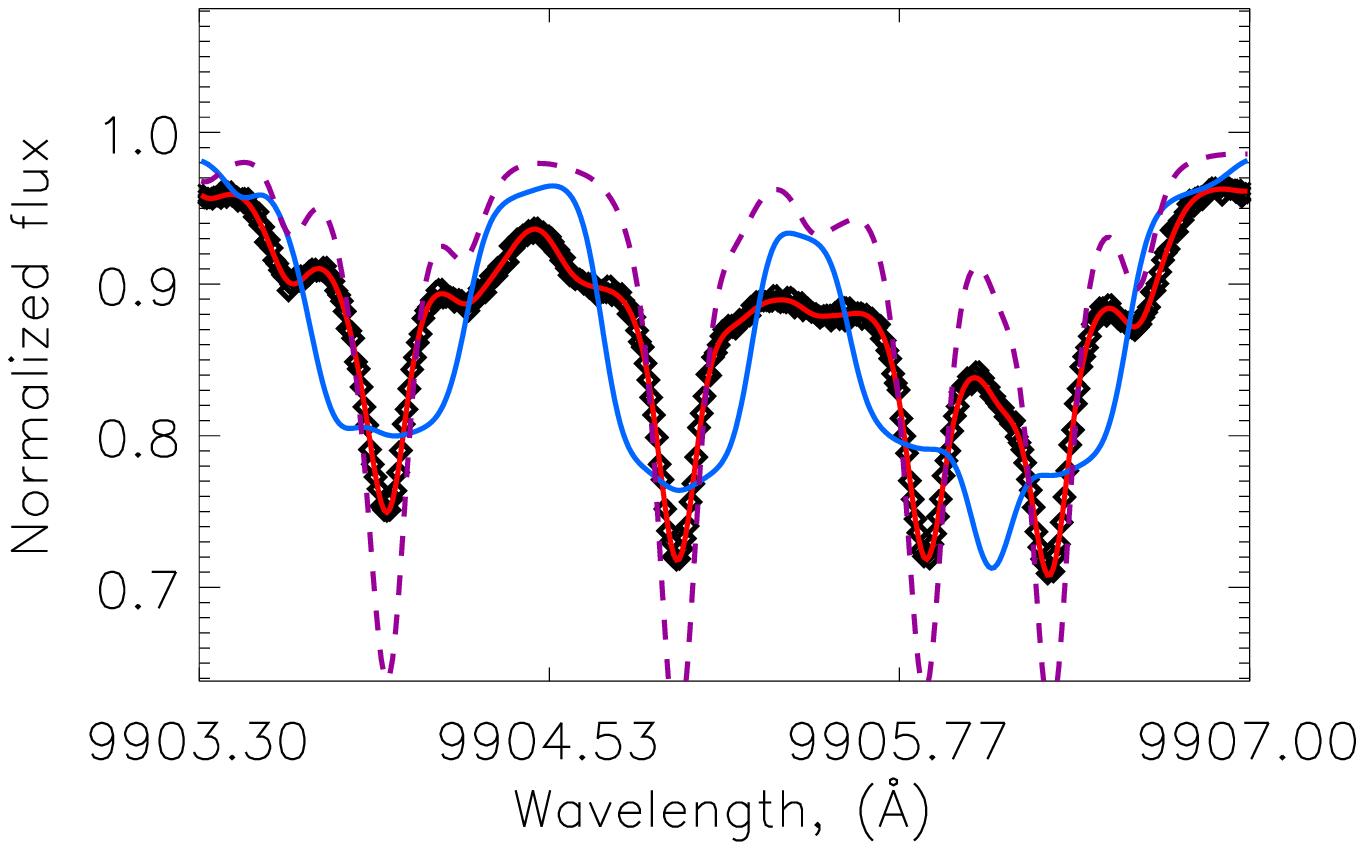}
}
\centerline{
\includegraphics[width=0.25\hsize]{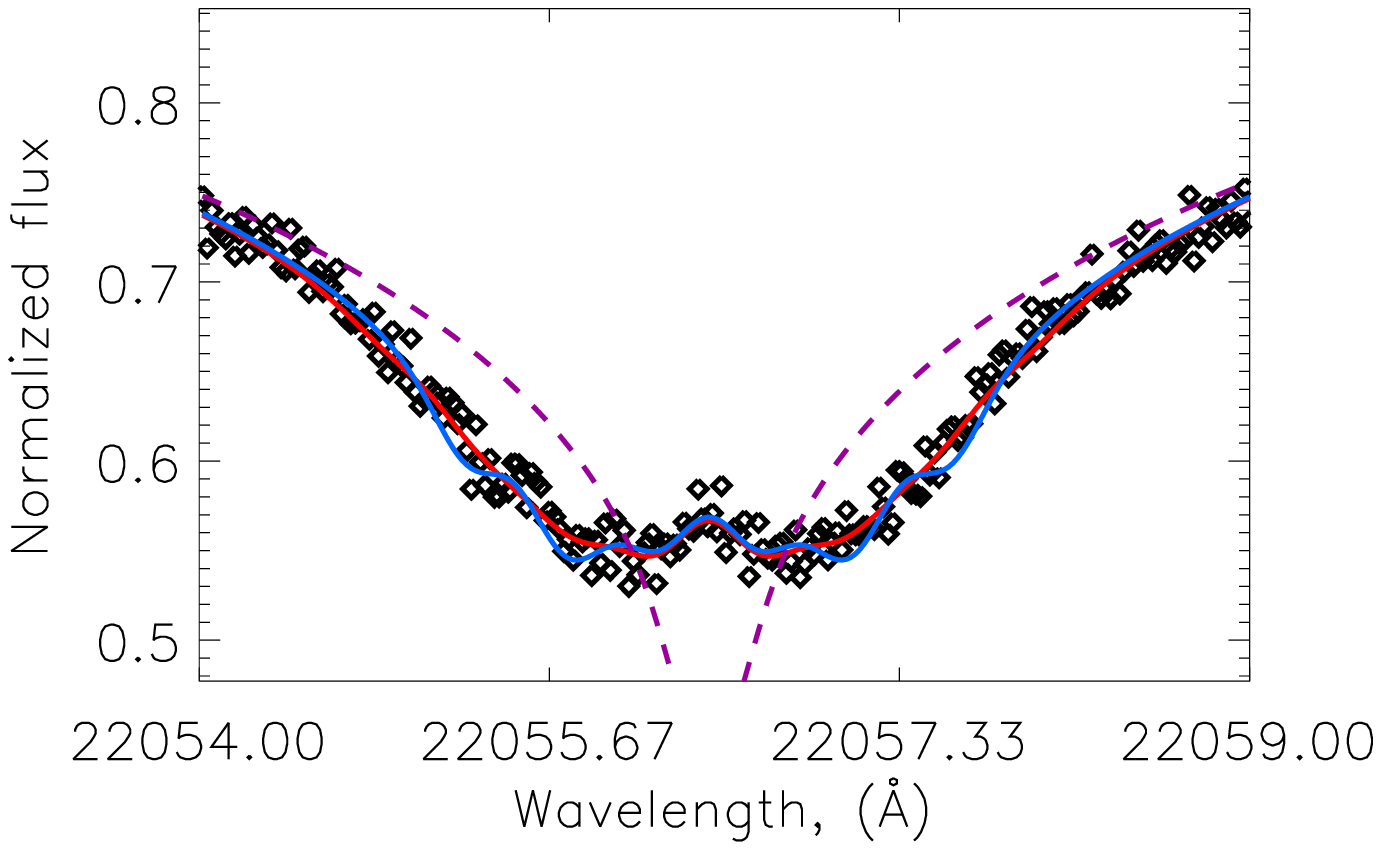}
\includegraphics[width=0.25\hsize]{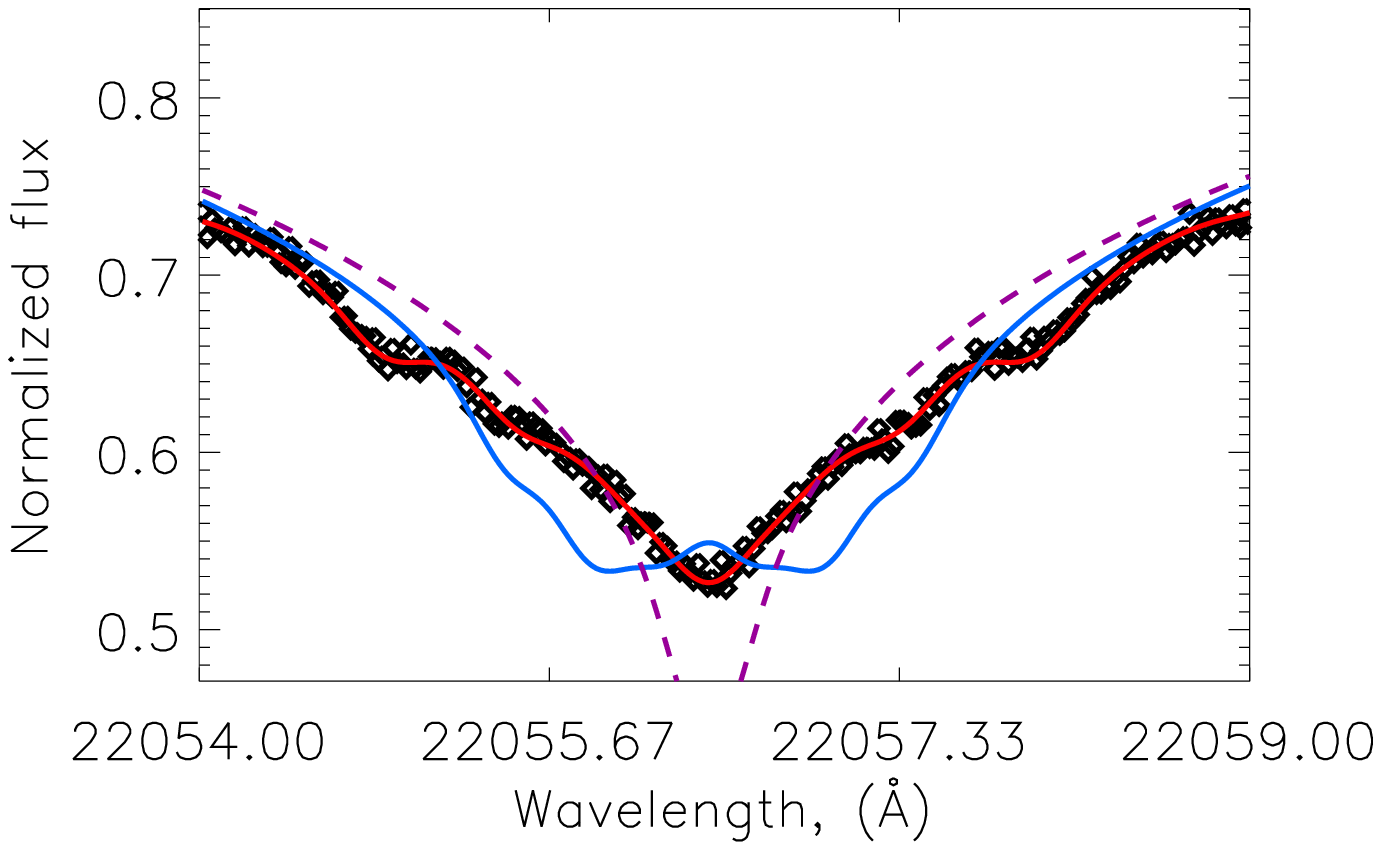}
\includegraphics[width=0.25\hsize]{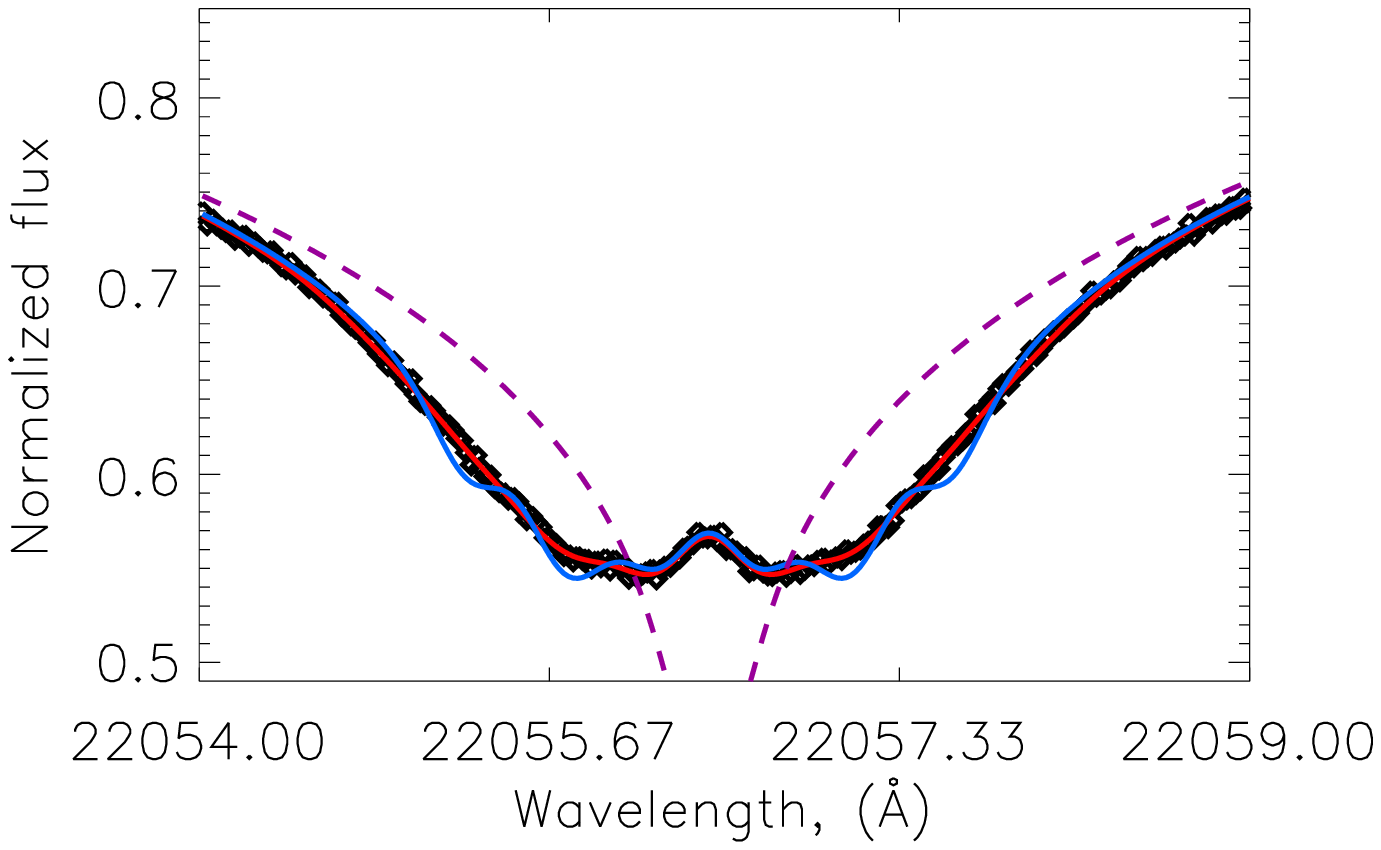}
\includegraphics[width=0.25\hsize]{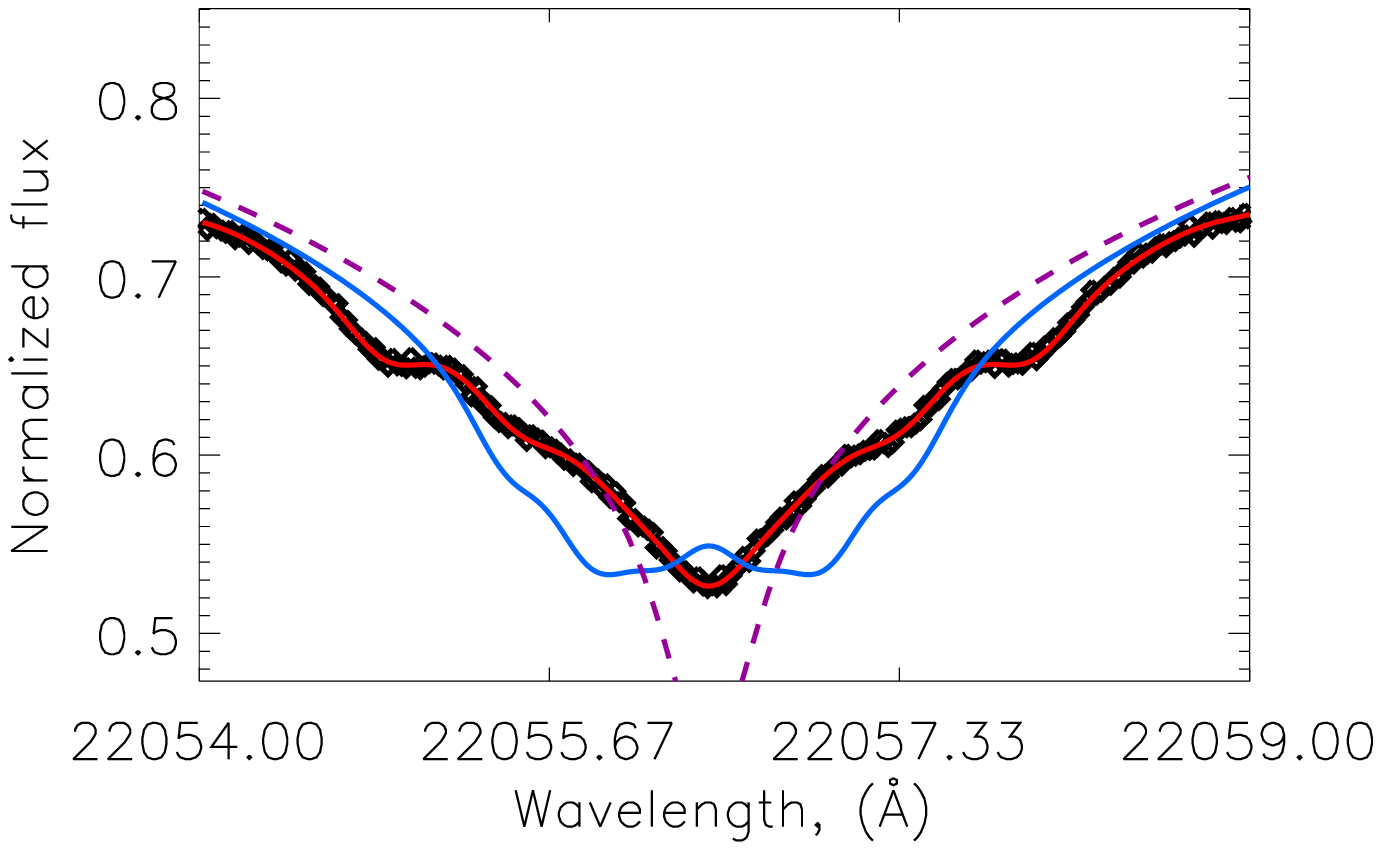}
}
\centerline{
\includegraphics[width=0.25\hsize]{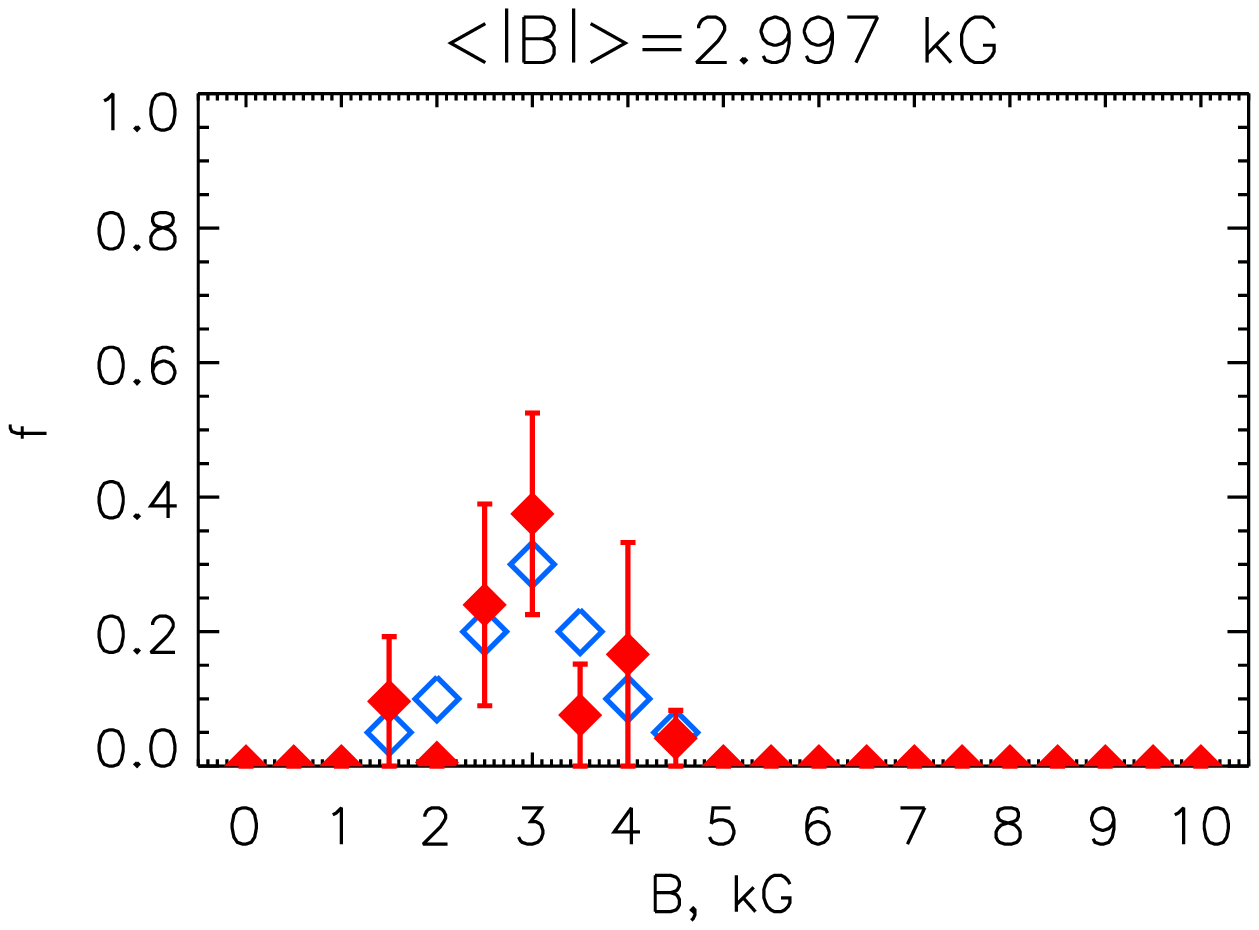}
\includegraphics[width=0.25\hsize]{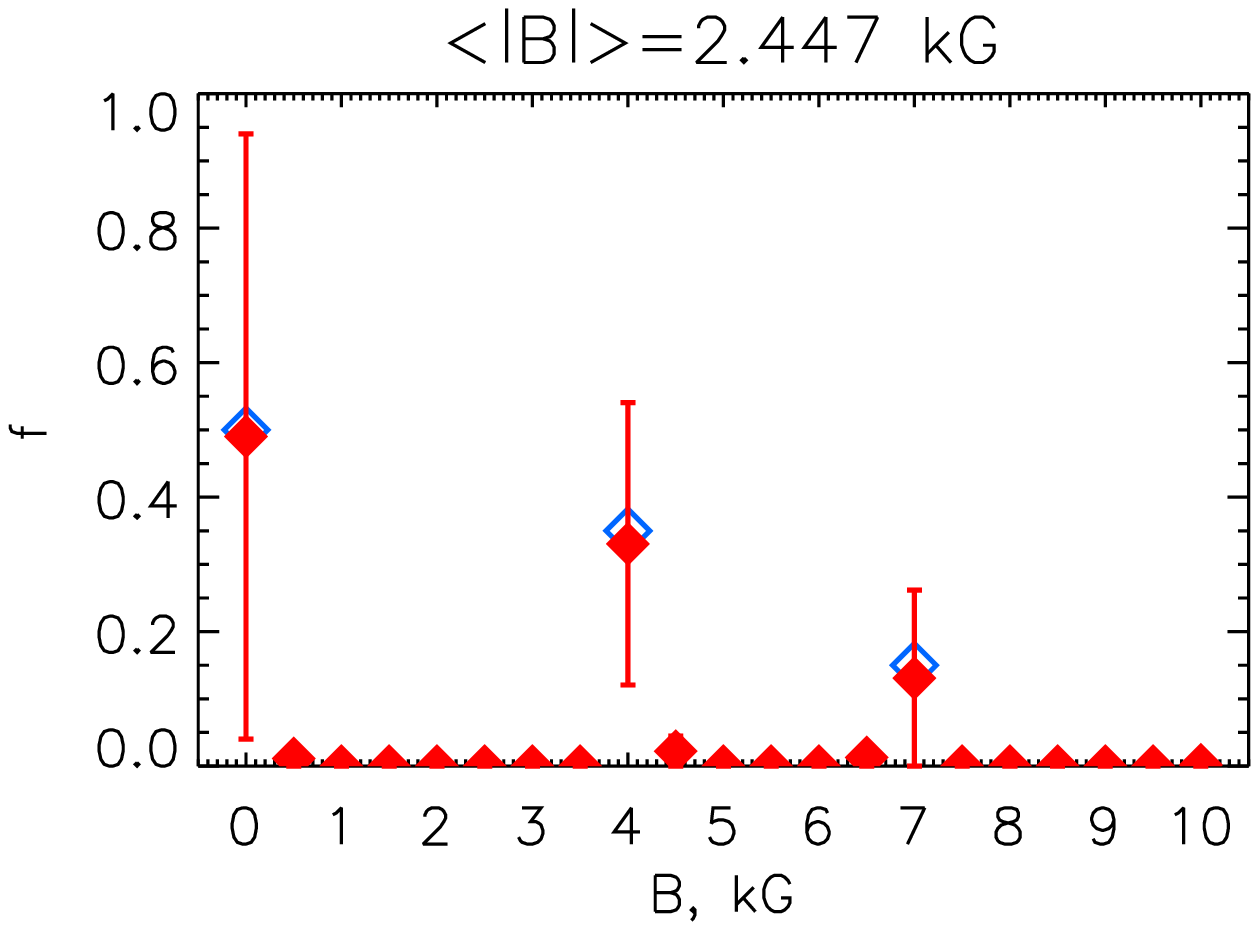}
\includegraphics[width=0.25\hsize]{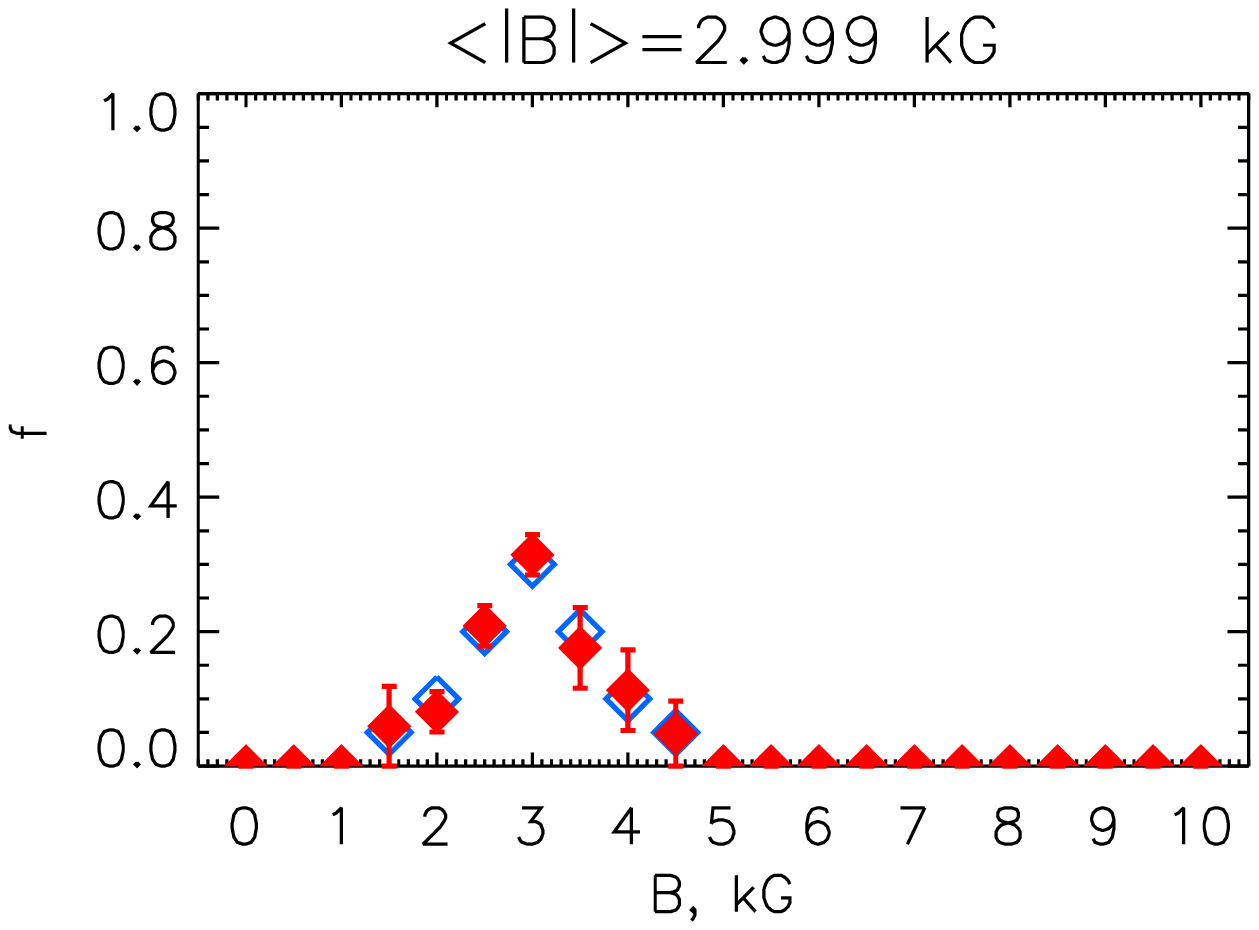}
\includegraphics[width=0.25\hsize]{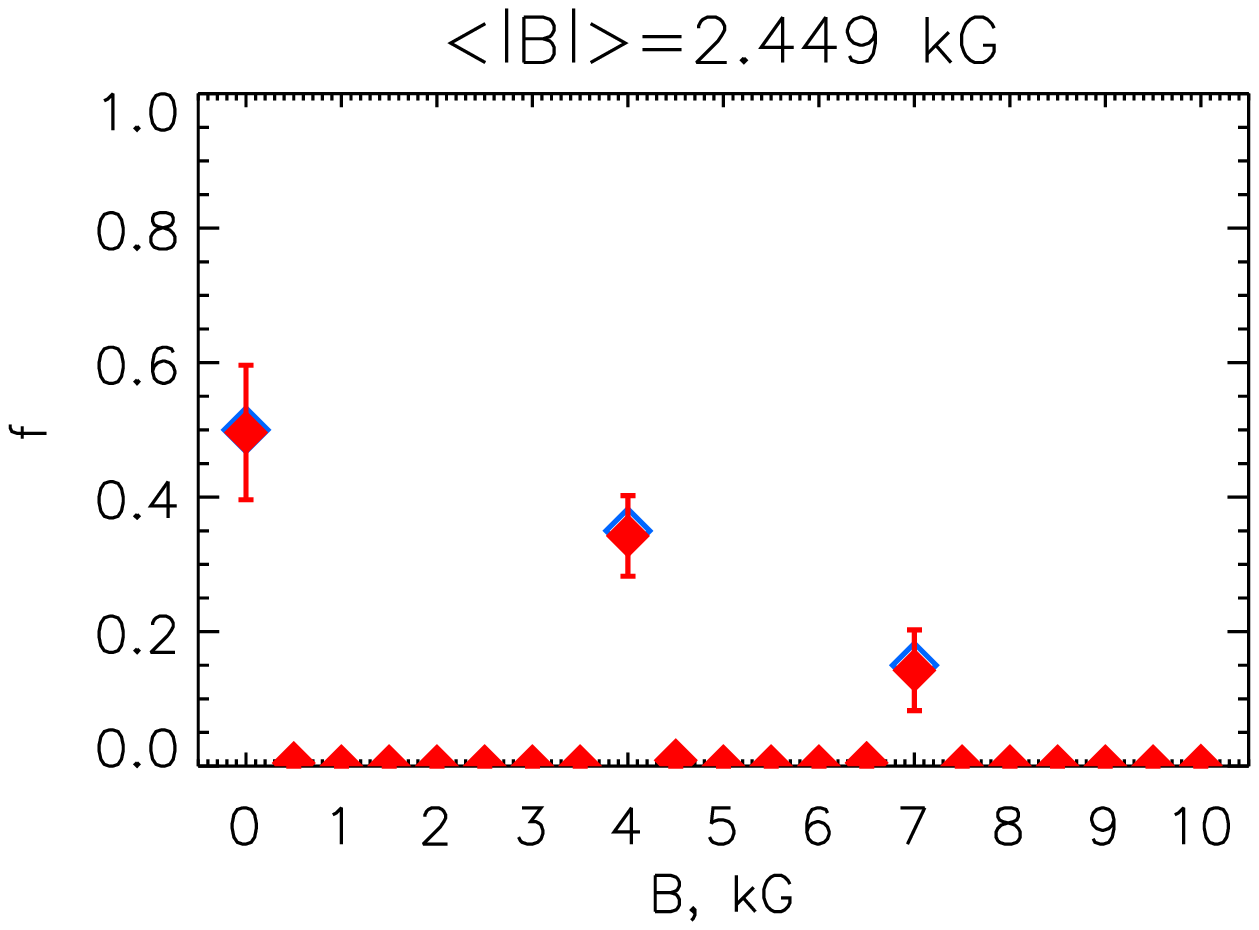}
}
\caption{Forward simulation of FeH and \ion{Na}{i} lines assuming two different distributions of surface magnetic fields.
Thick black symbols~--~simulated observed spectrum; violet dashed line~--~computation with zero magnetic field;
red line~--~computation with multi-component magnetic field shown on the bottom plots,
blue line~--~computations with homogeneous magnetic field  (i.e. $f=1$) 
of the same average intensity as multi-component magnetic field. Columns 1, 2  and 3, 4 correspond to the assumed
SNR of $100$ and $500$ respectively. Atmospheric parameters are the same for all simulations. Blue open diamonds on the bottom
plots show the true distribution of filling factors used to simulate observations.}
\label{fig:forward-fit}
\end{figure*}

In our investigation we employed the FeH line list of the Wing-Ford band ($F^4\,\Delta-X^4\,\Delta$ transitions{)}
and molecular constants taken from \citet{2003ApJ...594..651D}\footnote{http://bernath.uwaterloo.ca/FeH}.
Transition probabilities for some of these lines were corrected according to \citet{2010A&A...523A..58W}. 
Transition parameters for atomic lines were extracted from the VALD database \citep{1995A&AS..112..525P,1999A&AS..138..119K}.

To compute synthetic spectra of atomic and molecular lines in
the presence of a magnetic field, we employed the \synmast\ code \citep{2007pms..conf..109K}. The code
represents an improved version of the \synthmag\ code described by \citet{1999ASSL..243..515P}. 
It solves the polarized radiative transfer equation for a
given model atmosphere, atomic and molecular line lists and magnetic field
parameters. Model atmospheres are from MARCS the grid\footnote{http://marcs.astro.uu.se} \citep{2008A&A...486..951G}.

In order to analyze the magnetic field via the spectral synthesis it is necessary to know
the Land\'e g-factors of upper and lower levels of a particular transition.
In case of atomic lines the necessary information was extracted directly from VALD.
As shown by, e.g, \citet{2002A&A...385..701B} lines of FeH exhibit splitting 
that is in most cases intermediate between pure \hunda\ and \hundb\ and for which no analytic expression is possible.
Therefore, to compute g-factors we implement an approach described in \citet{2010A&A...523A..37S} that is based on 
numerical libraries from the MZL (Molecular Zeeman Library)
package originally written by B.~Leroy \citep{mzl}, and adopted by us for the particular case of FeH.

We attempt to measure the intensity and complexity of surface magnetic fields by carrying out
a detailed synthetic {spectral} fitting. 
The lines of FeH are selected in such a way that their Zeeman patterns are accurately reproduced
in the reference sunspot spectra for which both temperature and magnetic field are known independently from
the fit to atomic lines, as shown in \citet{2010A&A...523A..37S}. To fully characterize the magnetic field one needs to derive 
a) the surface averaged field modulus $\bs$ that affects the magnitude of the splitting between individual
Zeeman components of a given line, and b) the geometry of the magnetic field that influences the intensity and
shape of magnetically split lines. Note that a true and unique picture of the surface magnetic field
can be restored by surface imaging techniques like Zeeman Doppler Imaging which relies on the
{rotationally} modulated signals in polarized light or at least phase-resolved longitudinal magnetic field measurements with
subsequent modelling of spectroscopic lines.
Because only unpolarized Stokes~$I$ spectra are available to us, however, it is impossible to draw conclusions about
the geometry of the surface magnetic field in target stars. That is, no information about vector magnetic field
and positions of magnetic areas on stellar surfaces can be derived. Nevertheless, even Stokes~$I$ spectra alone contain
information about the magnetic field structure or what we call the complexity of the magnetic field,
i.e. the minimum number of magnetic field components required to fit the observed line profiles.
The {latter} depend on the geometry of the magnetic field
through the intensities of Zeeman $\pi$- and $\sigma^\pm$-components. 
This allows one to investigate how complex the magnetic field is compared to the
case when it can be described by a single component having a fixed strength.
For instance, a homogeneous magnetic field of a fixed strength 
and a configuration where the field is concentrated only in small surface areas (i.e. spots) 
would result in line profiles that look different.
This can then be described in {terms} of magnetic filling factors ${f_i}$ which are nothing but a measure of the
area on the stellar surface covered by a magnetic field of intensity ${\b_i}$. In case of a homogeneous field
($f=1$) the mean surface magnetic field is $\bs=\b$, while in case of more complex fields $\bs=\btimesfi$.
The aim of the present work is to measure the distribution of filling factors by fitting theoretical and observed
line profiles and to derive {the} minimum number of magnetic field components which provides 
{an acceptable} agreement between
theory and observation. This approach is similar to those used by, e.g., \citet{2000ASPC..198..371J} with a difference
of using many magnetically sensitive FeH lines instead of only a few atomic lines 
{as well as accurate treatment of thousands of blends in the spectrum synthesis code.}

The fitting procedure consists of the following steps. For each spectrum we apply a chi-square 
Levenberg-Marquardt minimization algorithm with filling factors $f_i$ as fit parameters. 
We consider $21$ filling factors which correspond {to magnetic fields}
ranging from $0$~kG to $10$~kG in steps of $0.5$~kG. We sequentially start from $2$ filling factors and then add new
ones one-by-one computing the corresponding $\chi^2$ and mean deviation $\sigma$ between observed and predicted spectra. 
The whole procedure is applied for different sets of atmospheric parameters: 
$\teff$, $\abn{Fe}$, and $\vsini$. We assume the same surface gravity, $\logg=5.0$, for all M-dwarfs. 
The effect of varying atmospheric parameters on the derived distributions of filling factors is also investigated.
This is an important
part of our work since little is known about the surface structure of active M-dwarfs: do they exhibit large cool or hot
areas similar to sunspots and plage regions observed in the Sun? How is the magnetic field correlated with these regions, etc.? 
Measurements of stellar magnetic fields from spectroscopy requires that atmospheric parameters are accurately known,
which is unfortunately not always the case because of well-known problems in matching spectra of M-dwarfs 
where errors in, e.g., effective temperatures and metallicities can be large.
This {raises} another goal of our investigation, namely to see the impact of choosing different atmospheric parameters
on the derived magnetic field and filling factors.
{Errors in atmospheric parameters are not the only uncertainties
that affect} the measurements of the magnetic fields. There are
other potentially important effects like, e.g., temperature inhomogenities of active regions with bright and
dark areas similar to those observed on the Sun which we ignore in the present study. A more in-depth analysis
would be needed to fully account for complex morphologies of {stellar} photospheres. We plan to
address these questions step-by-step {in future} calculations, assuming at present that the surface
temperatures of sample stars are homogenous and can be accurately derived from magnetically insensitive lines.

The initial values of $\teff$ and $\vsini$ for all targets were taken from RB07. 
The iron abundance was then adjusted to fit magnetically-insensitive FeH lines located at $\ll9914$~--~$9916$,
$\ll9925$~--~$9928$, $\lambda9945$, $\lambda9957$, and $\lambda9999$ regions.
{Because of the high densitiy} of FeH lines, {magnetically} sensitive FeH lines are in most cases found in blends
that contain many individual lines. In the present study the
FeH features in the following spectral intervals have been used for the magnetic field measurements: 
$\ll9903.3-9907.0$, $\ll9908.3-9913.8$, $\ll9919.0-9921.6$, $\ll9928.0-9929.0$, $\ll9931.0-9933.5$,
$\ll9938.5-9939.3$, $\ll9941.5-9942.6$, $\ll9945.5-9947.8$, $\ll9956.1-9958.80$, $\lambda9961.3$. These intervals have been used
in all calculations presented below and are same for all stars studed.
As shown by 3D hydrodynamical simulations  carried out by
\citet{2009A&A...508.1429W}, the velocity fields in atmospheres of M-dwarfs with temperatures $\teff<3500$~K
are well below $1$~\kms. This has only weak impact on line profiles leaving Zeeman effect and rotation
to be the dominating broadening mechanisms. {Therefore we assumed zero micro- and macroturbulent velosities
in all calculations.}

To test whether our fitting method is capable of restoring the original solution under conditions when all atmospheric
parameters are accurately known, we performed forward calculations of selected FeH and one \ion{Na}{i} $\lambda22056$ line 
using two distributions of filling factors: one with only three components of 
$0$~kG ($f=0.5$), $4$~kG ($f=0.35$), and $7$~kG ($f=0.15$),
and another with seven components between $1.5$~kG and $4.5$~kG with a smooth Gaussian-like distribution of filling factors.
The simulated ``observed'' spectra were computed for $\teff=3400$~K, $\vsini=0$~\kms, and convolved with $R=100\,000$. 
We find that the code always finds a unique distribution of filling factors.
However, because of implicit normalization
requirement $\sum f_i=1$ the filling factors are highly correlated variables. Therefore, no meaningful estimate
of error bars is possible based on the covariance matrix provided by the minimization technique.
Instead, we provide estimates of a significance of every magnetic field component.
For each magnetic field distribution the following procedure is applied:
\begin{enumerate}
\item
\label{item:obs}
{Noise is added to the simulated spectra. The noise amplitude ($1\sigma$) ranges  from 2\% to 0.001\%}
which covers the SNR of spectra of target stars in our sample.
\item
\label{item:ref}
For each simulated spectrum with a given noise level a distribution of magnetic fields and their filling factors are derived.
The spectra computed from these distributions of filling factors are {then taken as a reference to compute
the significance} for each magnetic field component.
\item
\label{item:bi}
For each magnetic field component $f_i>0$ its filling factor is decreased by a chosen amount $\Delta f$.
At the same time, the filling factor of the field component left {of} it (e.g., $f_{i-1}$) is increased  by the same amount 
so that the condition $\sum f_i=1$ always holds.
\item
{The adjusted distribution of filling factors obtained at the previous step is then used to compute a synthetic spectrum
and its deviation from the reference spectrum computed with best fit distribution in step~\ref{item:ref}. }
\item
The adjustment of a given filling factor is repeated until the maximum deviation from the reference spectrum 
at any fit point is larger than {the} $1\sigma$ error which was used to generate 
the simulated spectrum on step~\ref{item:obs}.
\end{enumerate}

The described procedure allows us to estimate a significance of
a particular magnetic field component relative to its closest neighbor and to verify whether
there is another solution possible within {a specified SNR of the} simulated observations.
For the sake of consistency {we will call these significane intervals error bars}, though one should
remember that they do not represent statistical error bars provided by {the} minimization procedure.
{Despite its} simplicity, the proposed method provides  
information about the uniqueness of solutions for filling factors and thus gives a robust
overview of field distributions in individual stars. It also has an advantage of reducing
computation time considereably compared to, e.g., direct Monte-Carlo simulations which are
preferred but currently very challenging to run because of complex spectra of FeH lines.

It is clear that the solution and corresponding error bars are strongly dependent upon SNR of observations.
As an example, Fig.~\ref{fig:forward-fit} shows results of two simulations with different distributions of filling factors and
SNR values of $100$ and $500$. The simulations were done assuming the magnetic field model with a dominating radial
field component (see below).
Two main conclusions can be drawn. First, the lowest SNR needed to restore the shape of the filling factors distribution is SNR$>300$ 
and the original amplitudes are restored with SNR$\approx500$, {as shown in
the last two columns} of Fig.~\ref{fig:forward-fit}. With SNR$=100$ it is
challenging to restore a complex distribution of magnetic fields similar to that shown on, e.g., lower left plots of Fig.~\ref{fig:forward-fit}
where many components are localized in a particular range of field strength. However, distributions where only a few field components are present
and well separated from each other are accurately reconstructed, although their significance is small. In other words, the lower
plot on the second column of Fig.~\ref{fig:forward-fit} shows that, e.g., {the component of $7$~kG strength can be replaced
by a component with a $6.5$~kG strength and still provide a similar quality of the fit to the line profiles within the error bars
of the observations. The only difference is that the $7$~kG component provides a smaller deviation between the observed and synthetic spectra. 
A second thing to note is that the restored mean magnetic field $\bs=\btimesfi$ is always equal to the original value even 
if the reconstructed amplitudes of filling factors deviate.}

Because only unpolarized Stokes $I$ spectra are analyzed in this paper, no information about vector magnetic field
could be retrieved. Therefore, we made use of a simple magnetic field model described with a small number of free
parameters. The field is homogeneous in the
stellar reference frame and is specified by the three vector components: radial 
$\br$, meridional $\bm$, and azimuthal $\ba$, reckoned in the spherical coordinate
system whose polar axis coincides with the line of sight. Then, the two field
components relevant for calculating the Stokes $I$ profiles are given by
\begin{equation}
B_{\rm l} = \br\cos{\theta} - \ba\sin{\theta}
\end{equation}
for the line of sight component and
\begin{equation}
B_{\rm t} = \left[ (\br\sin{\theta} + \ba\cos{\theta})^2 + \bm^2 \right]^{1/2}
\end{equation}
for the transverse component. In practice it is sufficient to adjust $\br$ and
$\bm$, keeping $\ba$ zero.

This approximation of the magnetic field structure is undoubtedly simplistic
and unsuitable for describing phase-dependent four Stokes-parameter profiles
of magnetic M~dwarfs. Nevertheless, as {shown} by the previous studies of 
Zeeman-sensitive lines in cool stars \citep{1996ApJ...459L..95J,2007ApJ...664..975J},
it is sufficient for modelling unpolarized spectra of M~dwarfs and T~Tauri stars
with strong fields.

In the following we assume that the stellar surface is covered by a dominating 
radial (RC model hereafter, $\bm=\ba=0$)
or meridional (MC model hereafter, $\br=\ba=0$) field component and perform
respective calculations for each target star.

\section{Testing molecular and atomic diagnostics}

\begin{figure*}
\centerline{
\includegraphics[width=0.33\hsize]{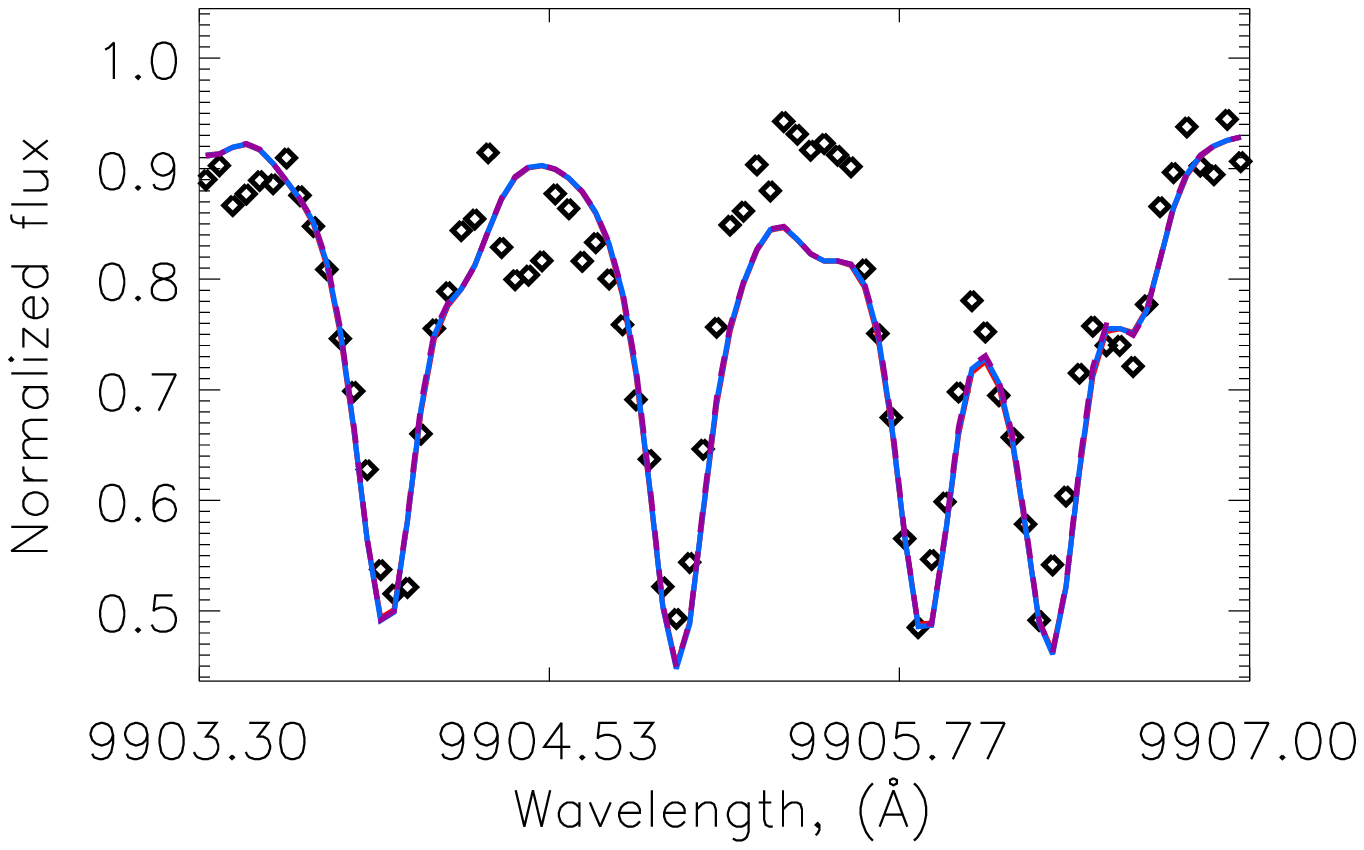}
\includegraphics[width=0.33\hsize]{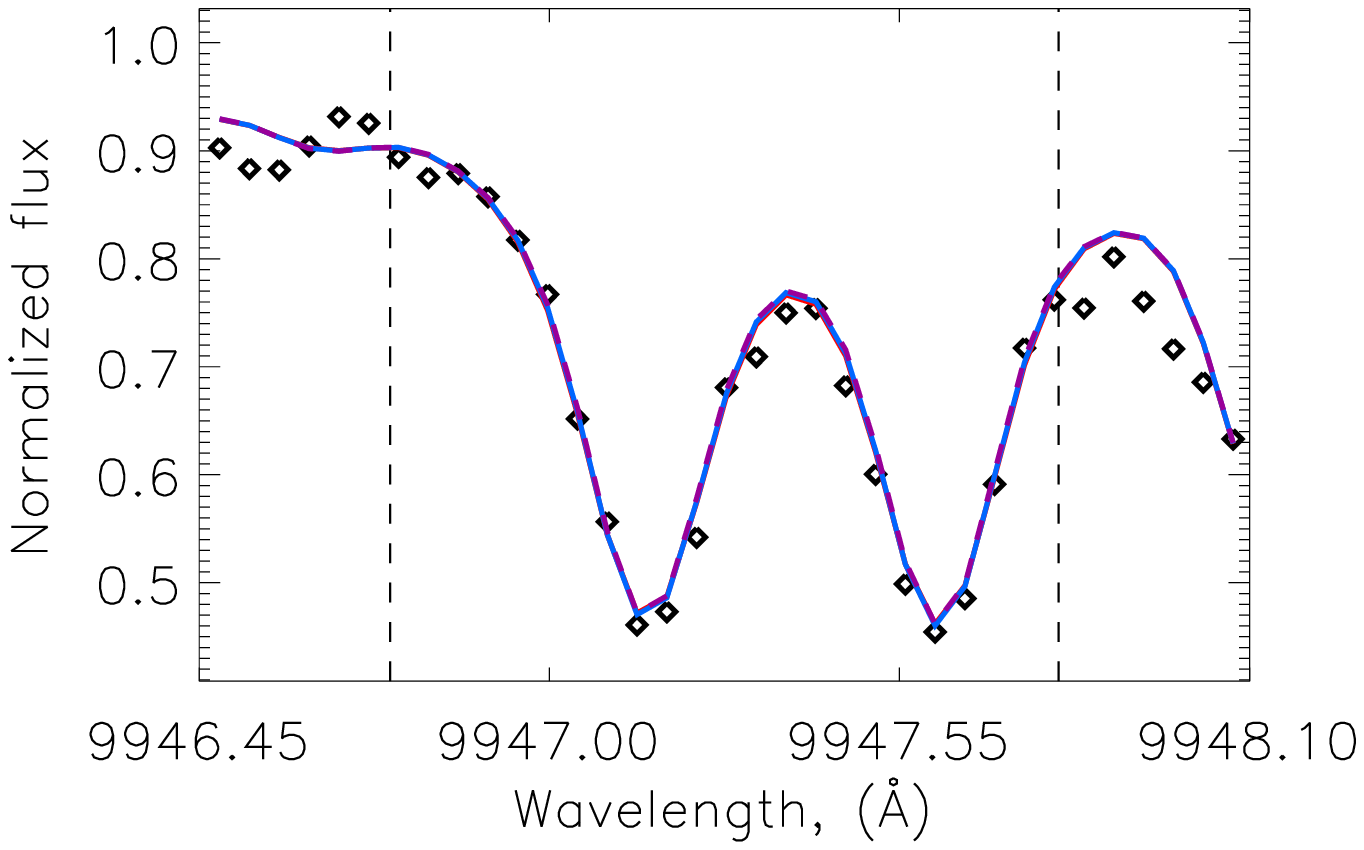}
\includegraphics[width=0.33\hsize]{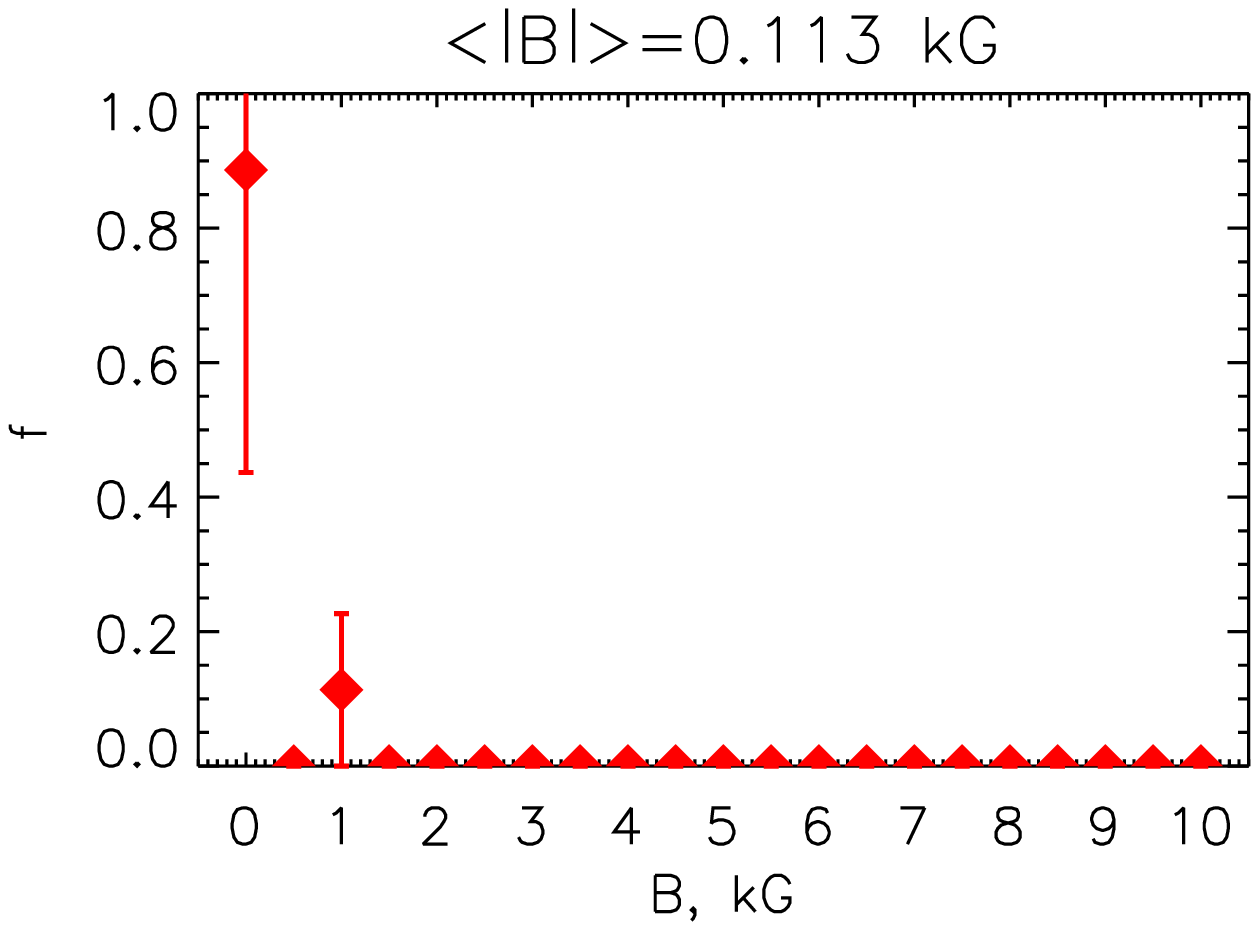}
}
\centerline{
\includegraphics[width=0.33\hsize]{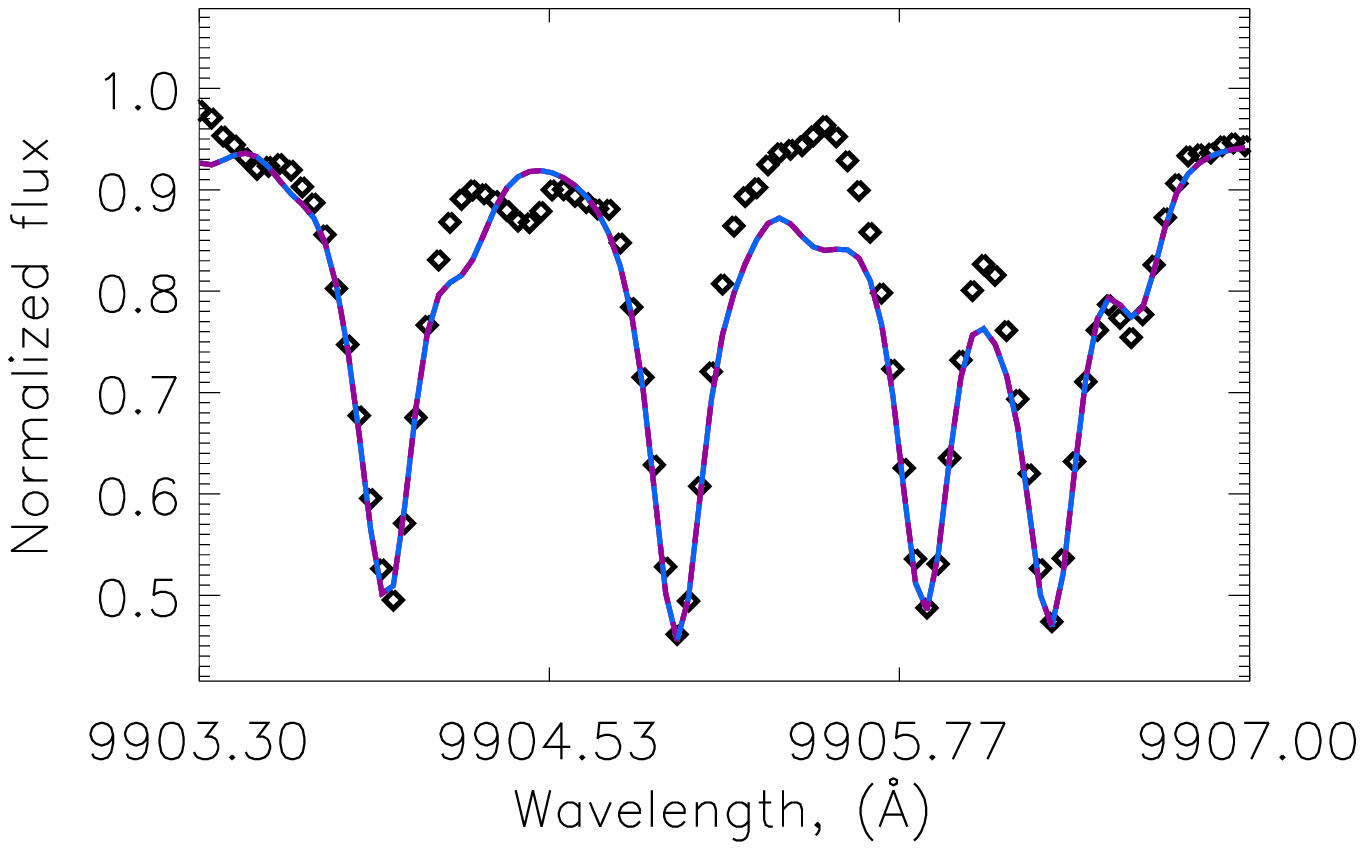}
\includegraphics[width=0.33\hsize]{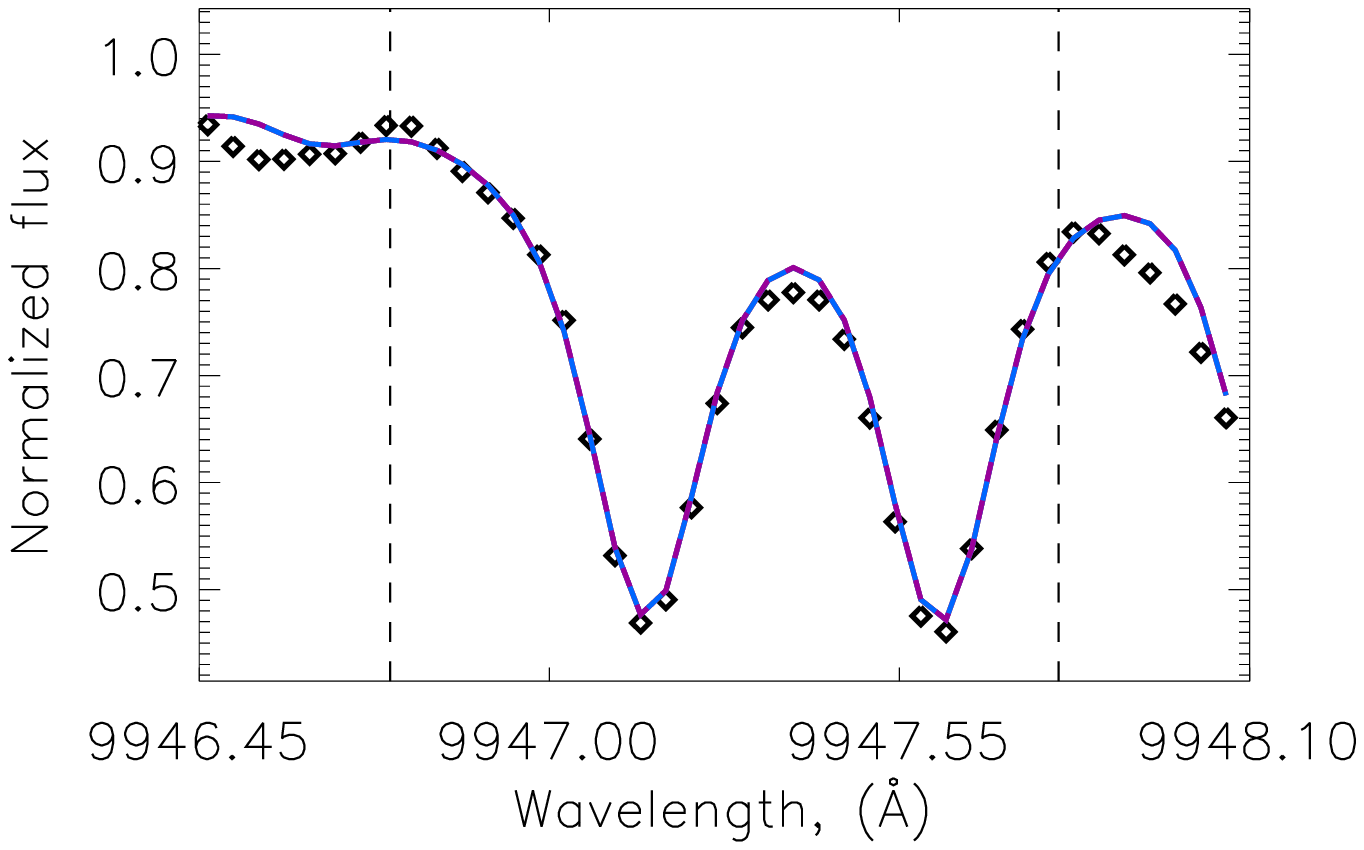}
\includegraphics[width=0.33\hsize]{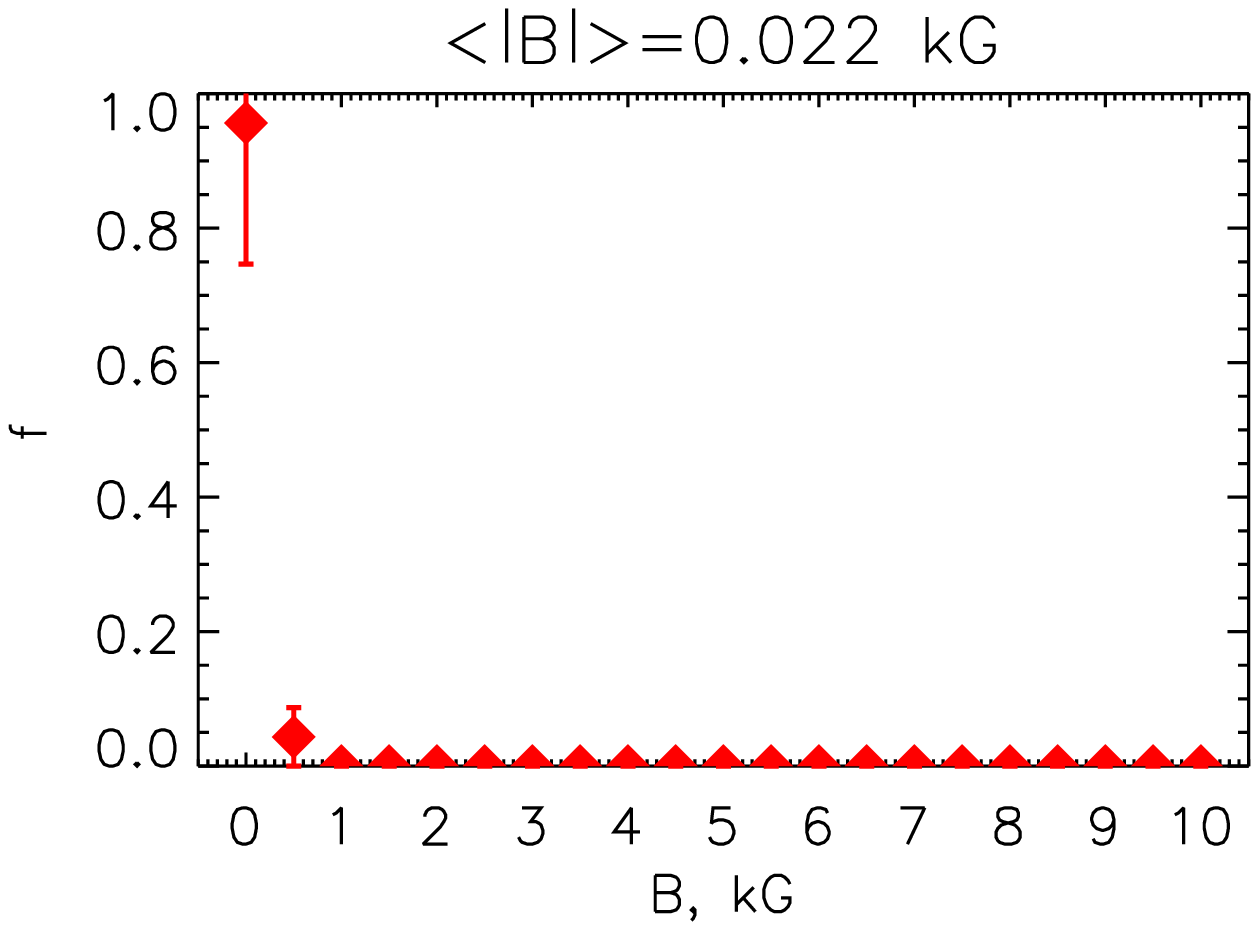}
}
\centerline{
\includegraphics[width=0.33\hsize]{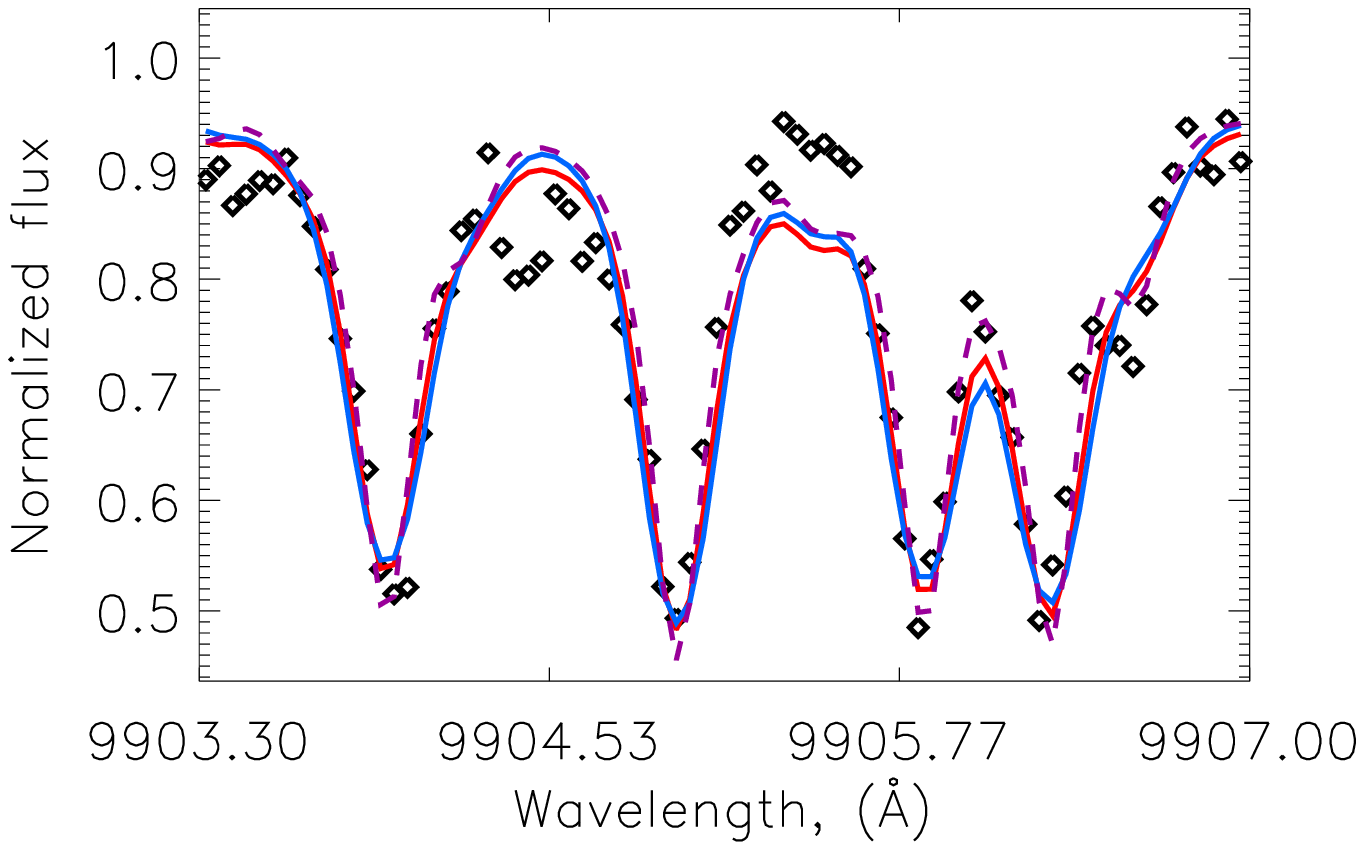}
\includegraphics[width=0.33\hsize]{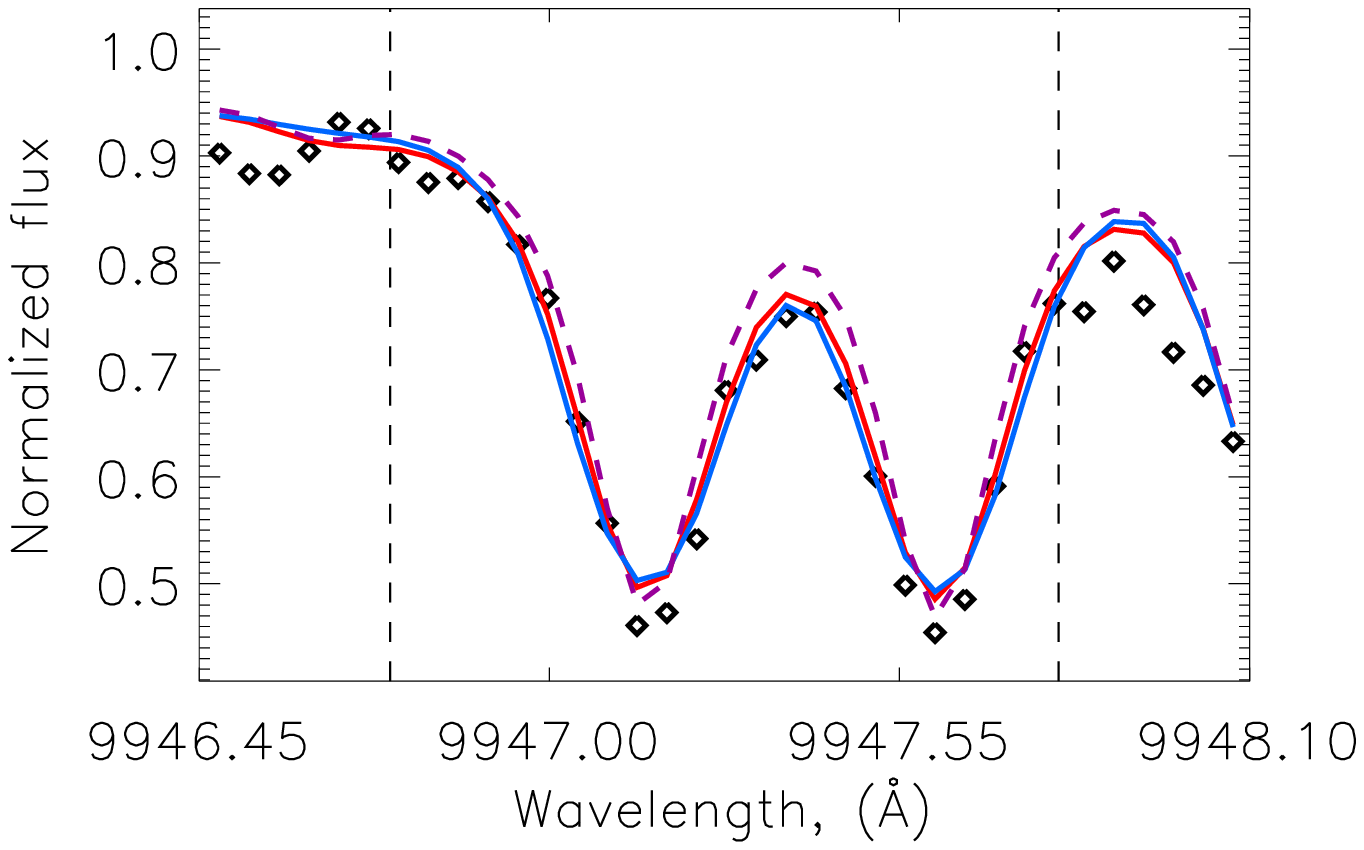}
\includegraphics[width=0.33\hsize]{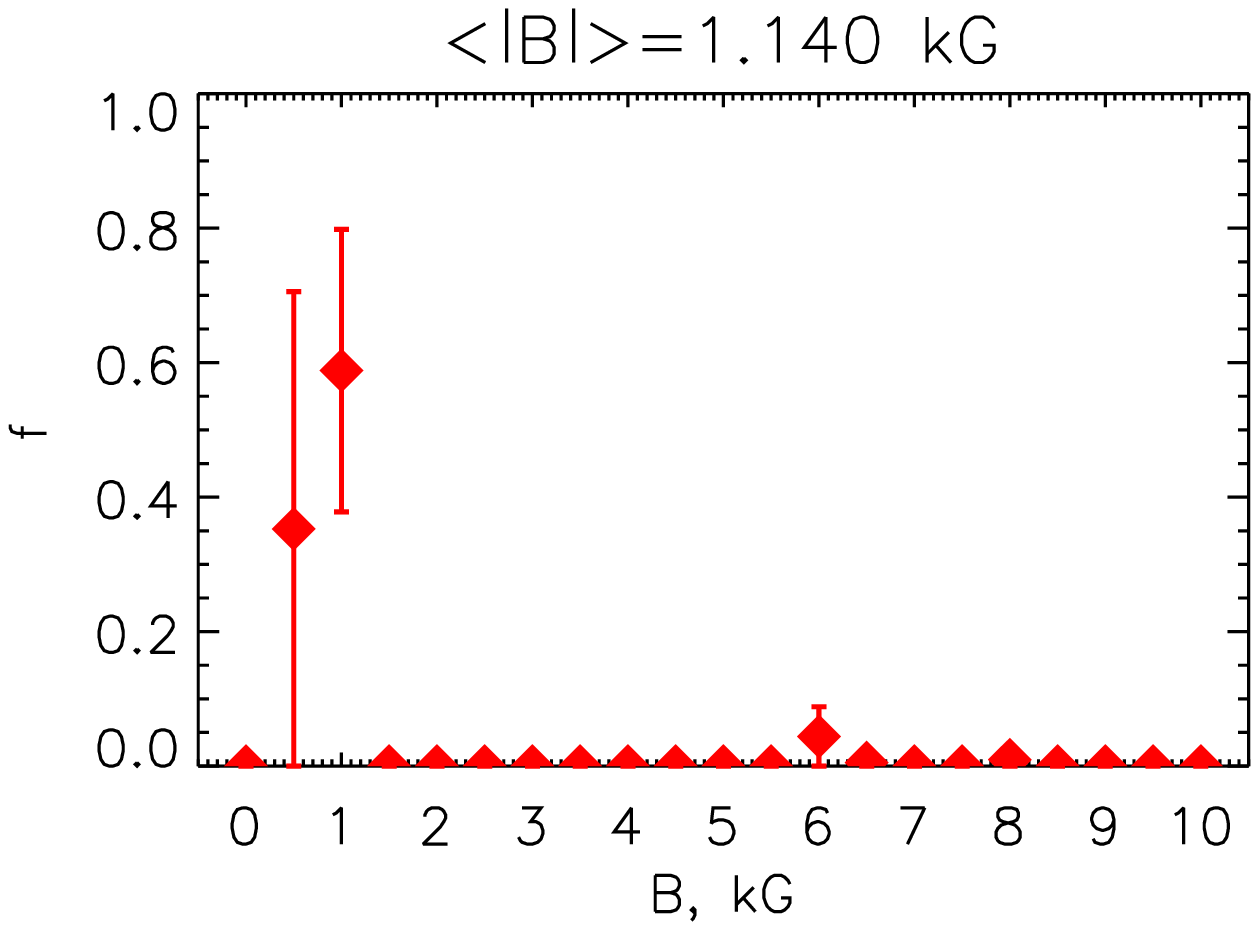}
}
\caption{Examples of theoretical fit to selected FeH lines and resulting distributions of filling factors for two spectra of
GJ~1002 taken in two different observing runs with CRIRES. Top panel~--~run 385.D-0273,  $\teff=3100$~K, $\abn{Fe}=-4.37$, $\vsini=2.43$~\kms.
Middle panel~--~ run 079.D-0357, $\teff=3100$~K, $\vsini=1.85$~\kms, $\abn{Fe}=-4.47$. Bottom panel~--~same as top panel but adopting atmospheric
parameters from middle panel{: implementation of these parameters results in a spurious detection of the magnetic field
as illustrated on the corresponding plot of filling factors (see text).}
Violet dashed line~--~computation with zero magnetic field, red line~--~computation with multi-component 
magnetic field shown on the right hand side, blue line~--~computations with homogeneous magnetic field  (i.e. $f=1$) 
of the same average intensity as multi-component magnetic field. Vertical dashed lines mark the edges of the actual fit region.}
\label{fig:gj1002-feh}
\end{figure*}

Our observations cover the range of FeH lines at $1$~\mum\ and strong \ion{Na}{i} lines at $2.2$~\mum. Presumably, the magnetic
fields derived independently from these two sets of lines must agree, if the atmospheric model structure and/or
physical conditions in the line formation depths are accurately predicted by model atmospheres and spectrum synthesis codes.
Sadly, we find this is not always the case: one needs noticeably different temperatures
to fit FeH and Na lines. In particular, wide wings of the later demand a lower $\teff$ compared to FeH lines. As a result,
magnetic fields derived from best-fit spectra assuming fixed $\teff$ deviate from one another. 
A good illustration can be seen from the analysis of a non-magnetic M$4.5$ dwarf GJ~1002. 
This star shows no detectable X-ray emission and thus the corresponding surface magnetic field
of the star is expected to be nearly zero or absent.
The atmospheric parameters were derived from the fit to magnetically insensitive FeH lines and are consistent
with estimates from previous independent studies: $\teff=3100$~K, $\abn{Fe}=-4.37$, $\vsini=2.4$~\kms. 
For instance, \cite{2010A&A...523A..37S} used  $\teff=3100$~K, $\abn{Fe}=-4.37$, $\vsini=2.5$~\kms. 
Note that values of $\abn{Fe}$ and $\vsini$ depend strongly
on {the} effective temperature adopted, but also on the continuum normalization and data quality. For instance
using the same spectra from \cite{2010A&A...523A..37S} but a slightly different continuum normalization procedure yields
$\vsini=1.85$~\kms, $\abn{Fe}=-4.47$. The magnetic field derived from the former set of parameters is only about $100$~G
(which we consider as no detection within observed error bars,
because such a small field causes very insignificant changes in theoretical line profiles
compared to precise zero field calculation), while it is zero if the later set of parameters was used.

Examples of theoretical fits to a few FeH lines assuming {the} RC model of the magnetic field and resulting
distributions of filling factors are illustrated on the first two panels of Fig.~\ref{fig:gj1002-feh}. 
The error bars {for the} filling factors were computed as described in the previous section.
Note that different combinations of
atmospheric parameters are allowed as long as they fit the observed magnetically insensitive lines. 
For instance, one of the unknown parameters in fitting FeH lines is the  constant of van der Waals broadening $\gammaw$. 
It was shown in \citet{2010A&A...523A..37S} that, in order to fit FeH lines in GJ~1002 with a given $\teff=3100$~K and $\vsini=2.5$~\kms, 
the value of $\gammaw$  must be increased by an enhancement factor of $f(\gammaw)=3.5$ compared to a classical value given in \citet{gray}.
Note that different combinations of atmospheric parameters such as iron abundance, $\gammaw$, and $\vsini$ 
could provide the same accurate fit to the observed spectra: a change in one of these parameter
could be compensated (to a certain extent) by an opposite change in the others.
For example, decreasing $f(\gammaw)=1.75$ would require $\abn{Fe}=4.30$,
$\vsini\approx2.0$ for the spectra of GJ~1002 from our run 385.D-0273. The bottom line of this example is that small 
inaccuracies in atmospheric parameters,
although they might be small when comparing theoretical fits to individual lines, might result in {a} spurious detection of  magnetic fields,
in particular for multi-component solutions.
An illustrative example is shown {in} the bottom panel of Fig.~\ref{fig:gj1002-feh} where we apply atmospheric parameters derived from the 
spectra of the two CRIRES runs. 
{In both these cases we detect nearly zero magnetic fields ($0.113$~kG and $0.022$~kG respectively)
as long as we apply individual atmospheric parameters derived from the fit to the magnetic insensitive lines
in the spectra of each run, but detect a spurious $\bs=1.14$~kG field when applying atmospheric parameters
from one run to fit the spectrum of another. This is a purely numerical artifact because in this particular case
the fitting algorithm increases the Zeeman broadening to compensate for the corresponding differences 
in $\vsini$ and $\abn{Fe}$ between the runs. Ideally the atmospheric parameters derived from the two
spectra of the same star must be the same. In reality, however, this is not the case mainly because of differences
in SNR and continuum normalization that affect depths of spectroscopic features.
This exercise highlights the importance of using magnetically insensitive lines for the determination
of atmospheric parameters that might need to be adjusted once new data, improved line lists, model atmospheres, etc.
are being used. Following this procedure helps to avoid false detections of magnetic fields.}

The main problem in using a multi-component approach for the magnetic field measurements is its sensitivity to the line shapes
that are largely affected by data processing algorithms and inaccuracies in atomic/molecular parameters. In the case of FeH lines,
no individual Zeeman components can be seen because of the large number of individual $\pi$ and $\sigma^\pm$ components relevant for every single
transition, which are smeared out and result only in line broadening. This is not the case for all FeH lines as will be described below,
but {is true for smaller} fields when Zeeman components of nearby strong lines do not overlap. 
Such broadening can easily be mimicked by an appropriate choice of the rotational
velocity. This degeneracy can be removed by the use of magnetically insensitive lines.

\begin{figure*}
\centerline{
\includegraphics[width=0.33\hsize]{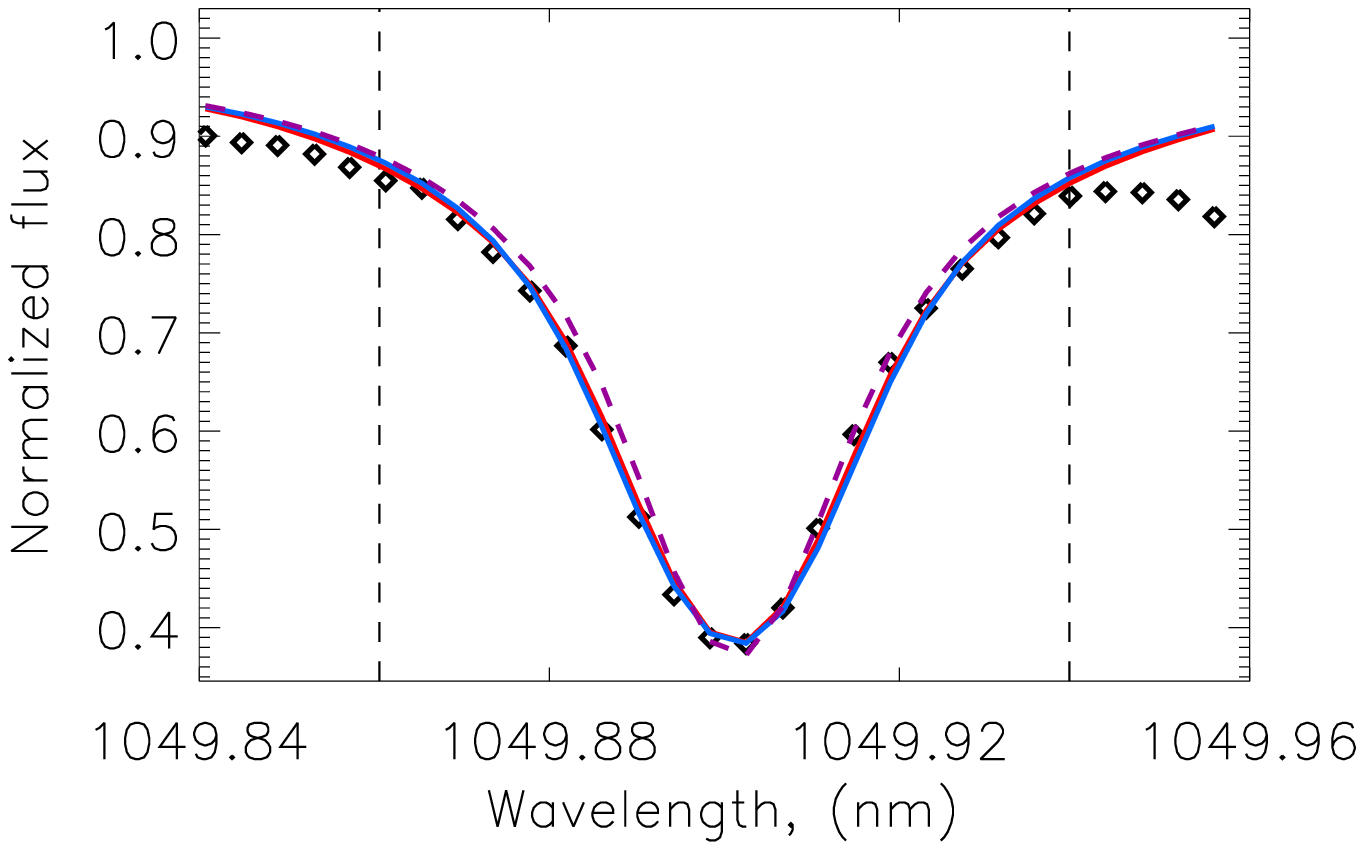}
\includegraphics[width=0.33\hsize]{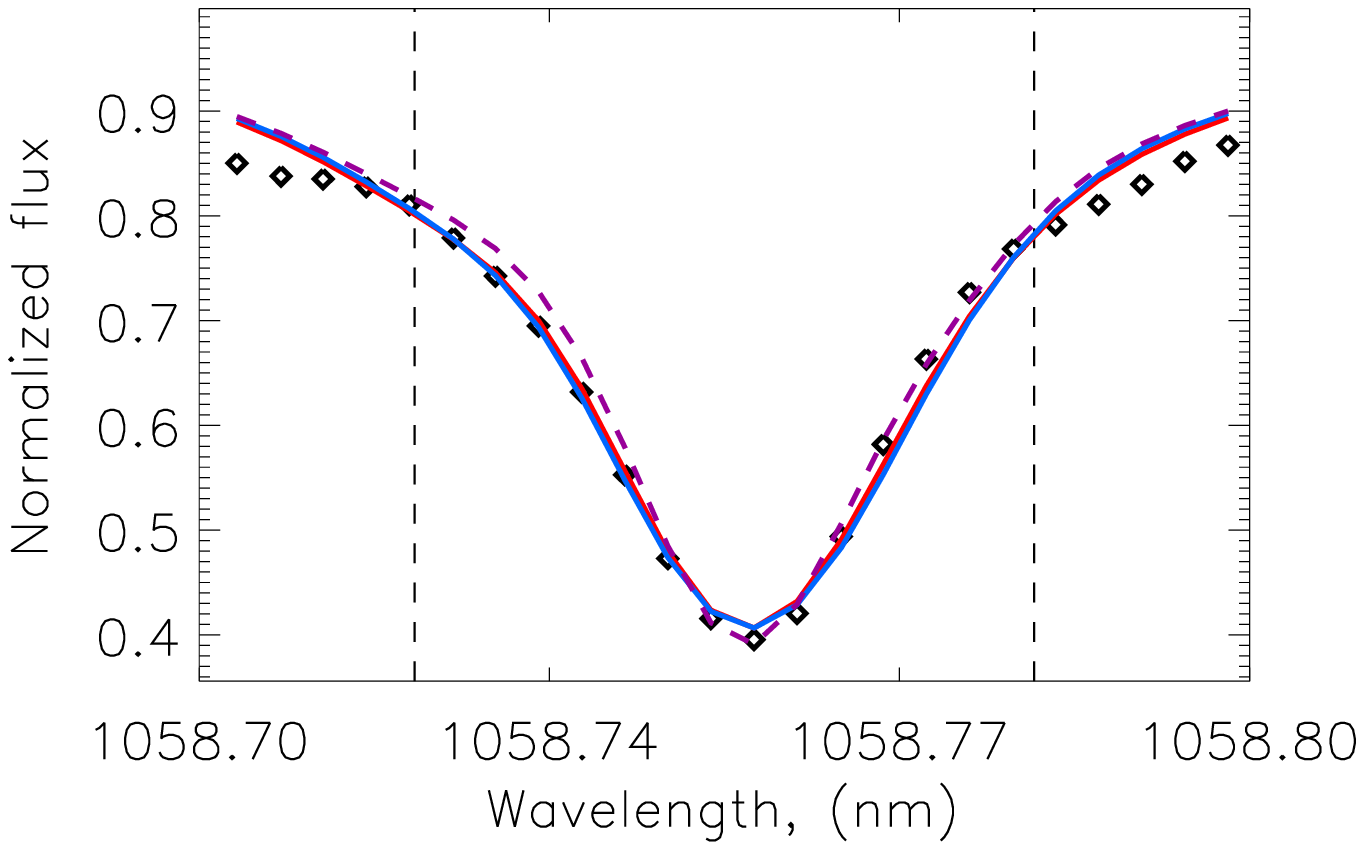}
\includegraphics[width=0.33\hsize]{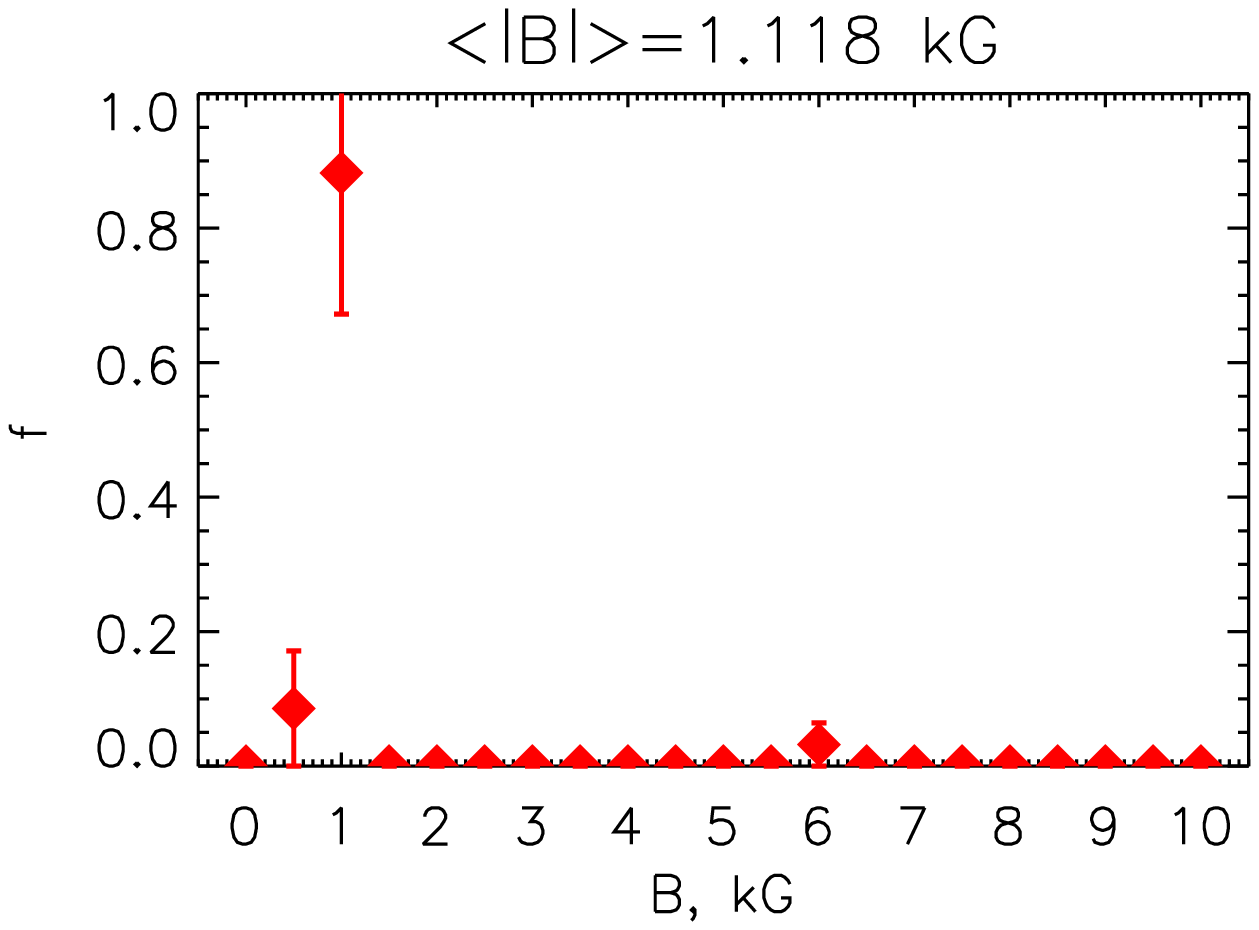}
}
\centerline{
\includegraphics[width=0.33\hsize]{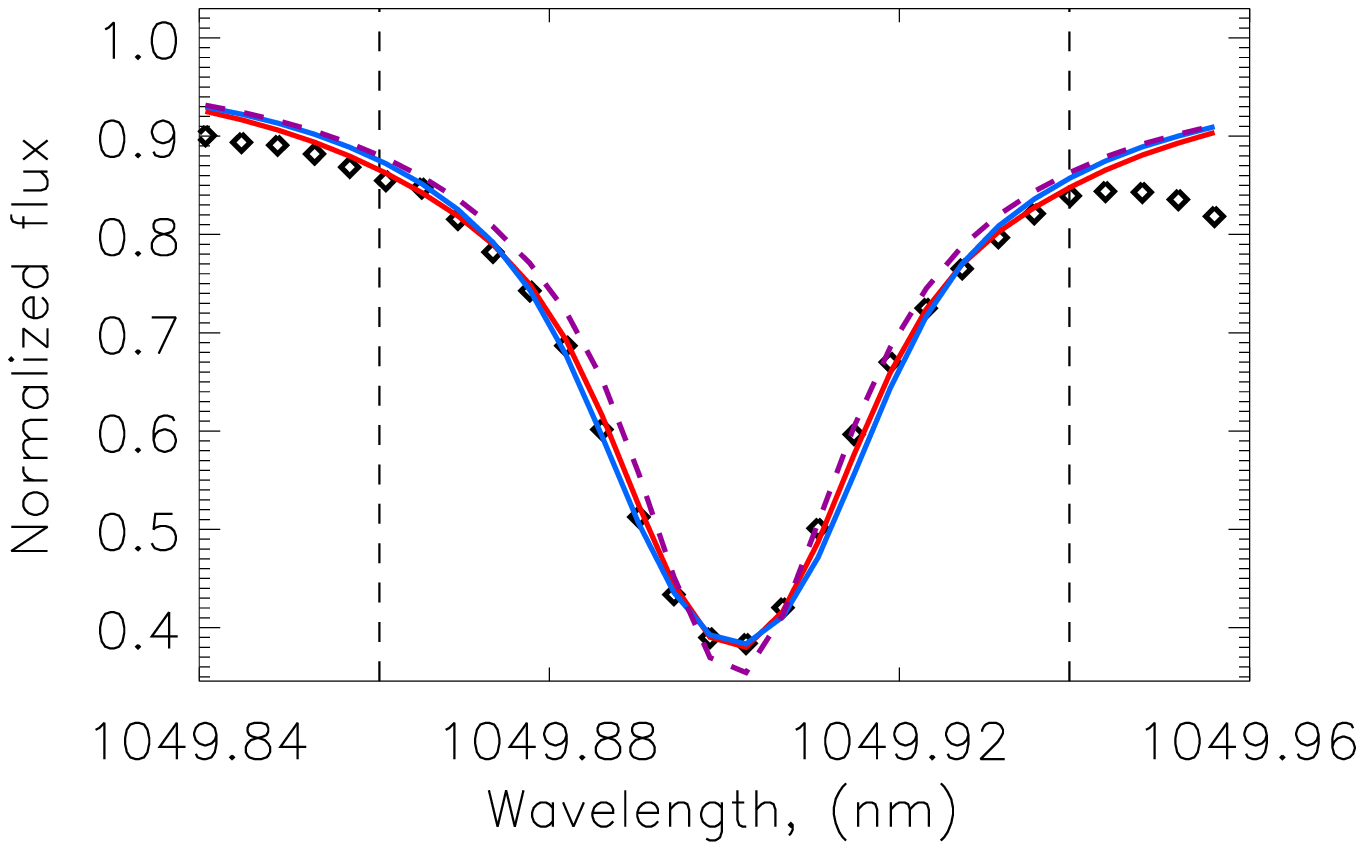}
\includegraphics[width=0.33\hsize]{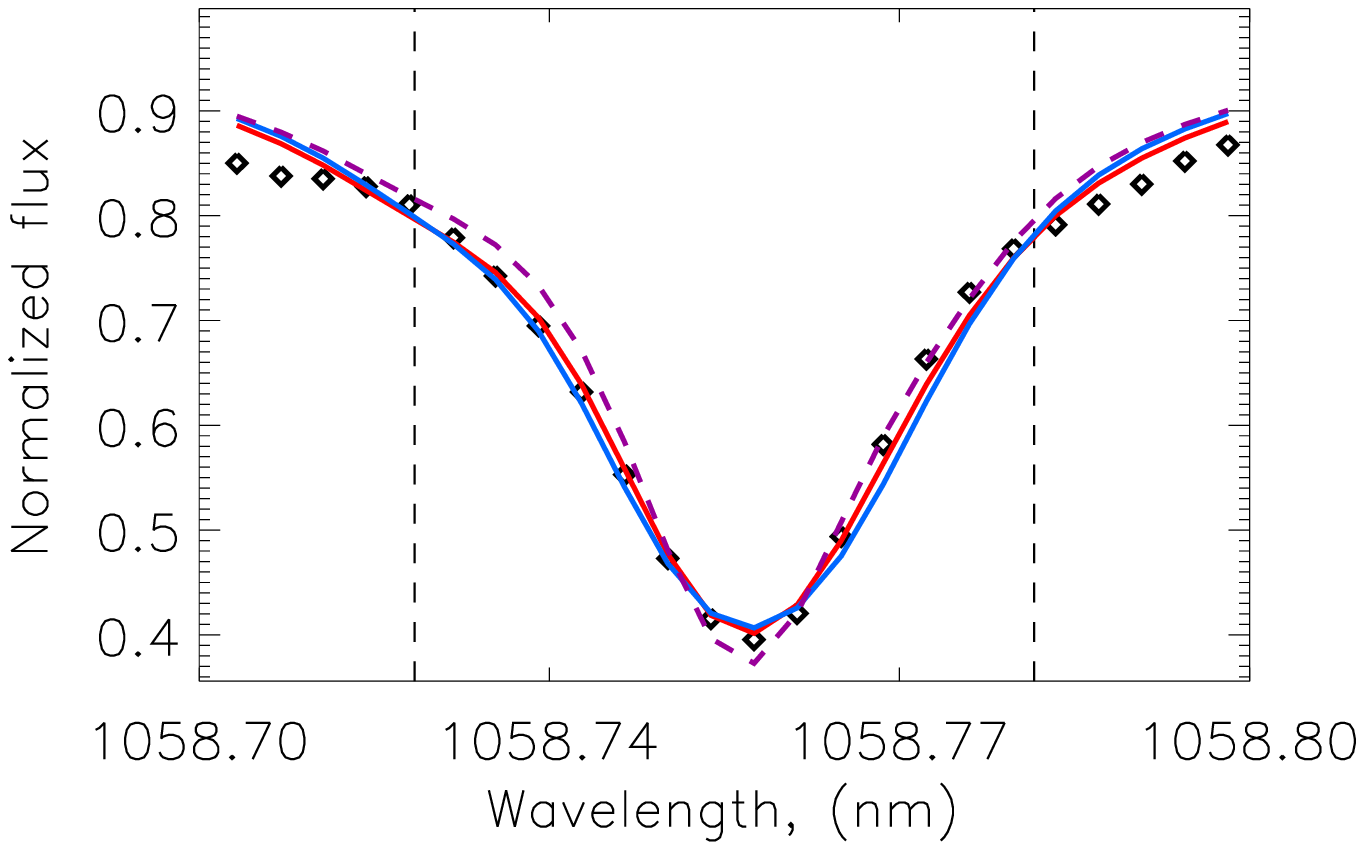}
\includegraphics[width=0.33\hsize]{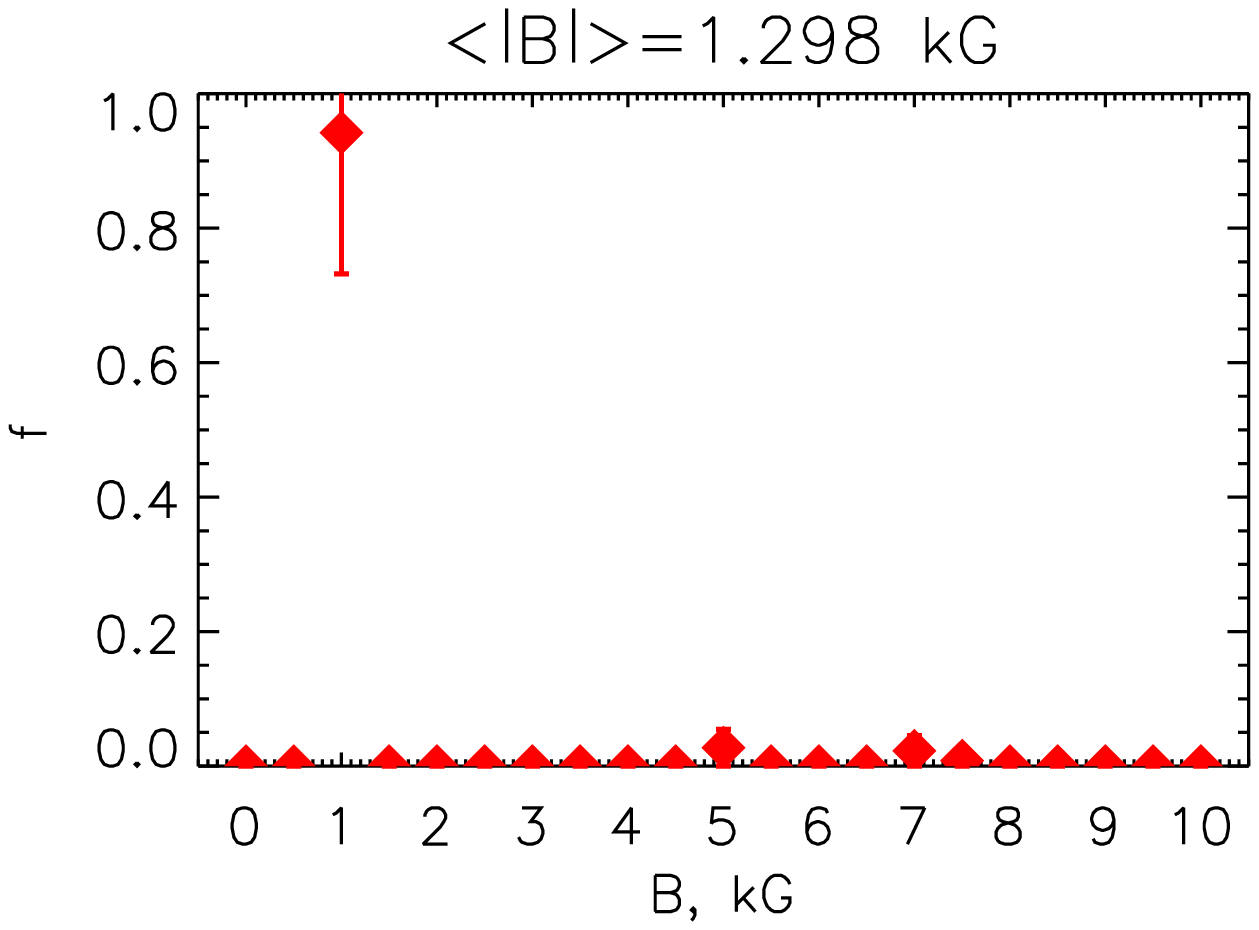}
}
\centerline{
\includegraphics[width=0.33\hsize]{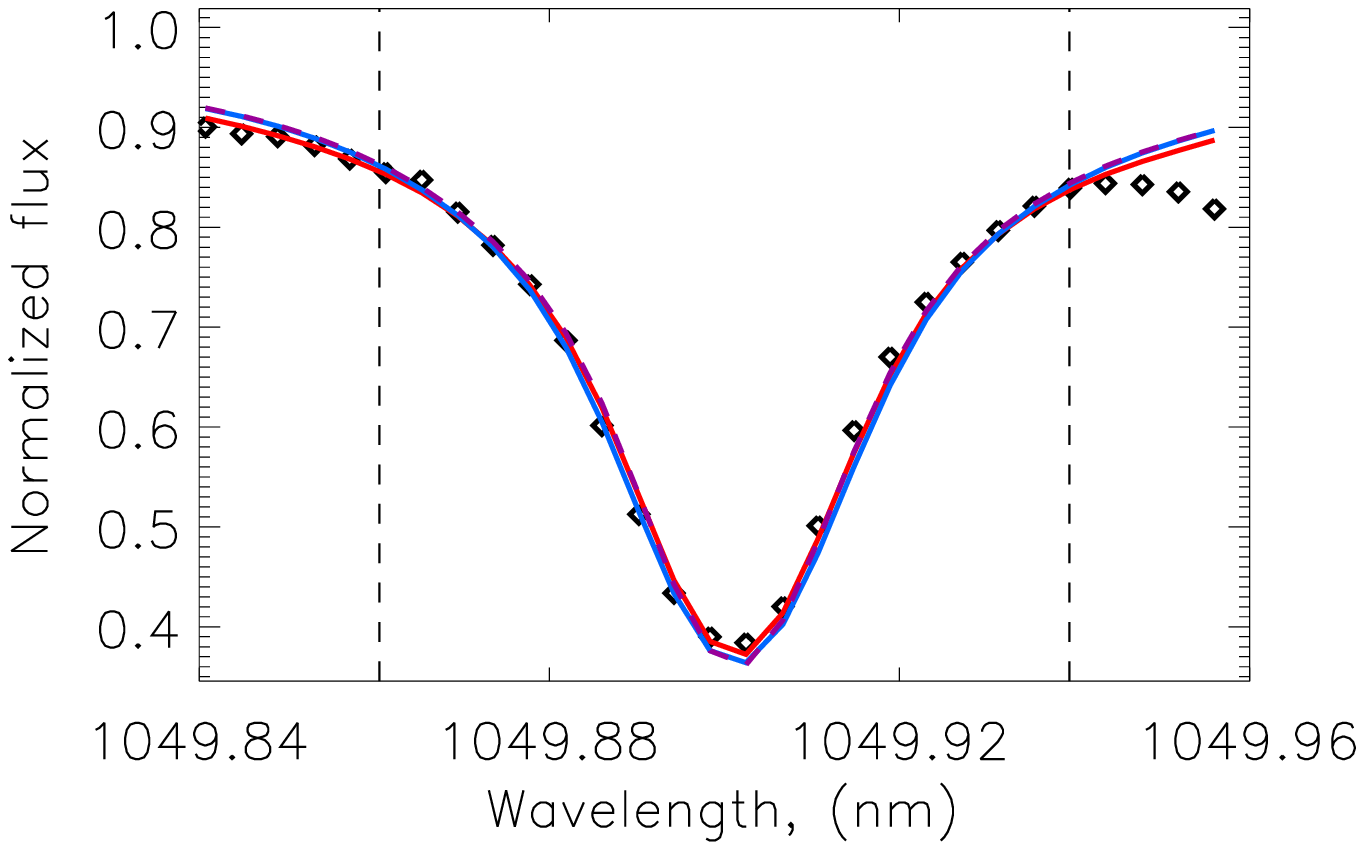}
\includegraphics[width=0.33\hsize]{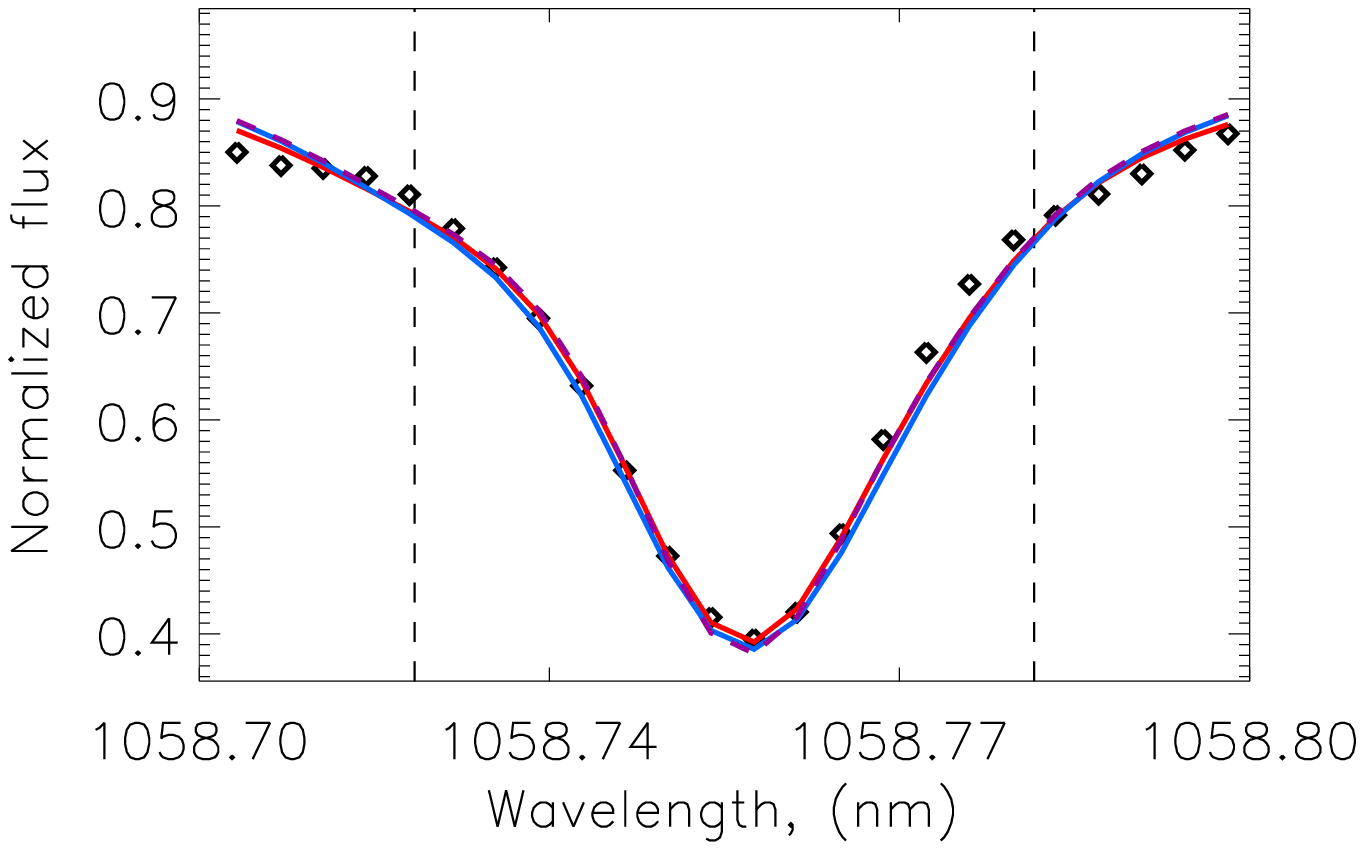}
\includegraphics[width=0.33\hsize]{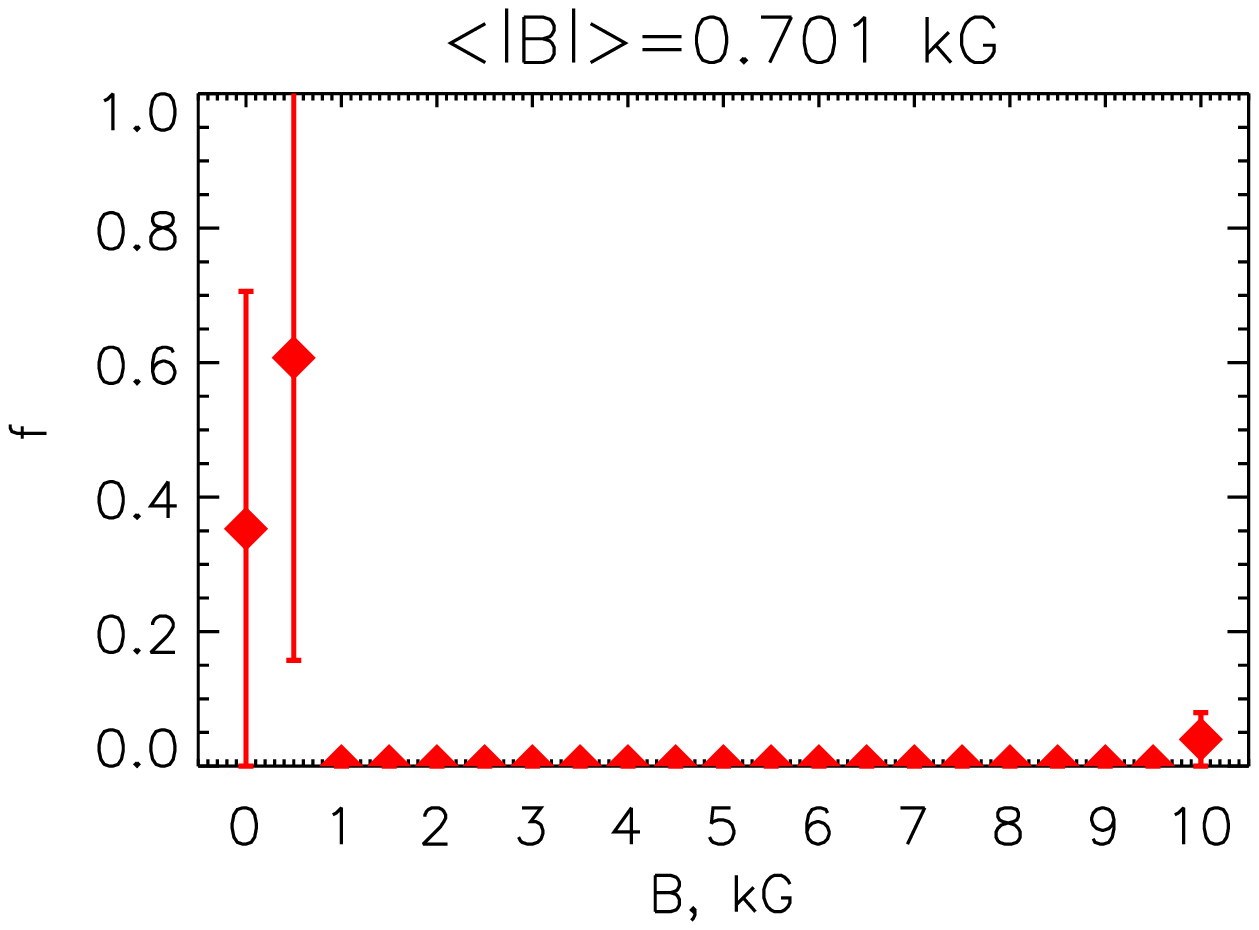}
}
\caption{Same as on Fig.~\ref{fig:gj1002-feh} but for \ion{Ti}{i} in GJ~1002.
Top panel~--~$\teff=3100$~K, $\abn{Ti}=-7.24$, $\vsini=2.43$~\kms.
Middle panel~--~$\teff=3100$~K,  $\abn{Ti}=-7.24$, $\vsini=1.85$~\kms.
Bottom panel~--~$\teff=3050$~K,  $\abn{Ti}=-7.24$, $\vsini=2.60$~\kms.}
\label{fig:gj1002-ti}
\end{figure*}

The spectrum of GJ~1002 observed with CRIRES in run 079.D-0357 also contains strong \ion{Ti}{i} lines in the $\ll10\,300-10\,700$ region.
We used less blended lines at $\lambda10399$, $\lambda10498$, and $\lambda10587$ to test our method when applied to atomic transitions.
Unfortunately, all sets of atmospheric parameters derived from FeH lines are consistent with non-zero magnetic fields. Adjusting
$\abn{Ti}$ for a fixed $\teff=3100$~K does not help to solve the problem of fitting the wings of Ti lines which we find to be the main reason
for the detection of non-zero magnetic field. The top panel of Fig.~\ref{fig:gj1002-ti} shows a fit to Ti lines with 
$\teff=3100$~K, $\vsini=2.4$~\kms, $\abn{Ti}=-7.24$. A $\vsini=1.9$~\kms\ (adopted from FeH lines) results in too deep
line cores, but almost in the same strength of the magnetic field detected (middle panel of Fig.~\ref{fig:gj1002-ti}).
In order to obtain a nearly zero magnetic field one has to decrease the effective temperature 
to $\teff\approx3050$~K and increase rotational velocity to $\vsini\approx2.6$~\kms, as shown on the bottom panel of Fig.~\ref{fig:gj1002-ti}.
It is seen that in the latter case magnetic field components as strong as $10$~kG appear, but this
is obviously an artifact of the method: such a high component occasionally provides a better fit to line wings though its
impact on the line profiles is very small.
A non-detection of the magnetic field is obtained assuming $\vsini=3$~\kms,
and the example of using $\vsini=2.6$~\kms\ shows that the highest field components may easily result from uncertainties in
rotational velocities and instrument resolving power. {Finally note that the tested Ti lines are not very magnetic sensitive,
with effective Land\'e factors $g_{\rm eff}\approx1$, therefore strong fields are needed to mimic small adjustments in $\vsini$ described above.}

\begin{figure*}
\centerline{
\includegraphics[width=0.33\hsize]{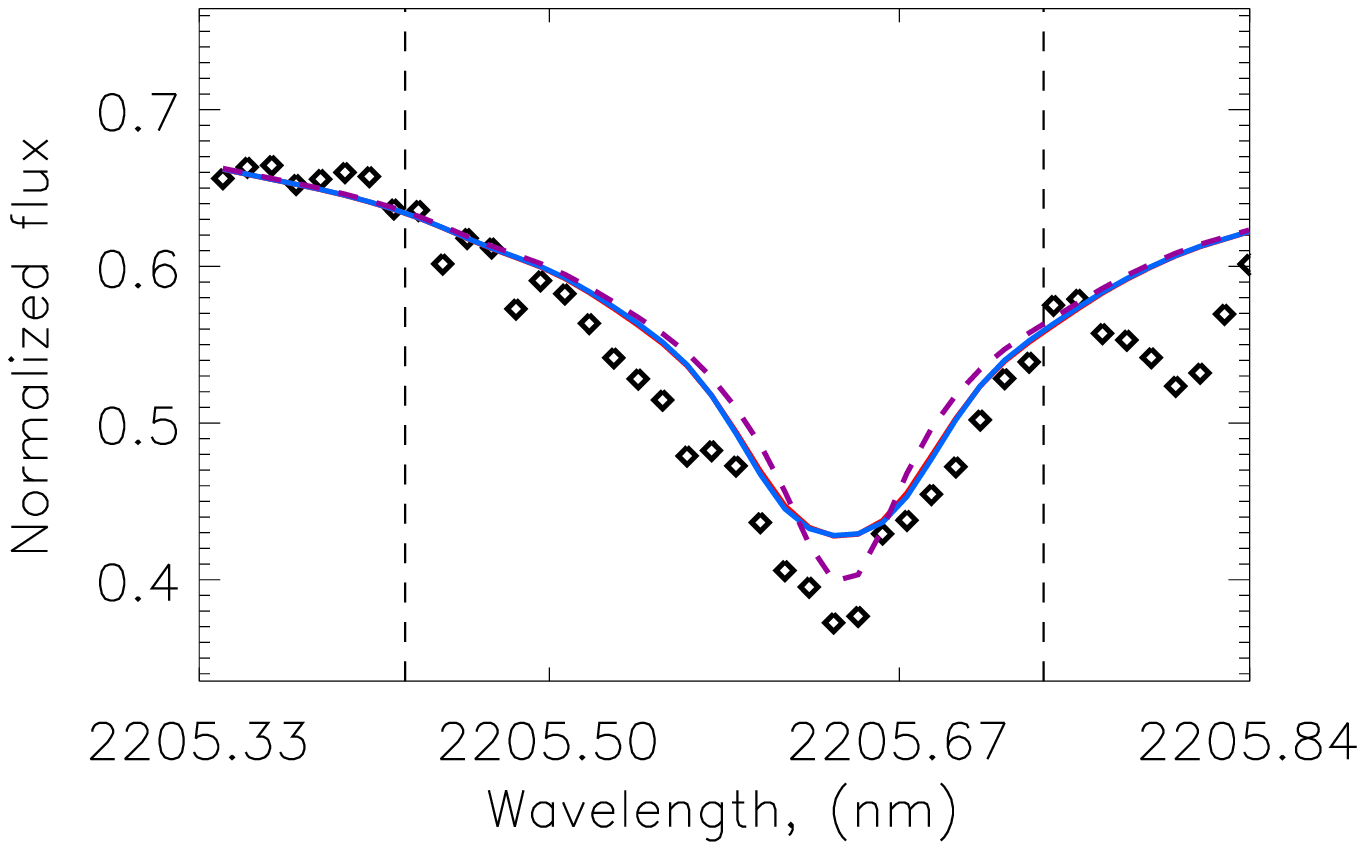}
\includegraphics[width=0.33\hsize]{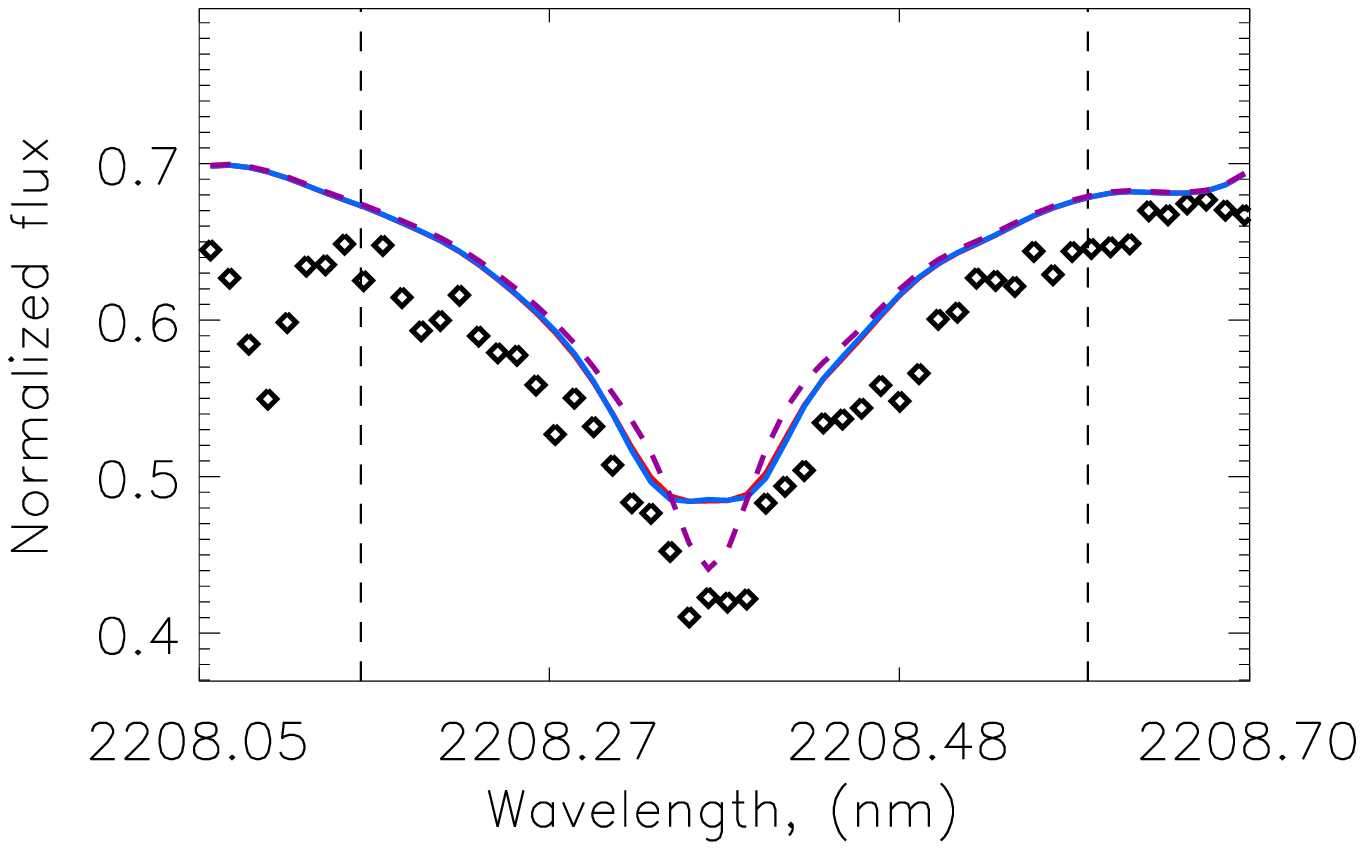}
\includegraphics[width=0.33\hsize]{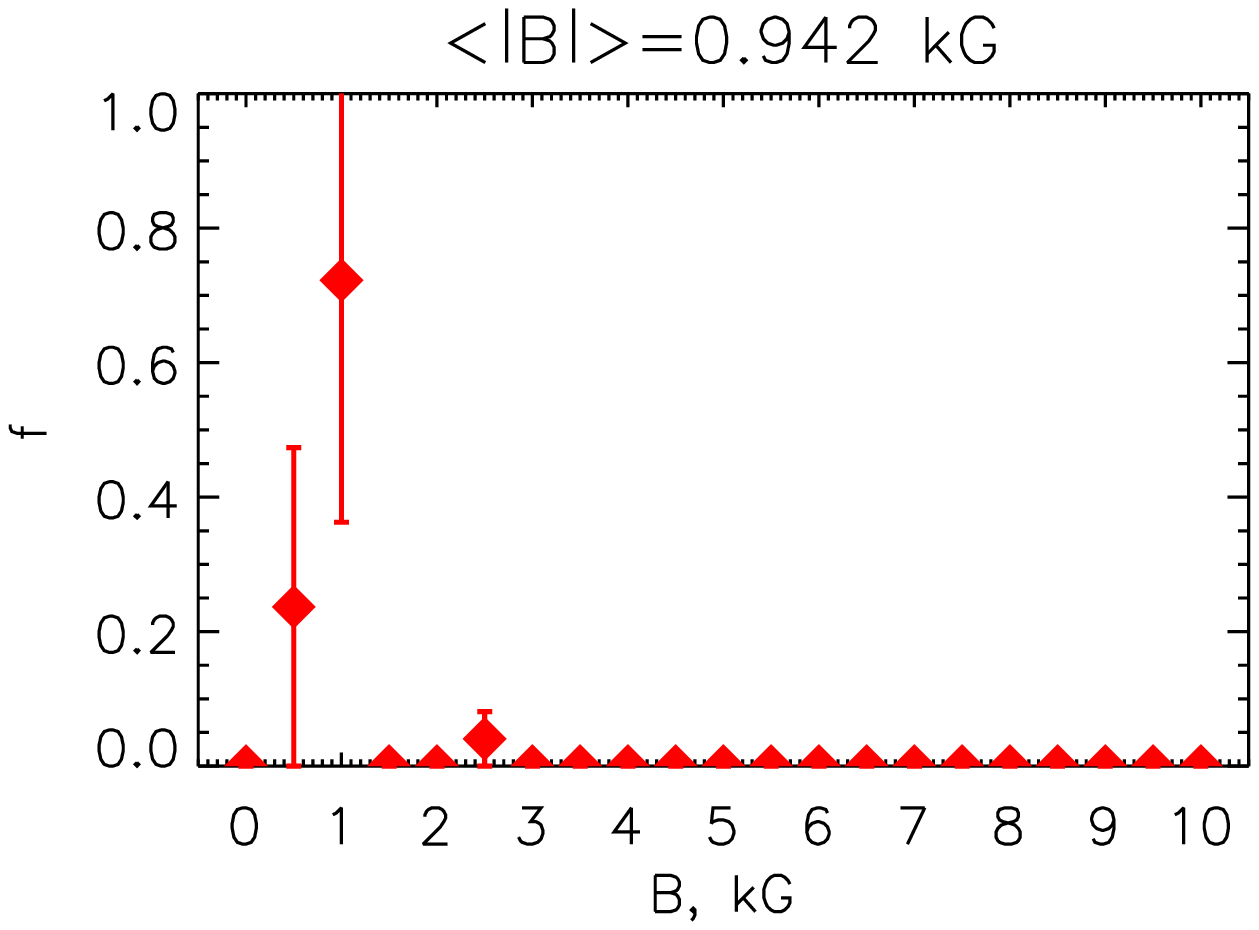}
}
\caption{Same as on Fig.~\ref{fig:gj1002-ti} but for \ion{Na}{i} lines.
Atmospheric parameters are: $\teff=3100$~K, $\abn{Na}=-5.87$, $\vsini=2.43$~\kms.}
\label{fig:gj1002-na}
\end{figure*}

No set of atmospheric parameters adopted from the above investigation provide a reasonable fit to \ion{Na}{i} lines
at $2.2$~\mum. These lines appear to be too wide and their profiles are strongly affected by stellar water lines to allow for
{an} accurate analysis. That is, there is always an unknown vertical offset if one attempts to synthesize those lines.
To improve the quality of the theoretical fit we used a list of water lines 
to determine the continuum offset and then shift each observed spectrum accordingly in such
a way that quasi-continuum levels in both observed and theoretical spectra match.
The list of water lines is taken from R.~Kurucz' website\footnote{http://kurucz.harvard.edu/molecules/H2O}. 
Still, this {does not} solve the problem completely, most probably because of intrinsic inaccuracies of the water list itself.

Figure~\ref{fig:gj1002-na} illustrates the problem. Even shifting each line vertically in the attempt to account
for the unknown continuum level would not help to achieve a zero magnetic field detection because lines remain too wide still.
{Note that depths and widths of Na lines (as well as Ti lines) are well fit in the spectrum of the
solar photosphere, and thus the problem that we encounter for these lines in cool M-dwarf spectra does not arise from
possibly inaccurate line transition parameters.}
Therefore we decided not to use Na lines for the magnetic field measurements. This is a disappointing result because Na lines
are strong features that probe a wide range of atmospheric depths. However, they can be used only when accurate telluric modelling,
improved molecular line lists, and more accurate continuum normalization methods will become available.

\section{Results}

In this section we present results of magnetic field determination for individual M-dwarfs, which are
also combined altogether in Table~\ref{tab:results}. As follows from the previous section, the \ion{Na}{i}
lines can not be used for field measurements and the region of {the} \ion{Ti}{i} lines (though they appear to be
good indicators of atmospheric structure) was not covered
by our observational settings. Therefore the results presented in this section are from FeH lines alone.

\subsection{GJ~388 (AD~Leo)}

\begin{figure*}
\centerline{
\includegraphics[width=0.33\hsize]{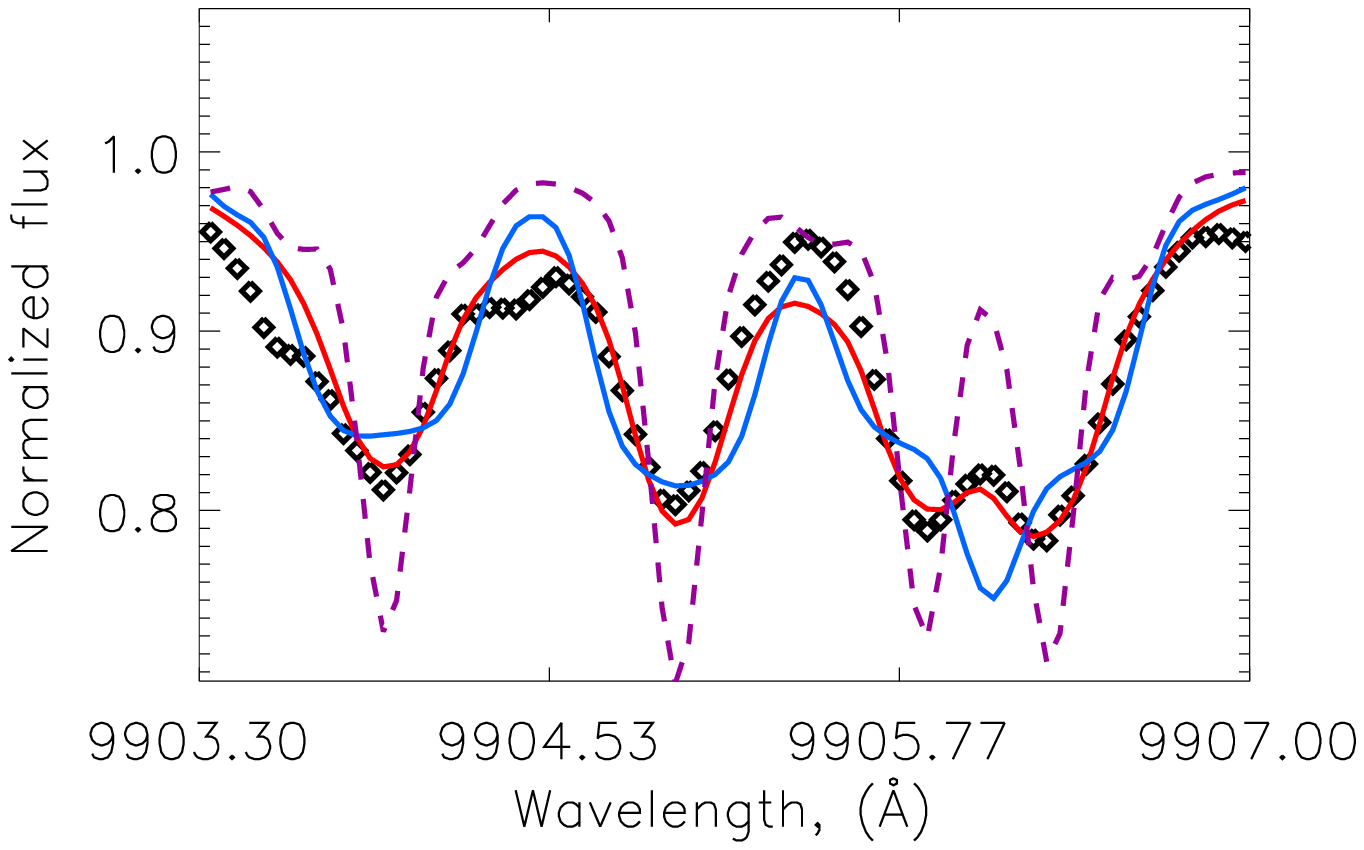}
\includegraphics[width=0.33\hsize]{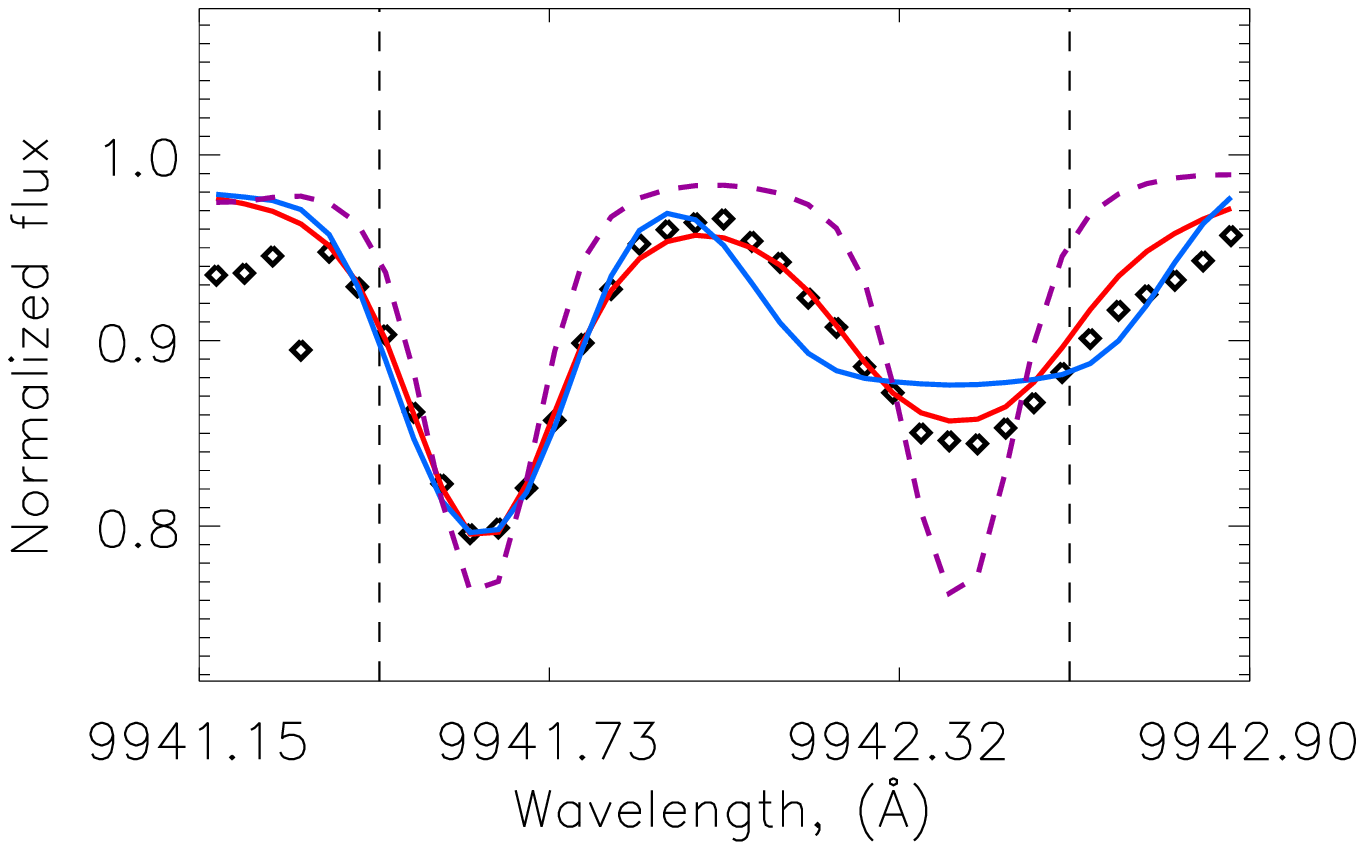}
\includegraphics[width=0.33\hsize]{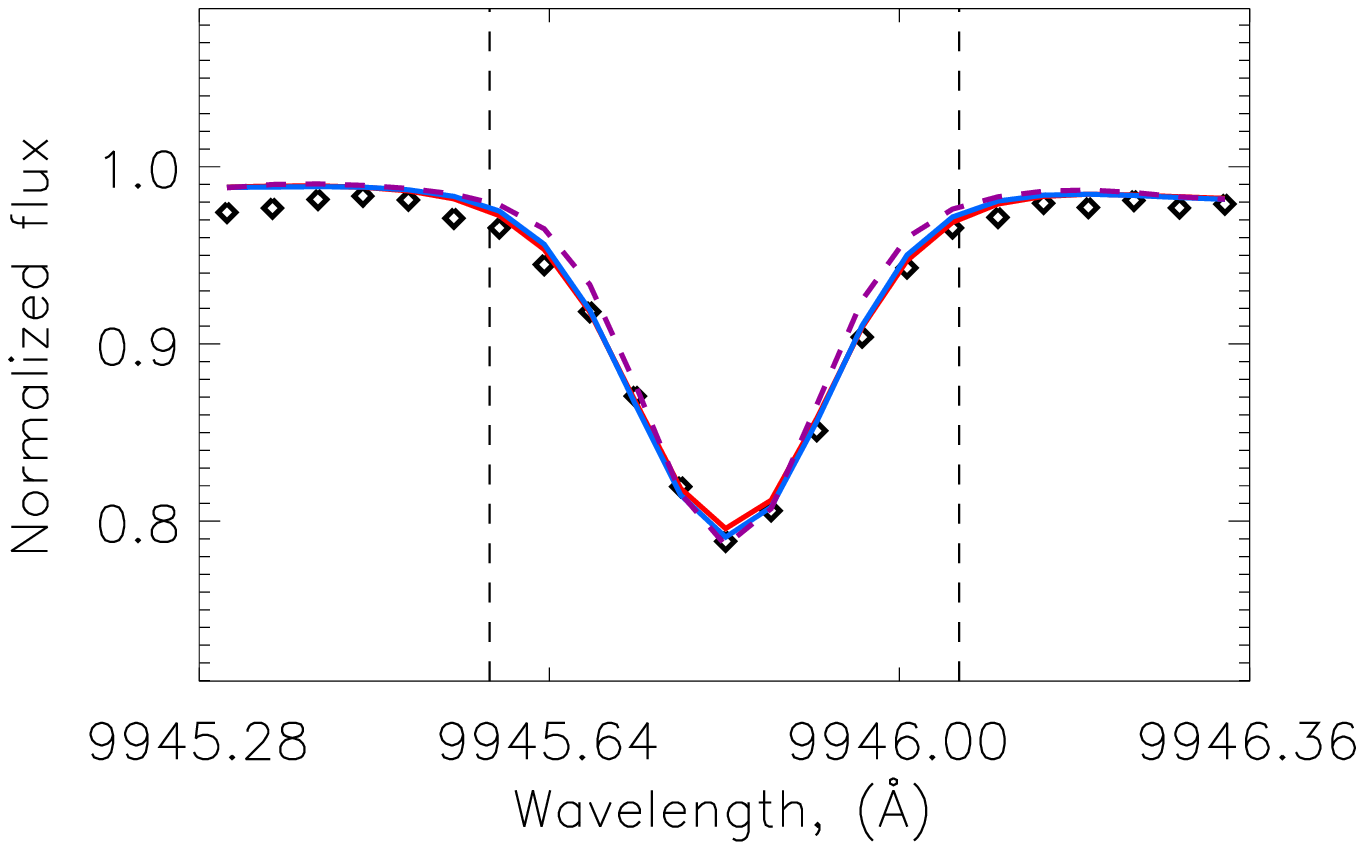}
}
\centerline{
\includegraphics[width=0.33\hsize]{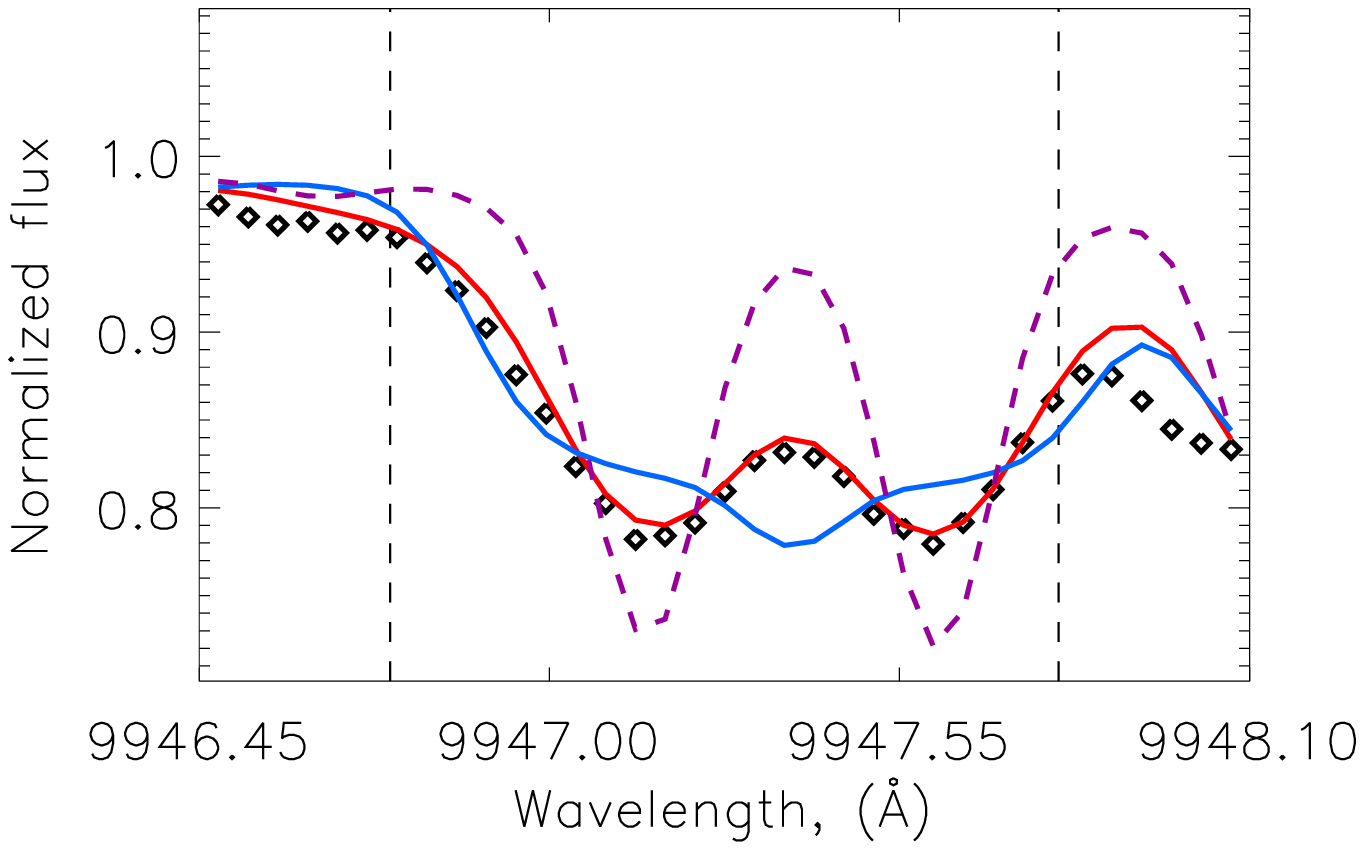}
\includegraphics[width=0.33\hsize]{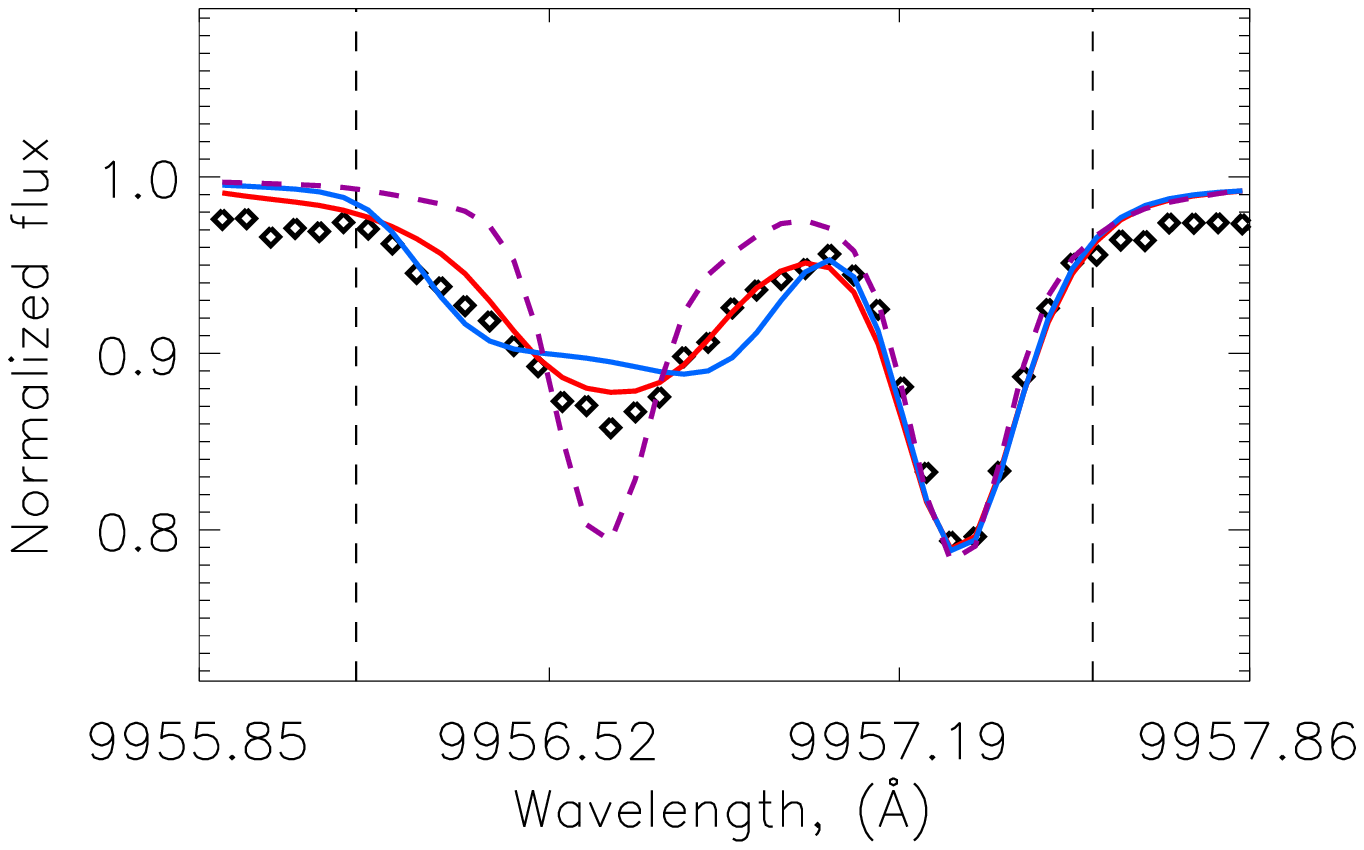}
\includegraphics[width=0.33\hsize]{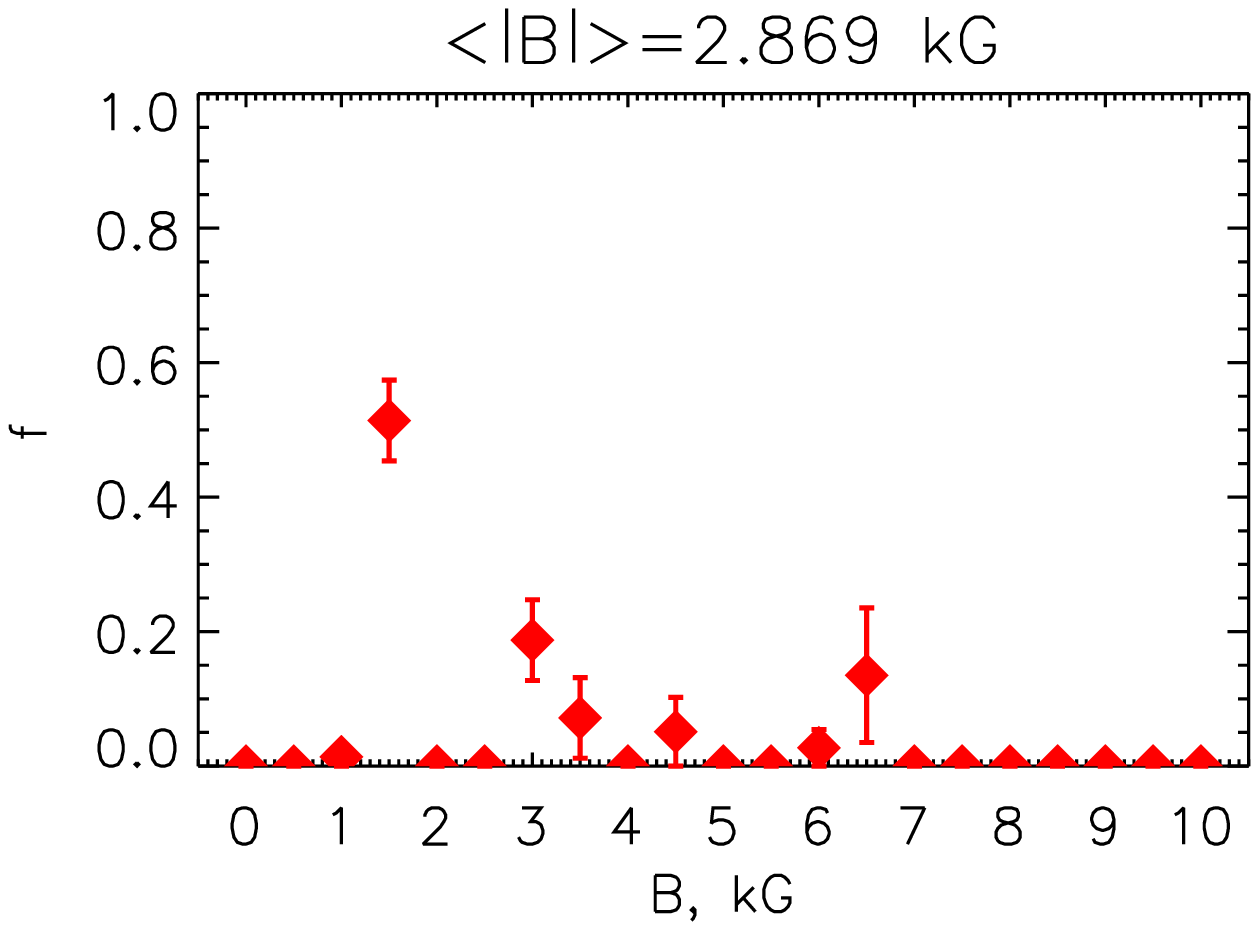}
}
\centerline{
\includegraphics[width=0.33\hsize]{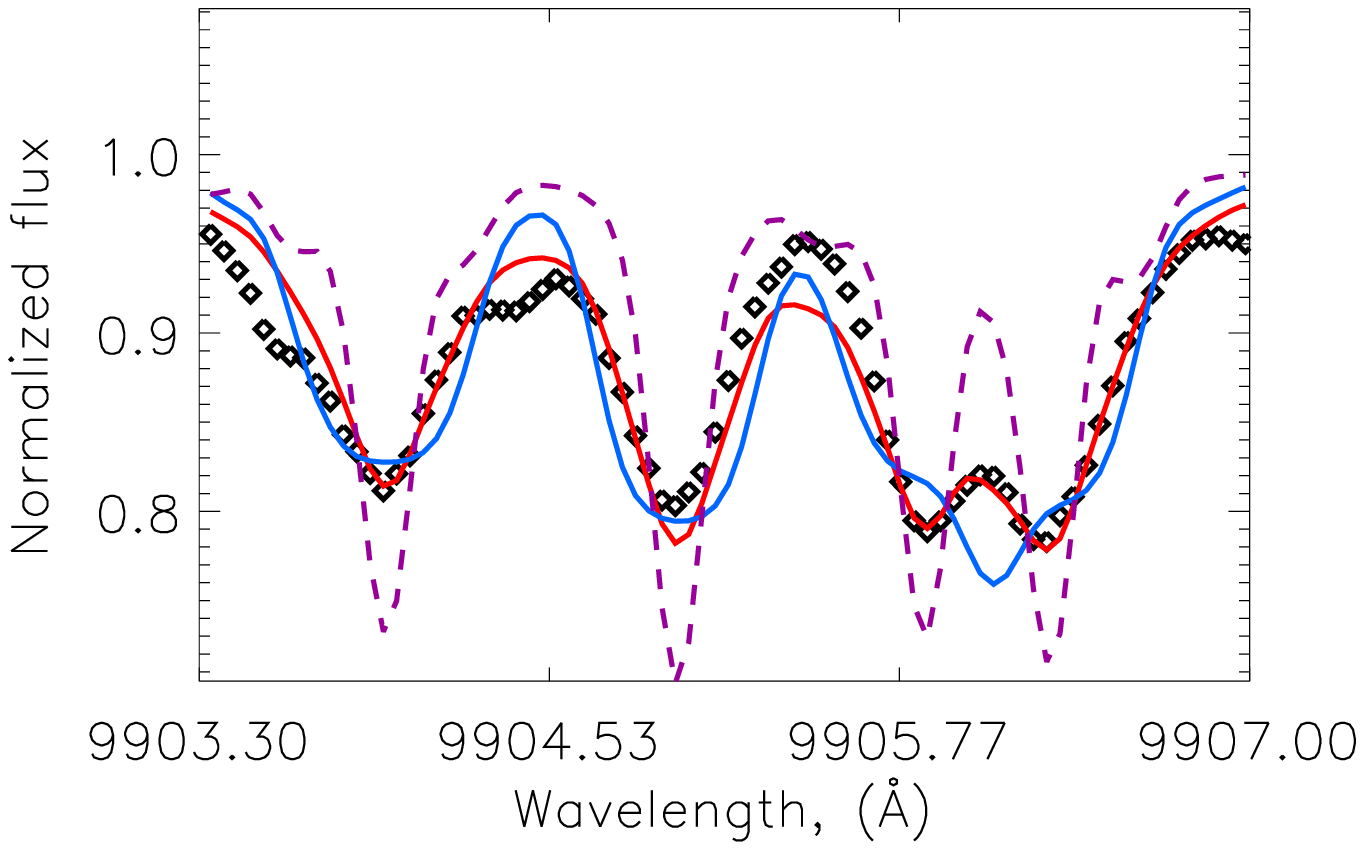}
\includegraphics[width=0.33\hsize]{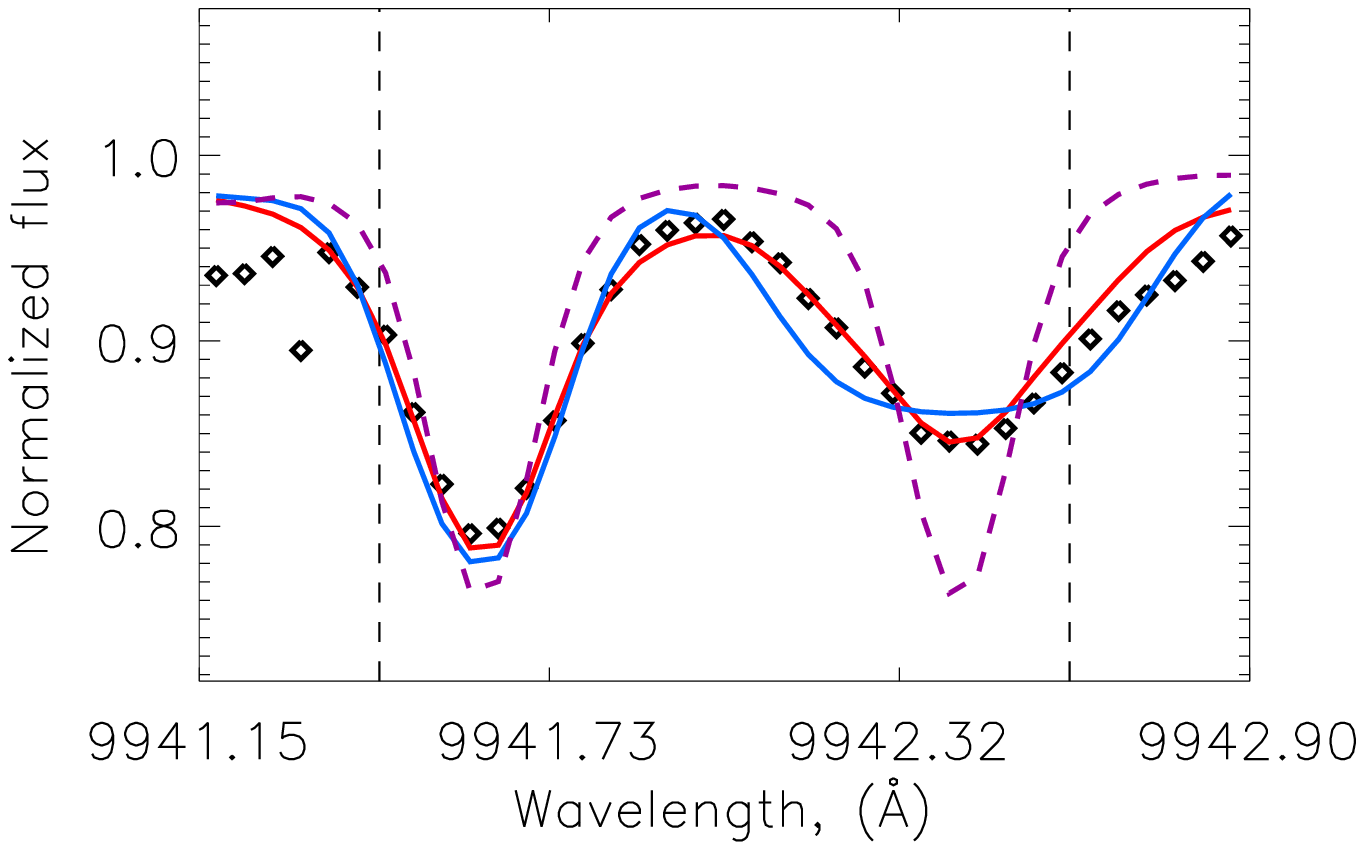}
\includegraphics[width=0.33\hsize]{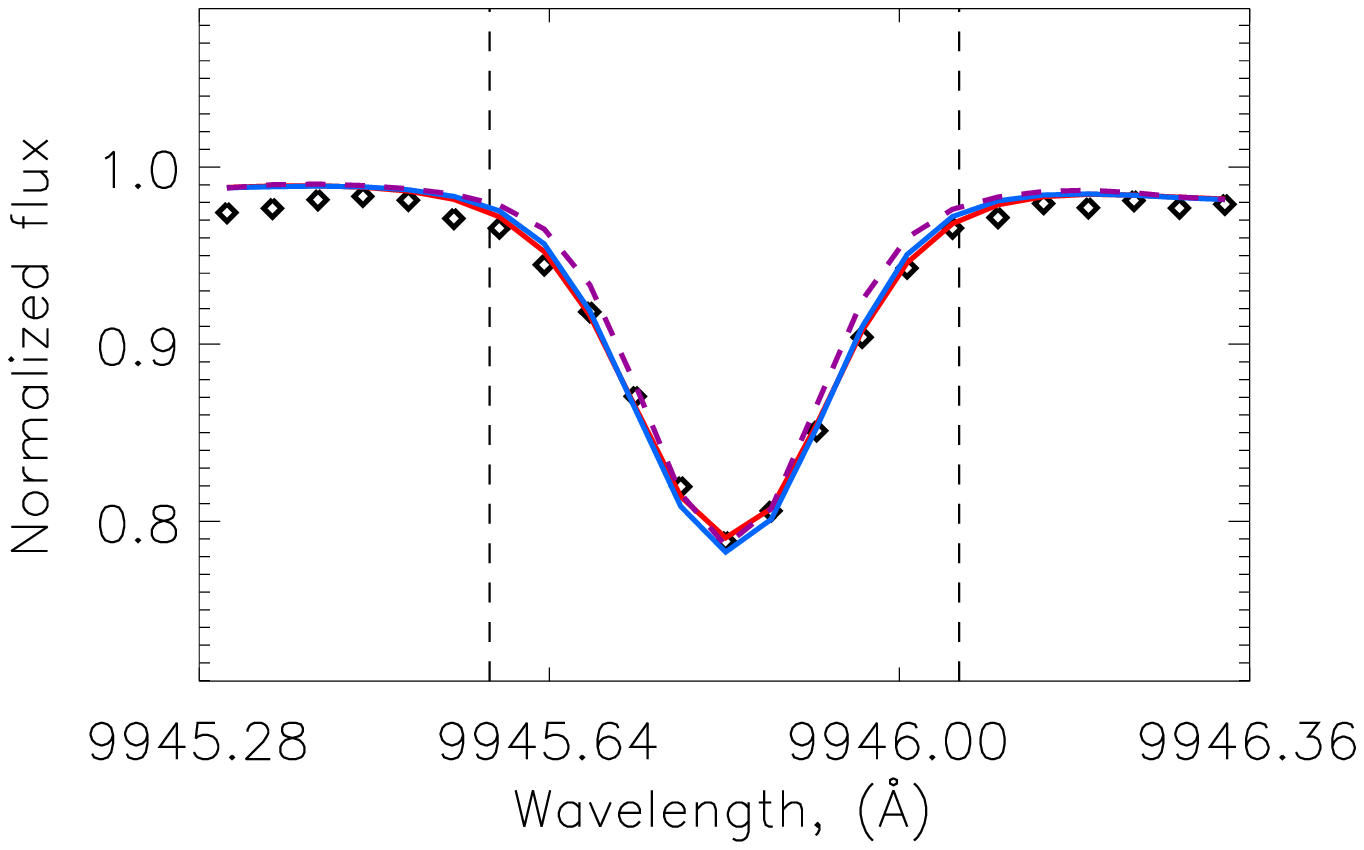}
}                                                             
\centerline{                                                  
\includegraphics[width=0.33\hsize]{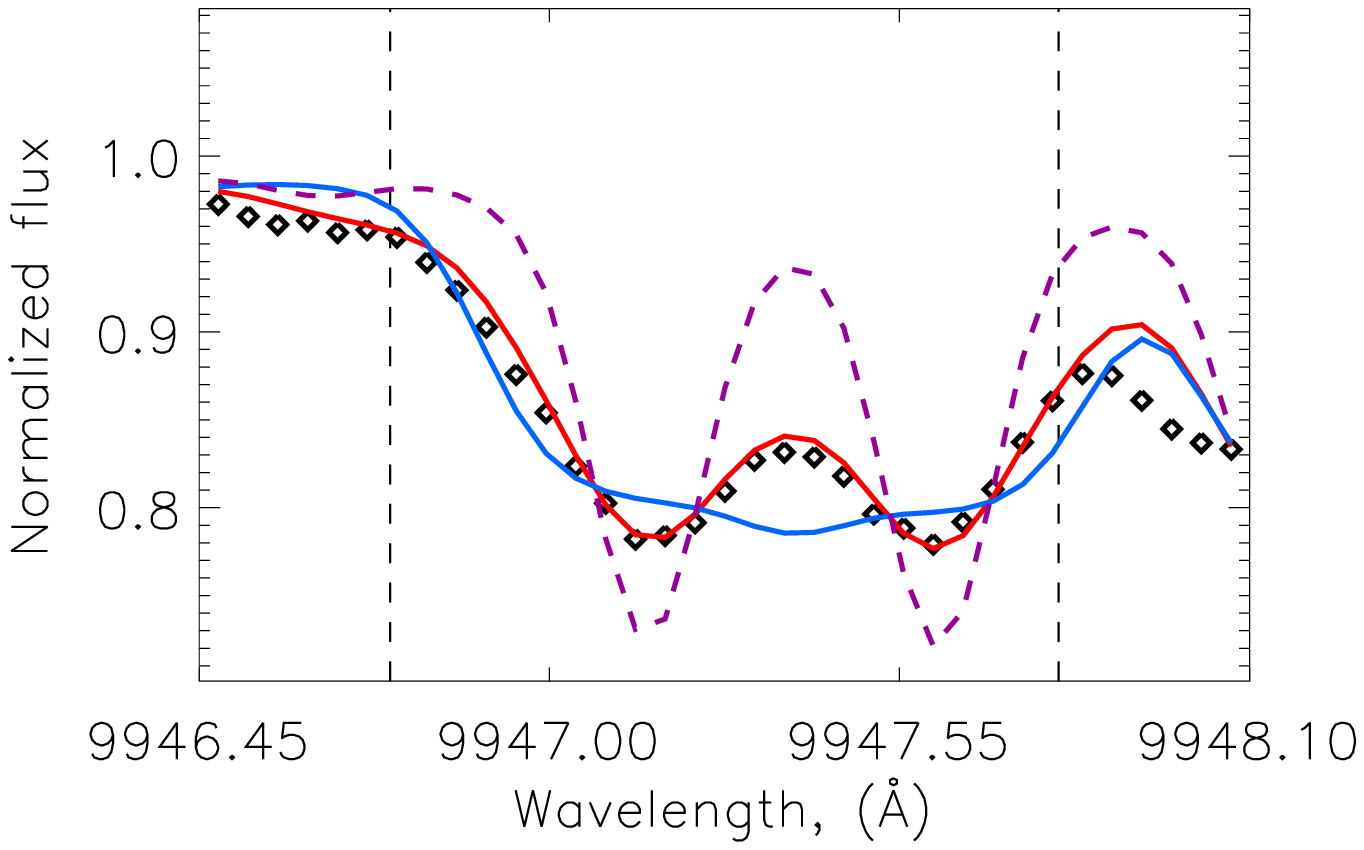}
\includegraphics[width=0.33\hsize]{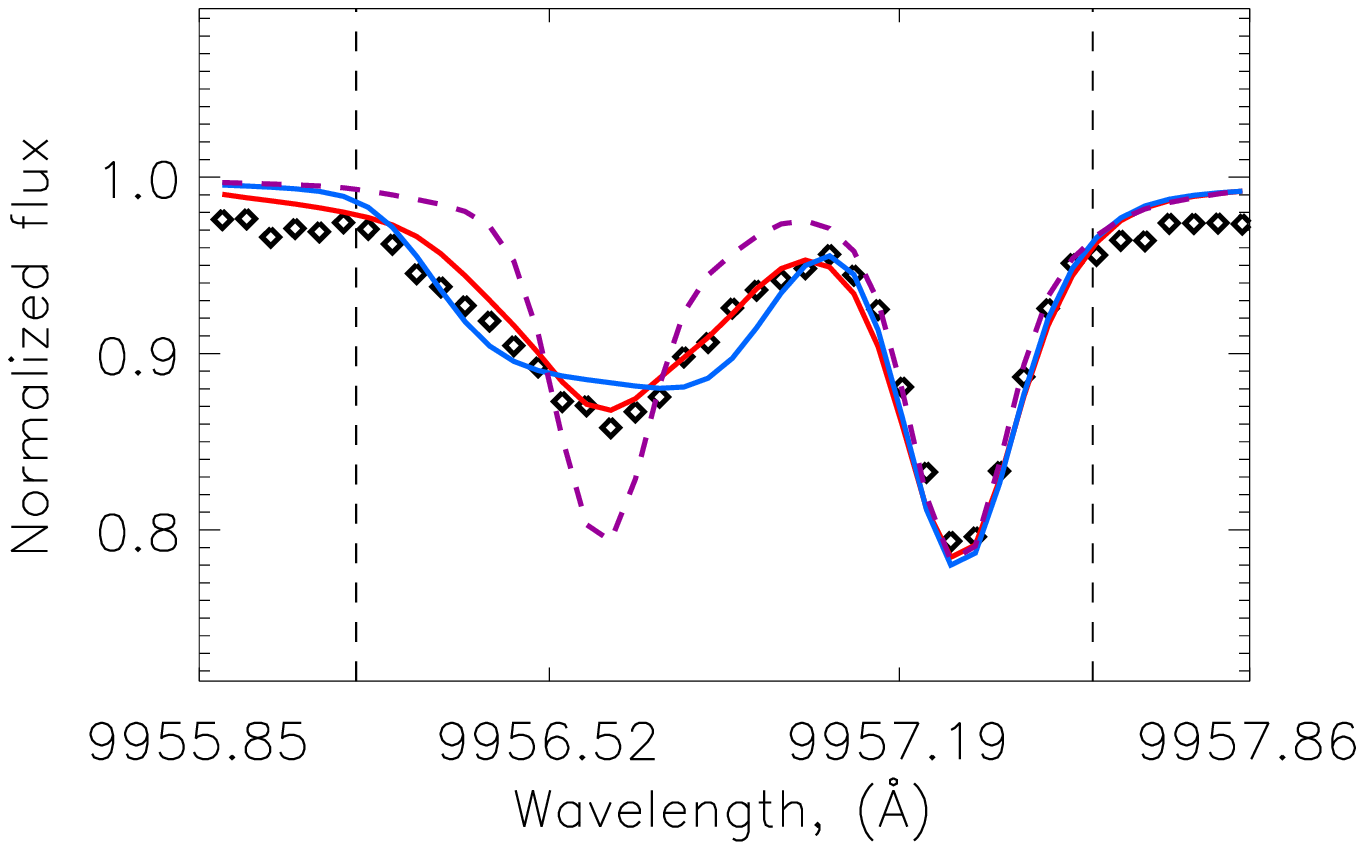}
\includegraphics[width=0.33\hsize]{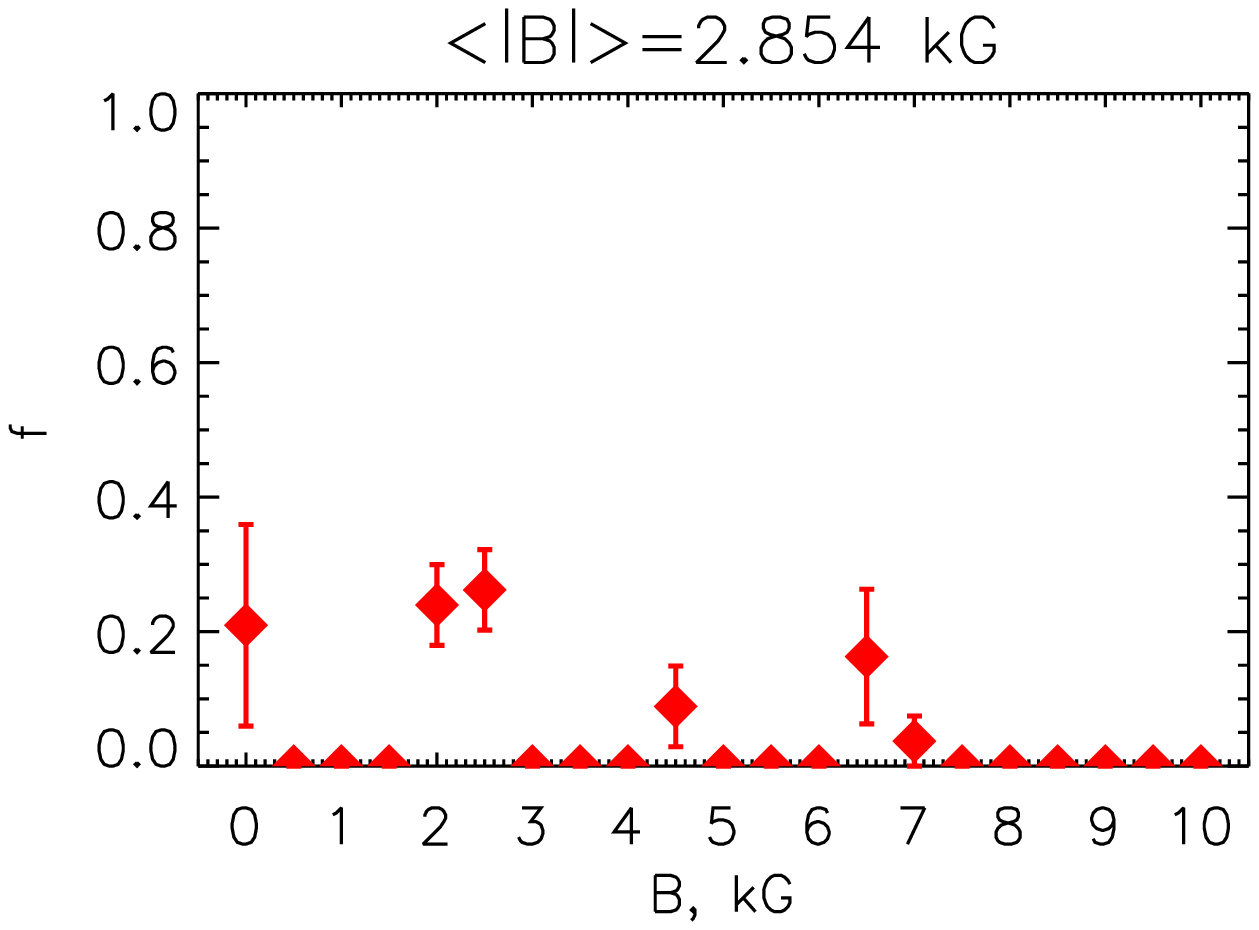}
}
\caption{Examples of theoretical fit to selected FeH lines and resulting distributions of filling factors for GJ~388 assuming
RC (first and second raws) and MC (third and forth rows) magnetic field models.
Atmospheric parameters $\teff=3400$~K, $\abn{Fe}=-4.64$, $\vsini=3$~\kms.
Violet dashed line~--~computation with zero magnetic field, red line~--~computation with multi-component 
magnetic field shown on the right bottom plot, blue line~--~computations with homogeneous magnetic field (i.e. $f=1$)
of the same intensity as multi-component magnetic field.}
\label{fig:gj388-fit}
\end{figure*}

GJ~388 is an M$3.5$ star with a rotational velocity of $\vsini\leqslant3$~\kms\, $\teff=3400$~K, and surface magnetic field
of $2.9$~kG as estimated in RB07. 

From magnetically insensitive FeH lines we estimated the iron abundance to be $\abn{Fe}=-4.64$.
We find that $\vsini$ does not influence the determination of the magnetic field if decreased below $3$~\kms: 
it can always be compensated by the respective increase of the iron abundance to obtain {a} similar quality of the fit.
The derived {mean} magnetic field strength {matches} the value given by RB07, that is $\bs=2.9$~kG.

Figure~\ref{fig:gj388-fit} illustrates the comparison between observed and predicted profiles of some FeH lines. 
It reveals the strong sensitivity of FeH features to the magnetic field configuration, 
i.e. to the presence of different magnetic field components. FeH lines at $\lambda9906$ and
$\lambda9947$ are the most illustrative examples of Zeeman broadening, (but see also lines at $\lambda9942$ and $\lambda9956$).

The distribution of the magnetic field that we recover depends critically upon the assumed magnetic field model 
used in spectrum synthesis.
Employing the RC model results in one strong component of $1.5$~kG
covering about $50$\% of the visible stellar surface. There are four other significant field components covering the rest of the
surface with intensities spread between $3$~kG and $6.5$~kG. These components appear from the necessity to obtain a match
between the narrow cores (requiring weak field components) and wide wings (requiring strong field components).
There is no significant zero field component, and we do not find any zero-field component when we applied different 
atmospheric parameters. For instance, using a higher temperature of $\teff=3500$~K, $\vsini=3$~\kms, and solar $\abn{Fe}=-4.54$ results in
a similar distribution of filling factors as well as $\bs$ value.
A zero-field component appears when the MC model is assumed. {In addition}, the components of
$2$~kG and $2.5$~kG of the same filling factors ($f\approx0.25$) appear. The strongest $6.5$~kG component 
{stays similar to that found in the} case of RC model. 
The detection of a zero-field component is
not surprising because Zeeman $\pi$-components are deep in MC model compared to RC one,
therefore they immediately provide a good fit to sharp line cores of such features as, e.g., 
$\lambda9942$ and $\lambda9956$.

Although the two magnetic field models result in distributions of filling factors that look different, 
we always find at least three distinct groups of magnetic field components and no smooth distributions
{are} ever detected. The derived mean magnetic fields $\bs$ are very similar too. By examining individual lines
we find that {the MC model provides}, on average, a better fit to observations.

\subsection{GJ~729}

\begin{figure*}
\centerline{
\includegraphics[width=0.33\hsize]{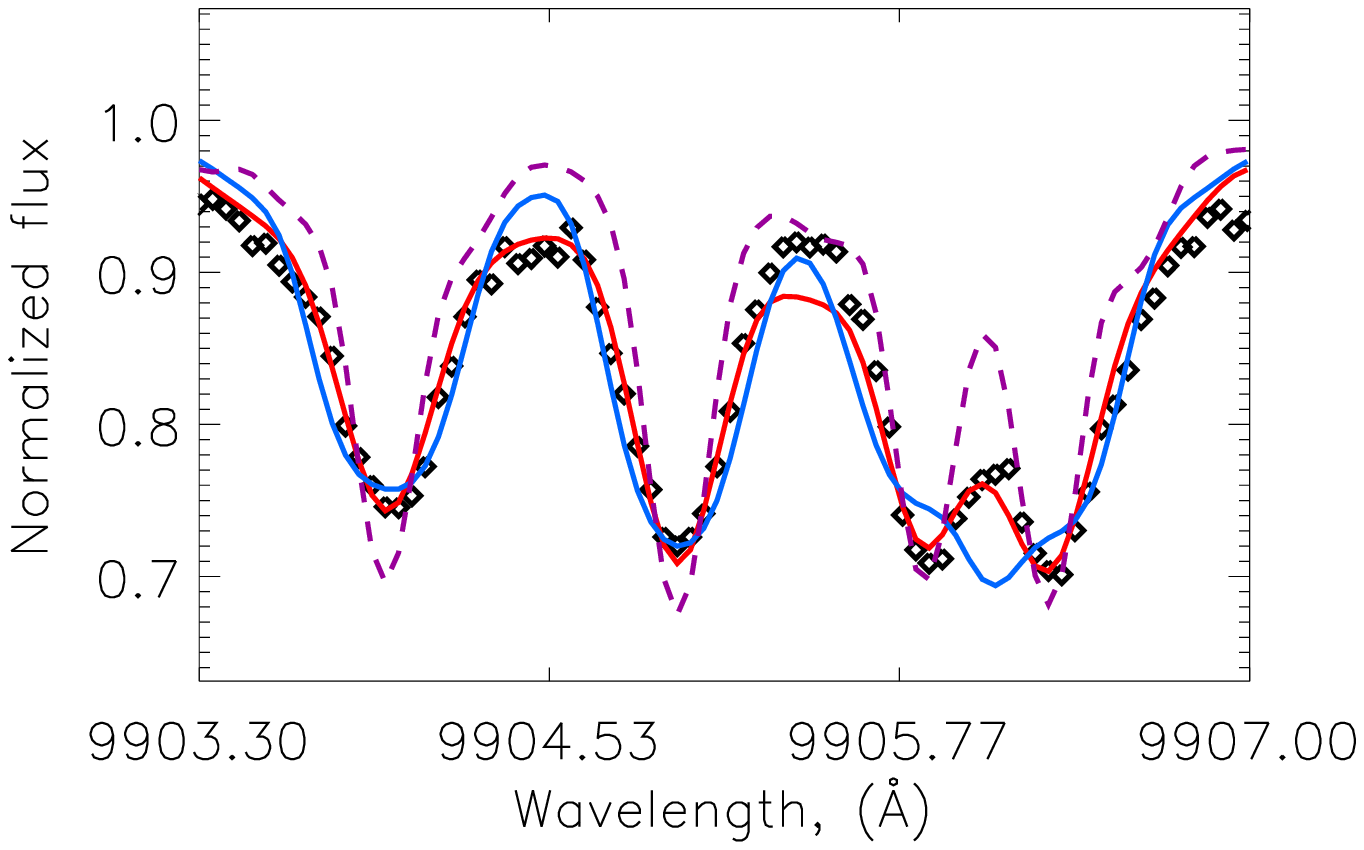}
\includegraphics[width=0.33\hsize]{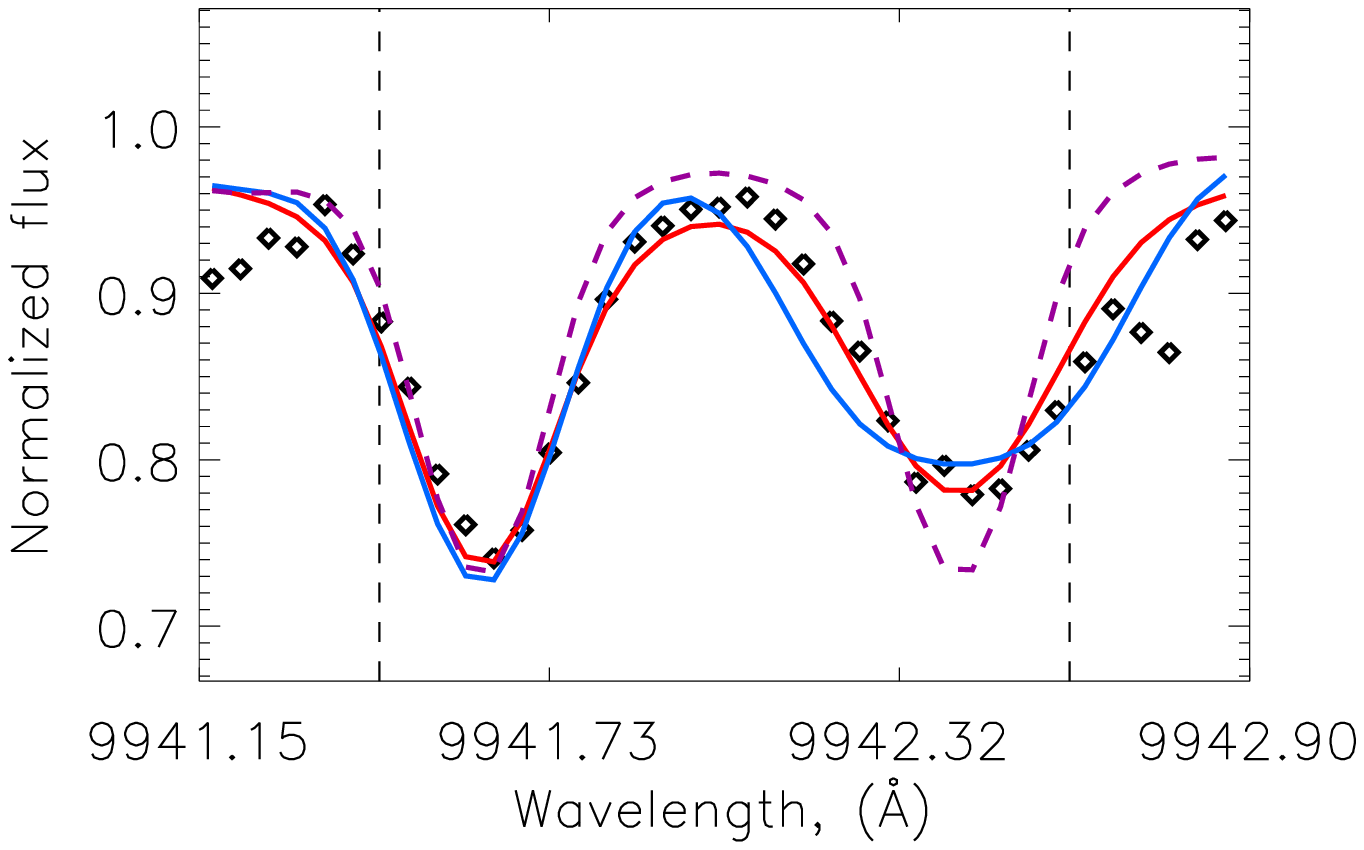}
\includegraphics[width=0.33\hsize]{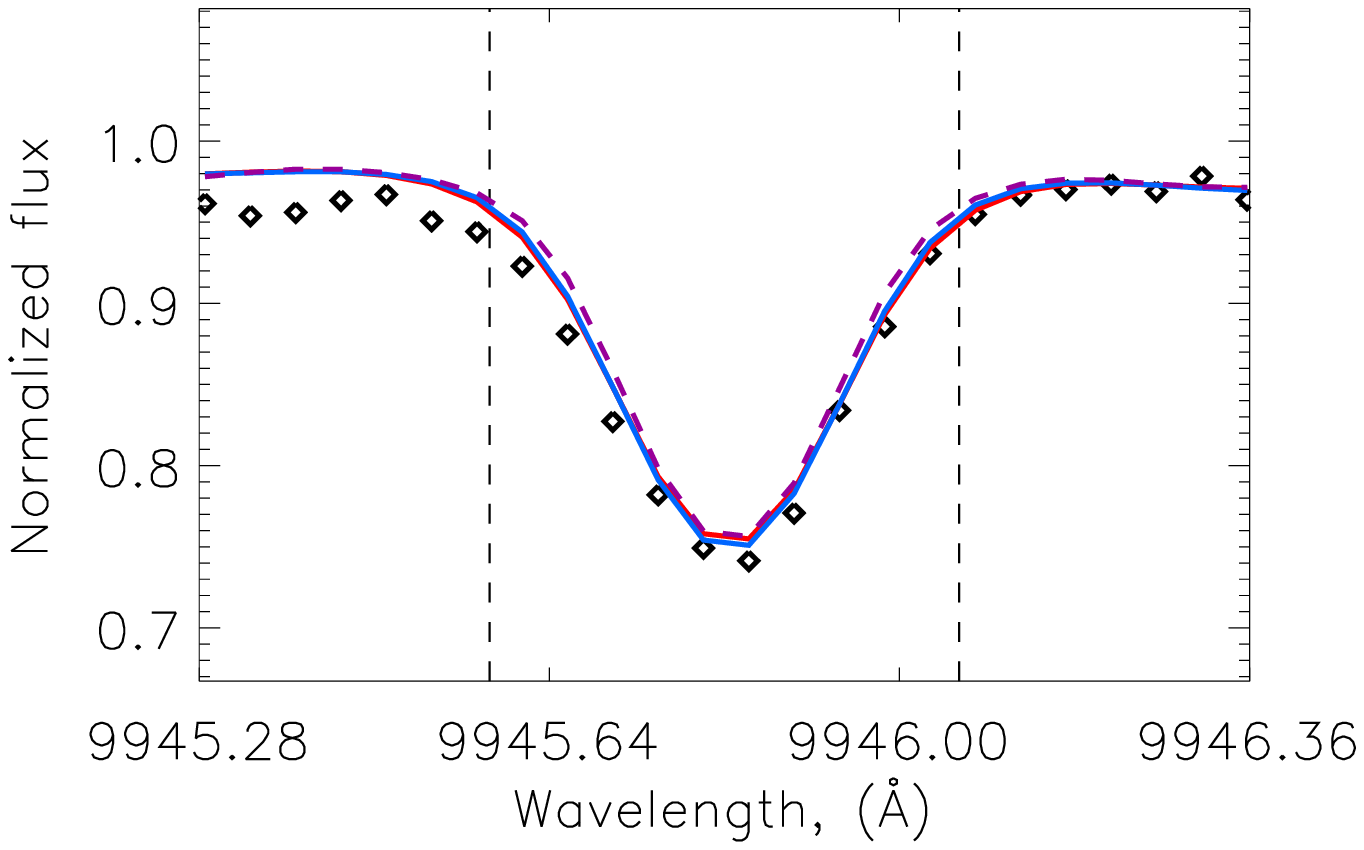}
}
\centerline{
\includegraphics[width=0.33\hsize]{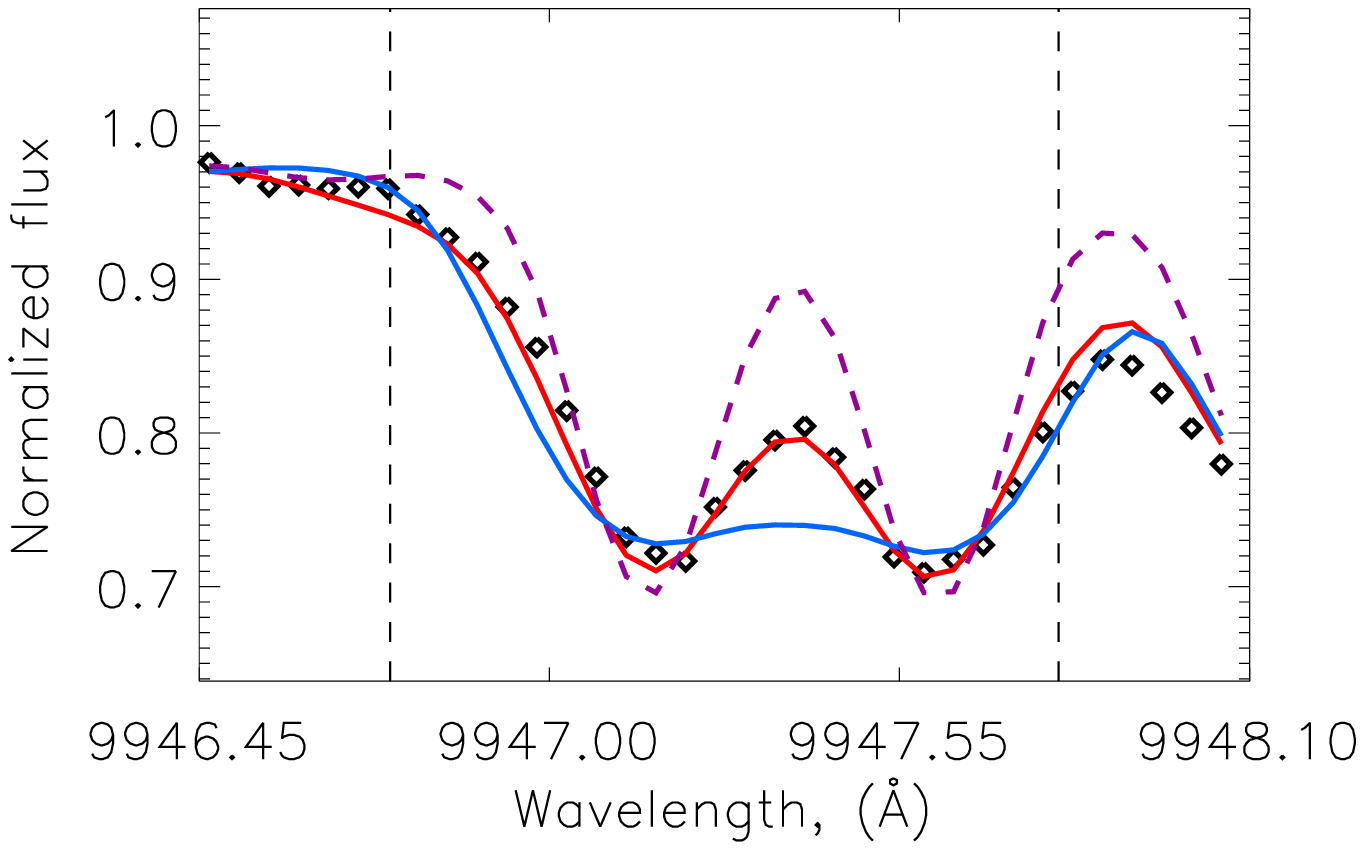}
\includegraphics[width=0.33\hsize]{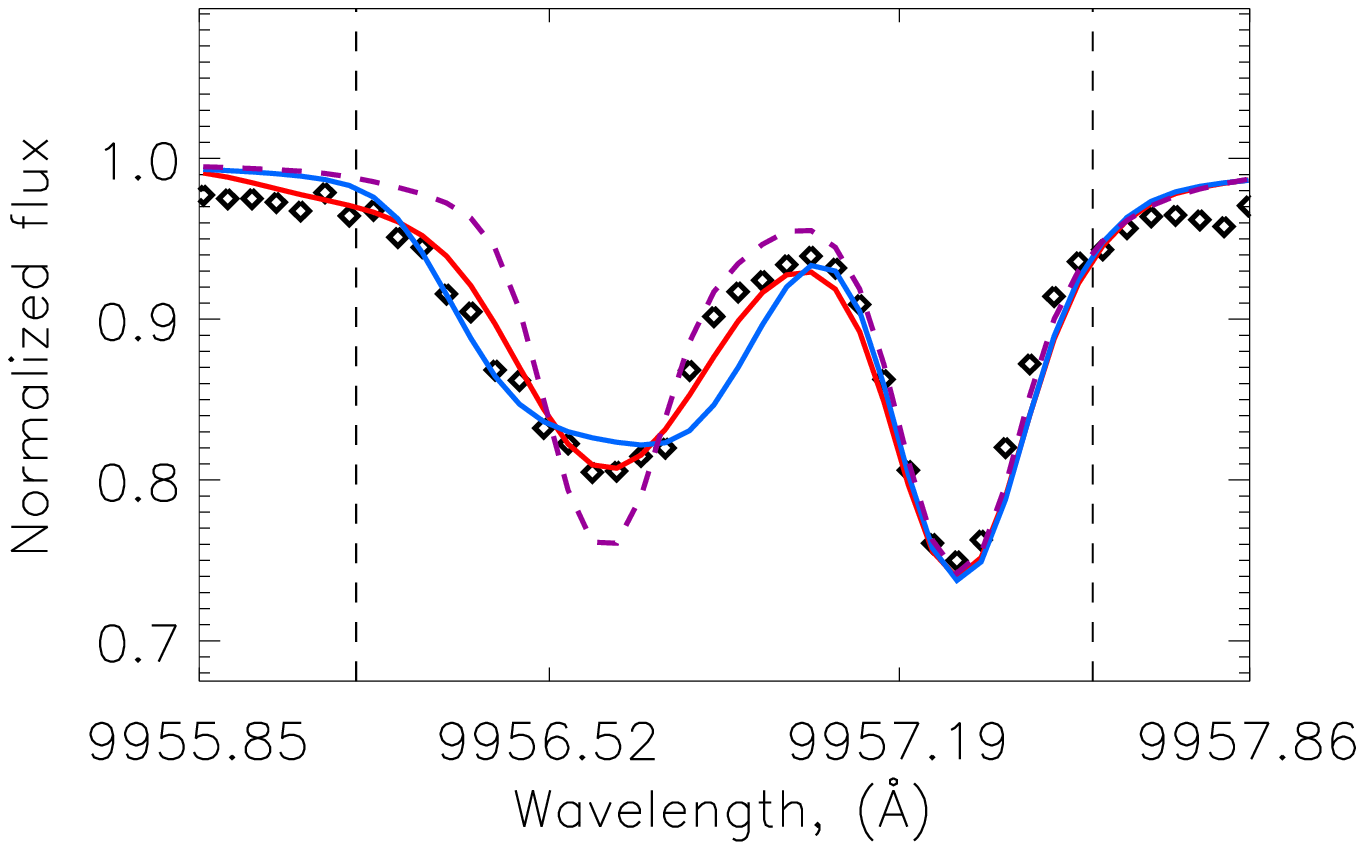}
\includegraphics[width=0.33\hsize]{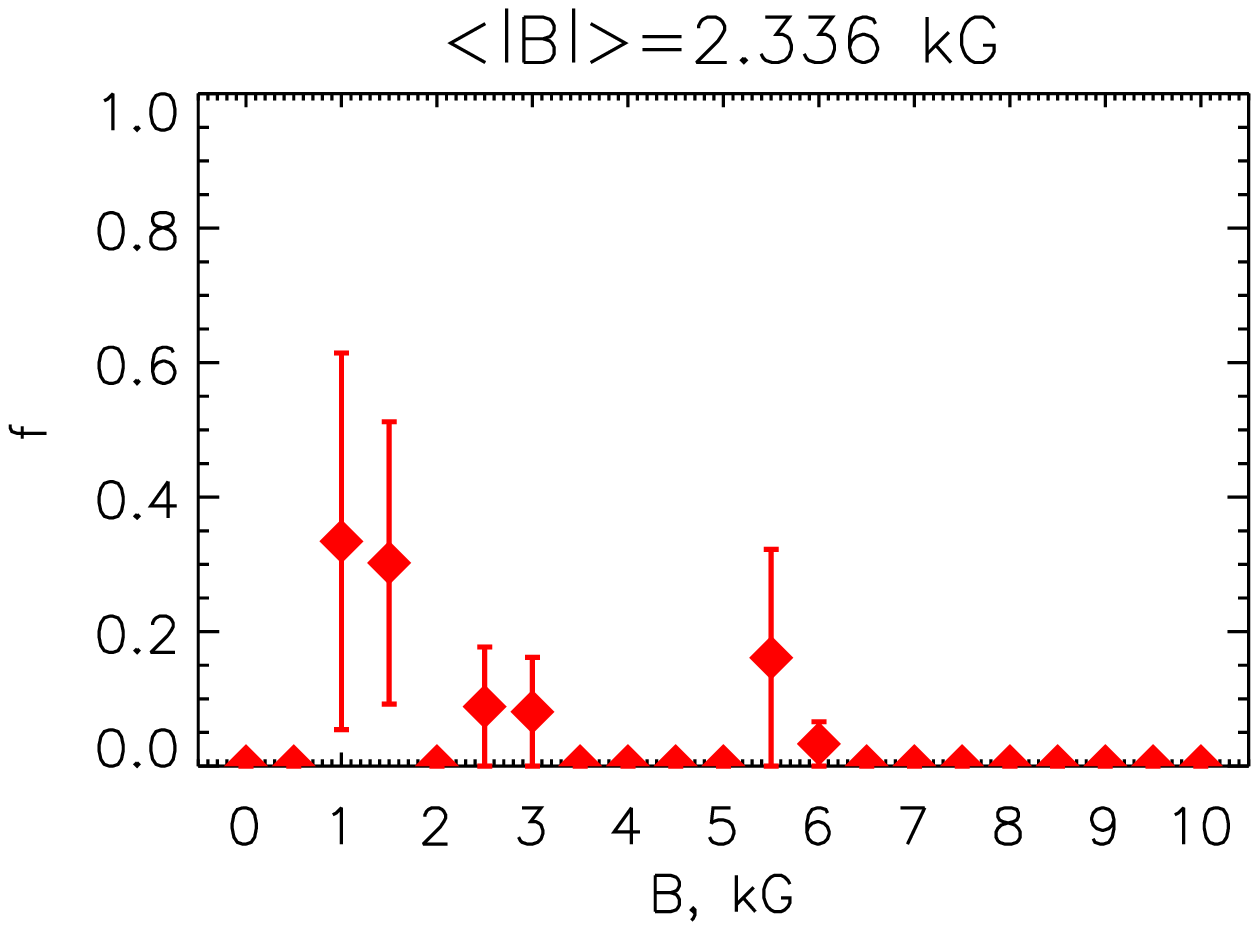}
}
\centerline{
\includegraphics[width=0.33\hsize]{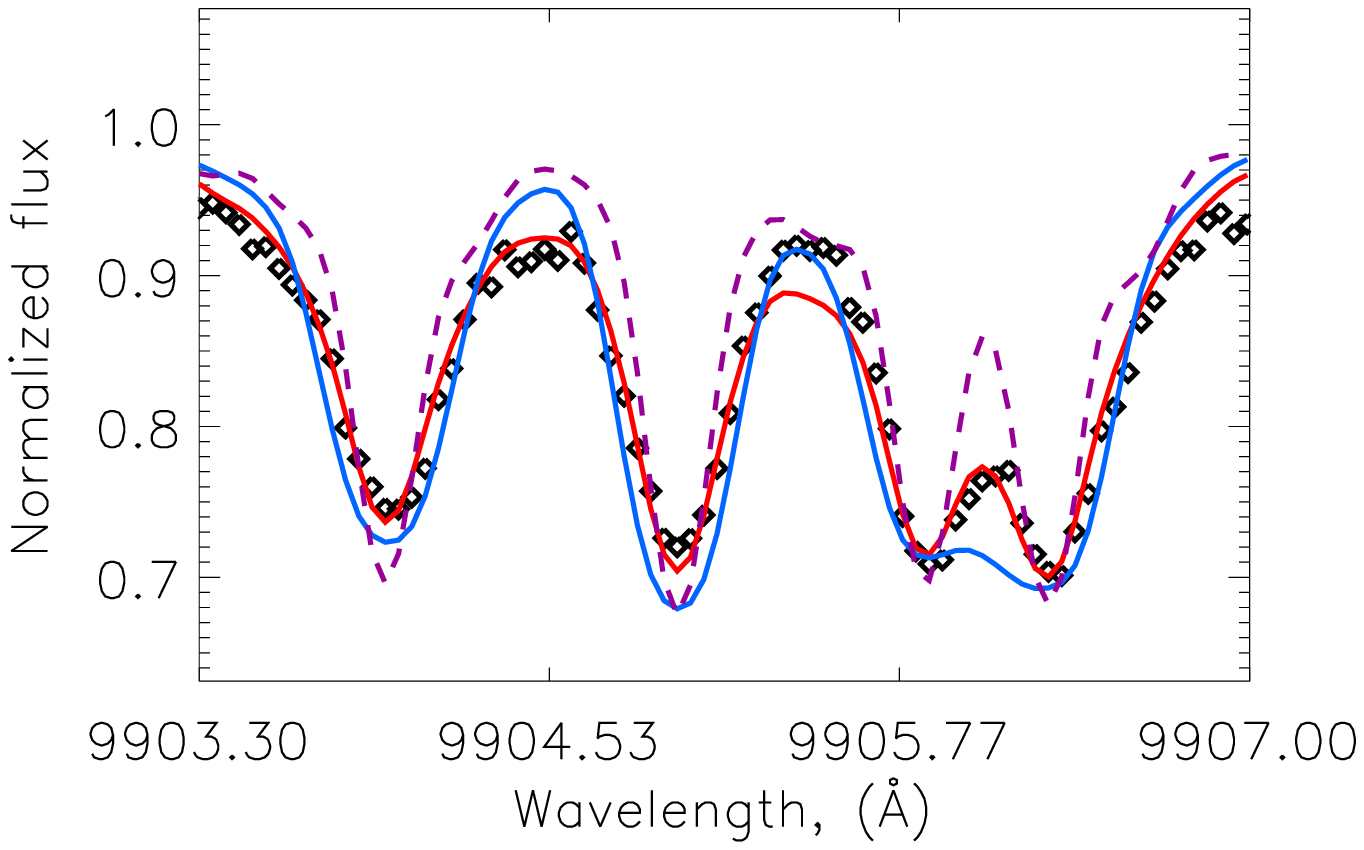}
\includegraphics[width=0.33\hsize]{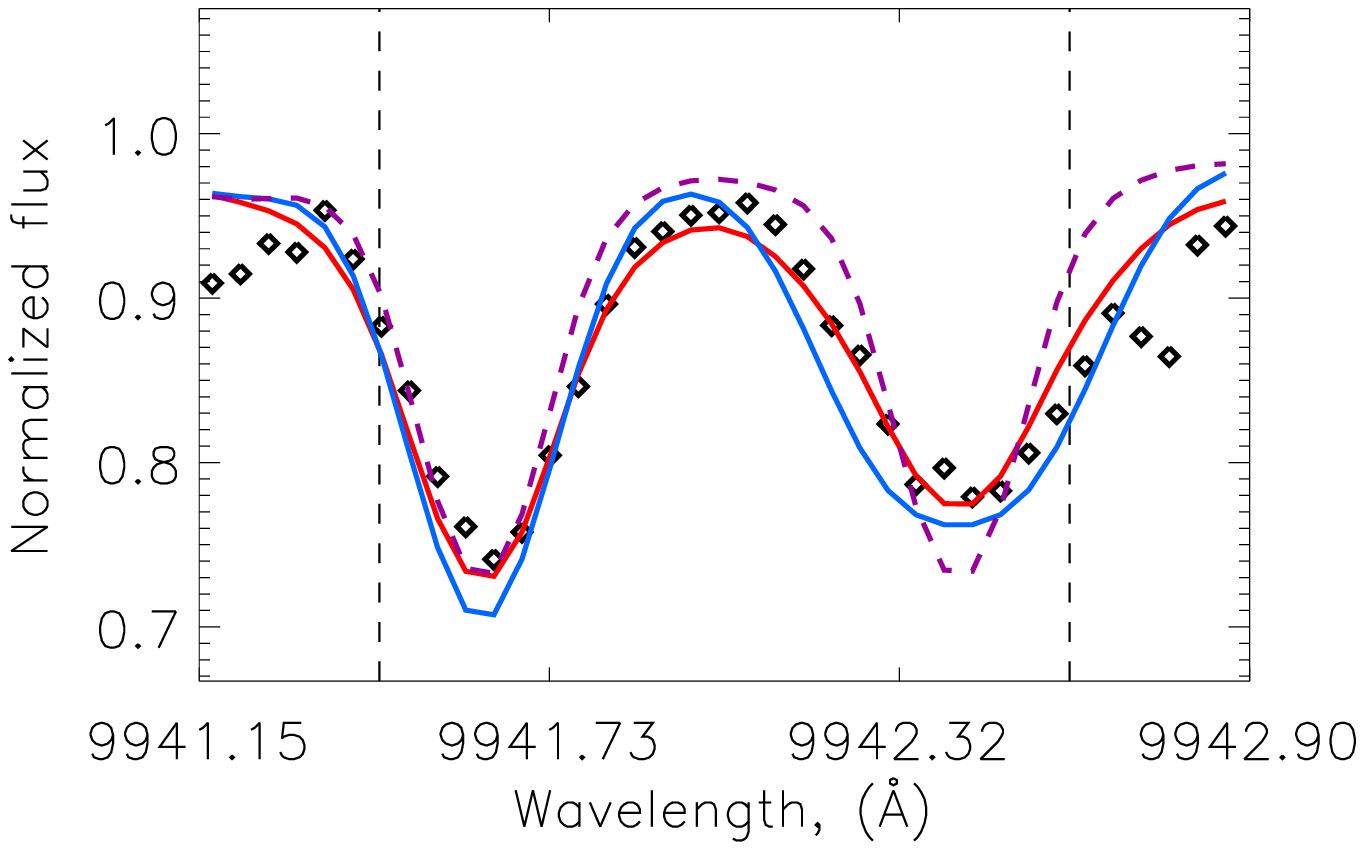}
\includegraphics[width=0.33\hsize]{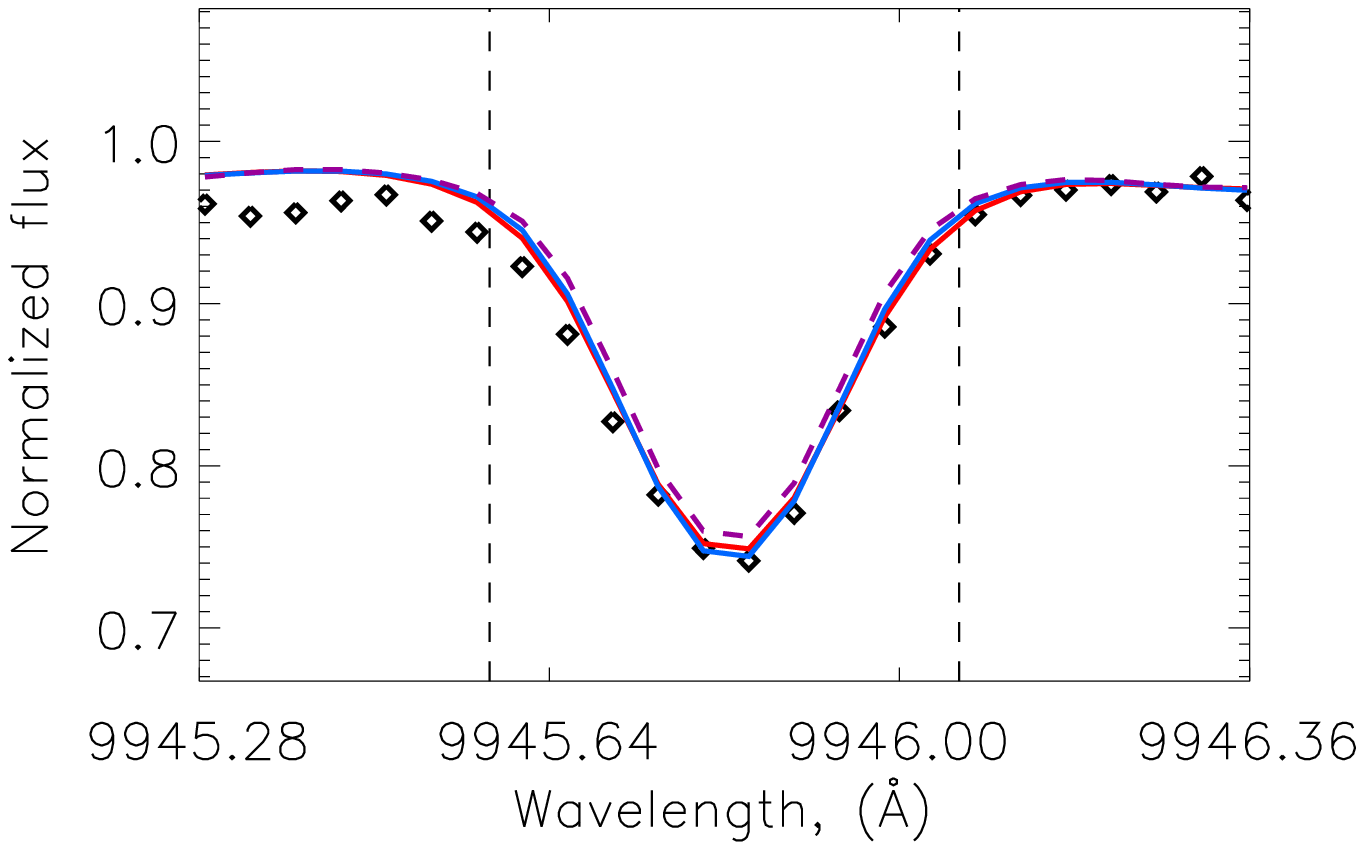}
}
\centerline{
\includegraphics[width=0.33\hsize]{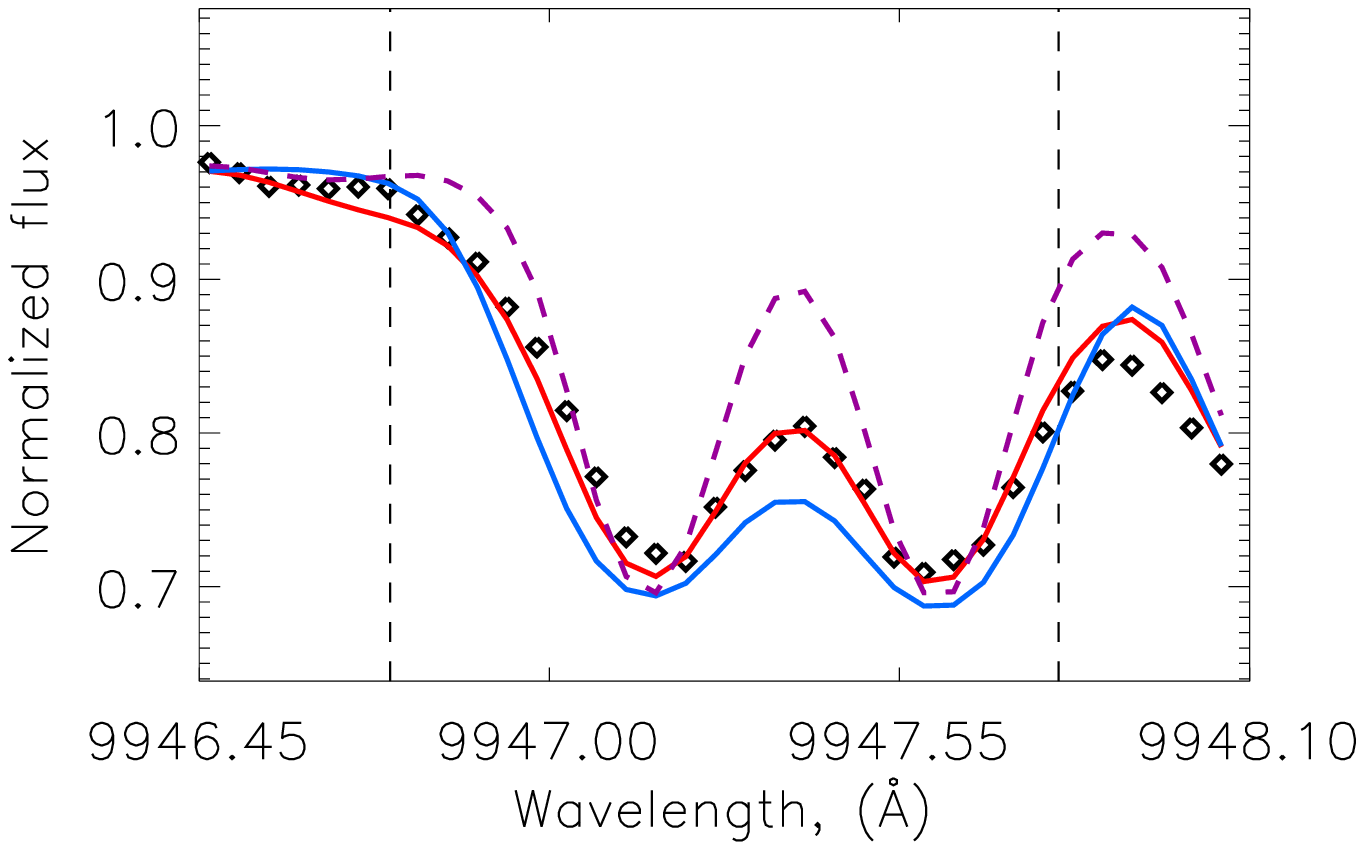}
\includegraphics[width=0.33\hsize]{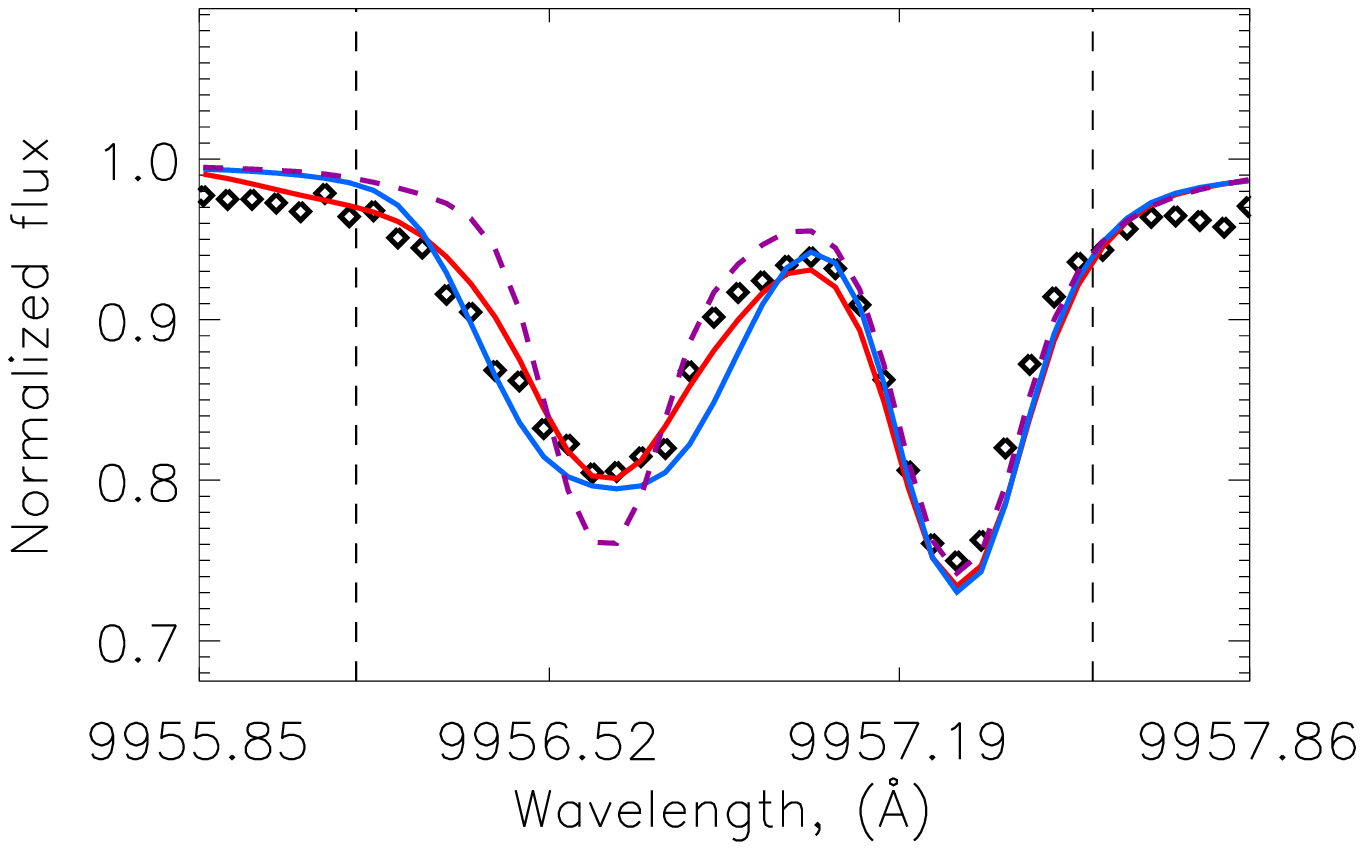}
\includegraphics[width=0.33\hsize]{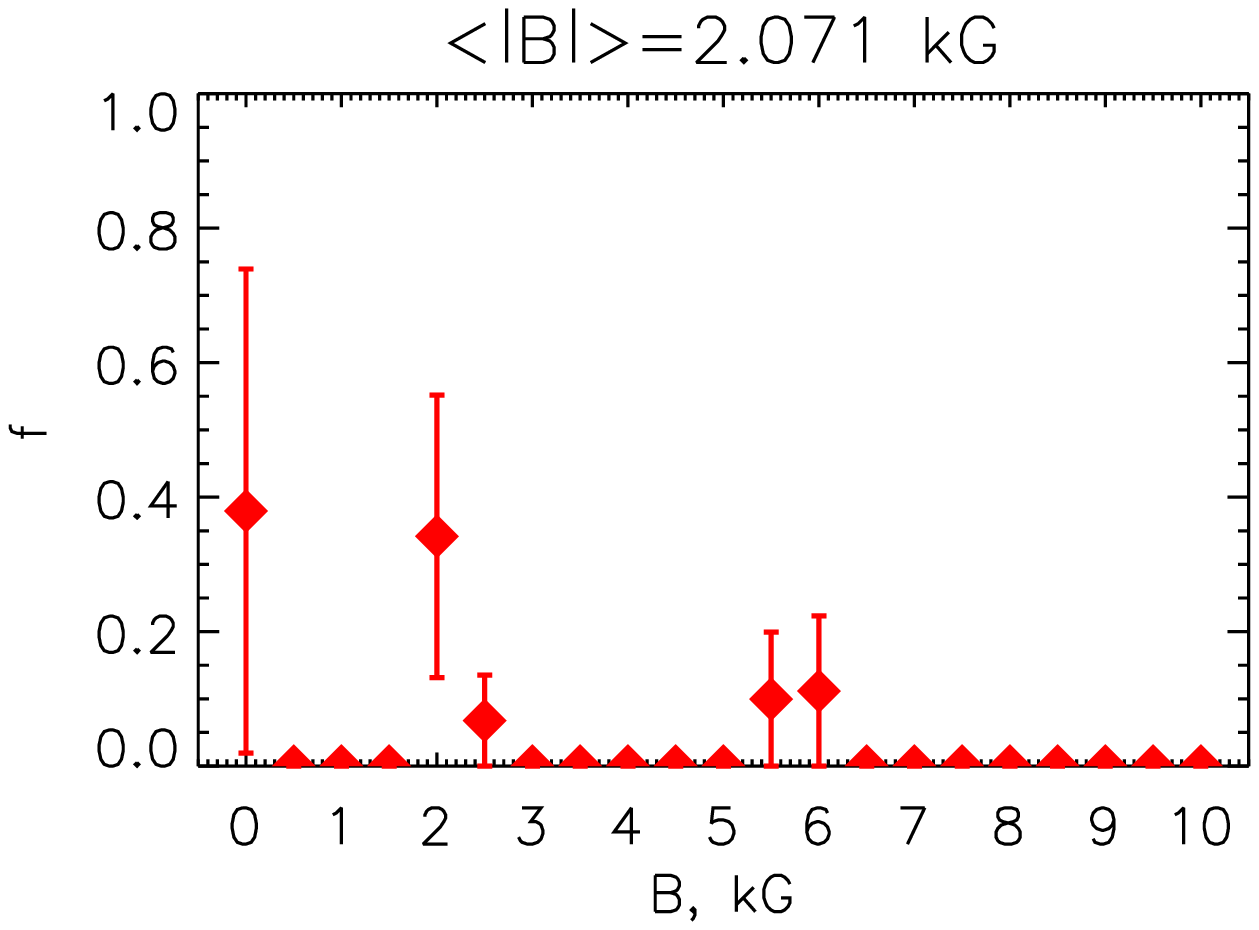}
}
\caption{Same as on Fig.~\ref{fig:gj388-fit} but for GJ~729.
Atmospheric parameters $\teff=3400$~K, $\abn{Fe}=-4.40$, $\vsini=4$~\kms.}
\label{fig:gj729-fit}
\end{figure*}

GJ~729 is classified as M$3.5$ star for which RB07 provide $\vsini\approx4$~\kms\ and $\bs=2.2$~kG.
The best fit between theory and observations was obtained employing atmospheric parameters
from RB07 and $\abn{Fe}=-4.40$. Assuming solar iron abundance $\abn{Fe}=-4.54$
results in {a} lower effective temperature of $\teff=3300$~K in order to obtain the same quality of the fit. 
The distributions of filling factors look very similar for these two sets of atmospheric
parameters. Because the first set provides formally a better fit, we prefer it to the latter.
Figure~\ref{fig:gj729-fit} illustrates {the} match between theory and observations.
The magnetic field  of the star appears to be distributed over three groups of components.
For the RC model we find components with the small $1-1.5$~kG, moderate $2.5-3$~kG,
and strong $5.5$~kG intensities. Although errors bars are large, the existence of such distinct features 
as FeH $\lambda9906$  demonstrate {that} well-separated deep cores only appear with $2.5-3$~kG
magnetic field components so that their significance is well constrained.
{A similar three-like component distribution is also recovered when assuming MC model,
but as in the previous case of the GJ~388, a strong zero-field component appears ($f=0.4$) along with 
a $2$~kG component} of the similar strength. The mean magnetic fields are also a bit different and amount to
$\bs=2.3$~kG and $\bs=2.1$~kG respectively. Finally, we {do not prefer either} of the two cases by looking
at individual line profiles.

\subsection{GJ~285 (YZ~CMi)}

\begin{figure*}
\centerline{
\includegraphics[width=0.33\hsize]{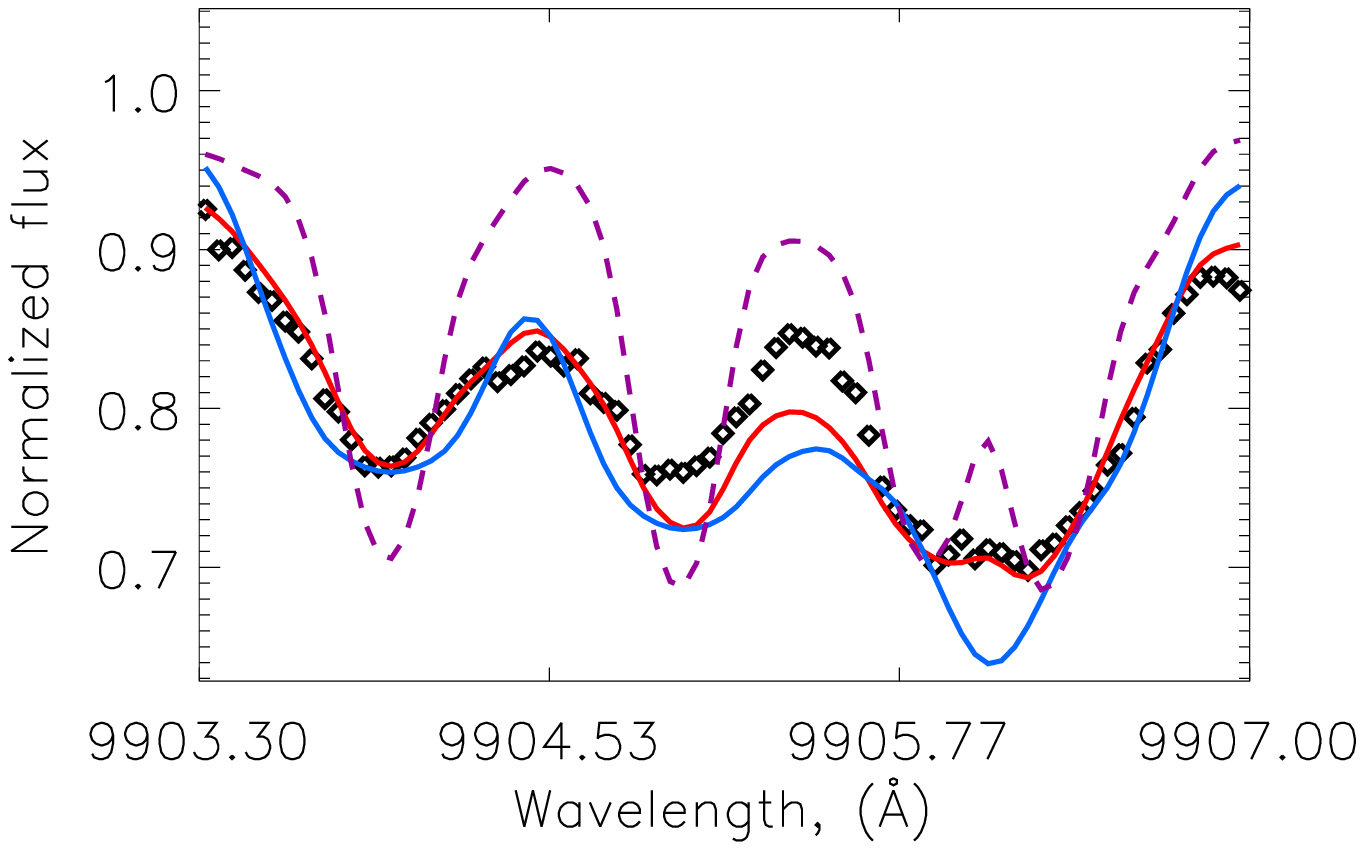}
\includegraphics[width=0.33\hsize]{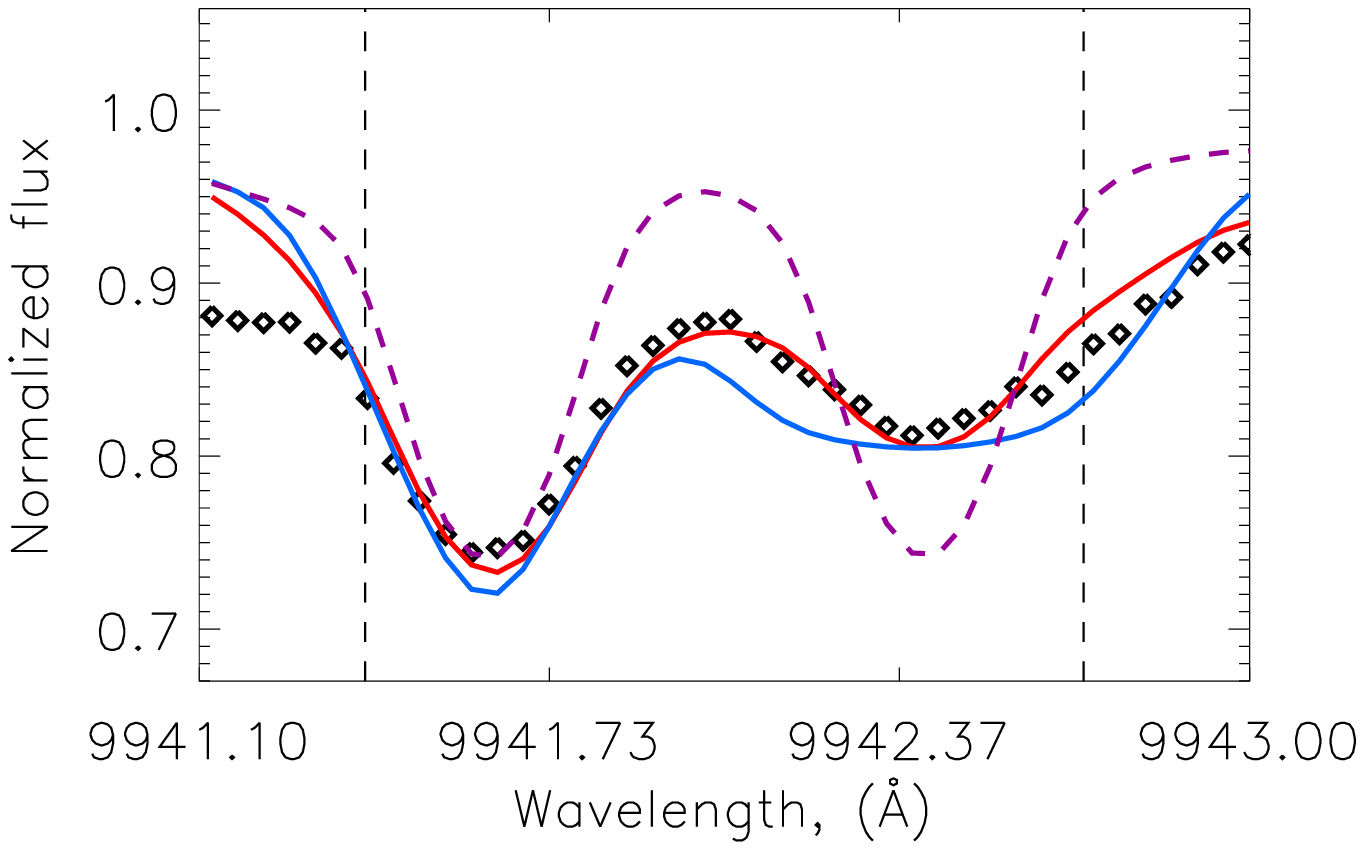}
\includegraphics[width=0.33\hsize]{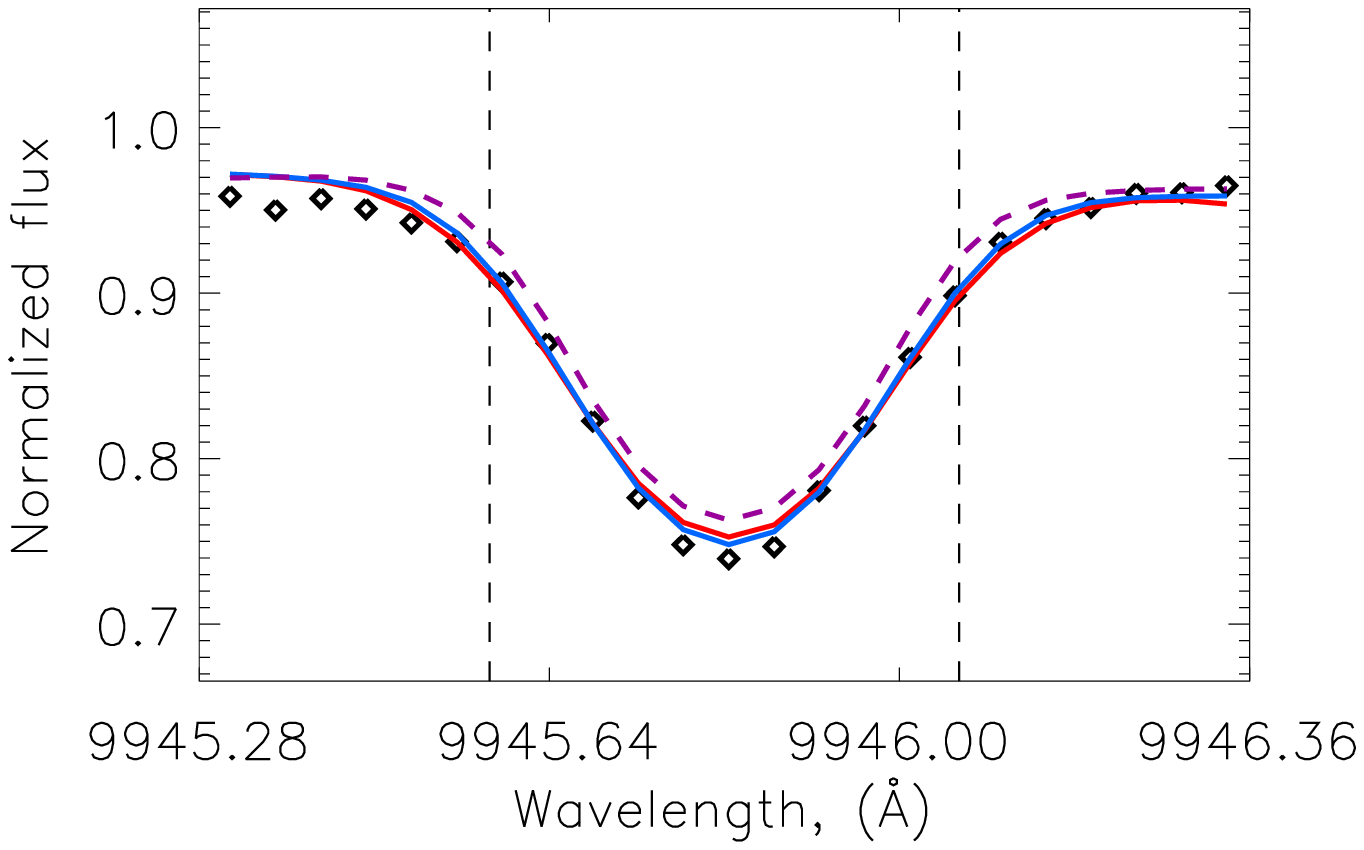}
}
\centerline{
\includegraphics[width=0.33\hsize]{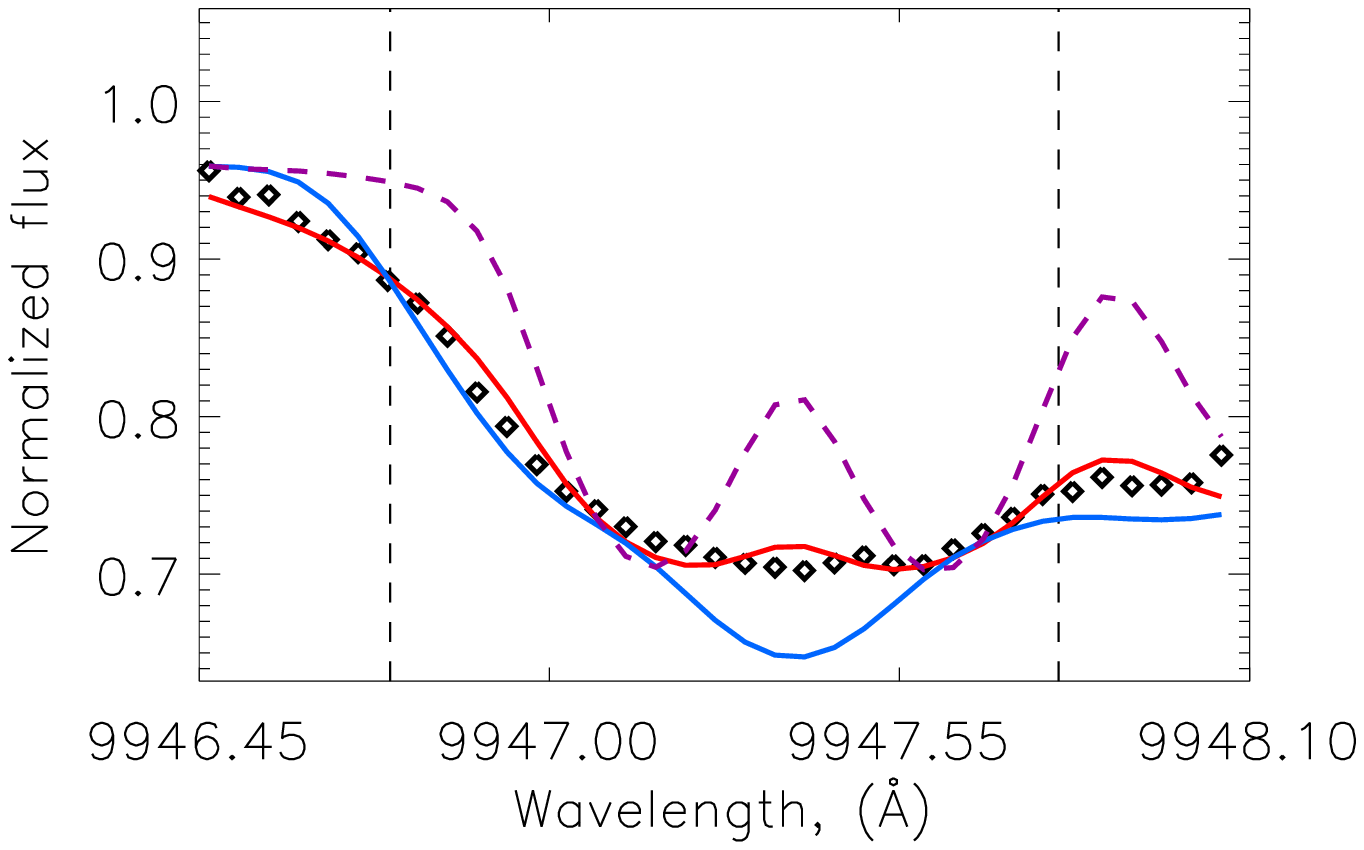}
\includegraphics[width=0.33\hsize]{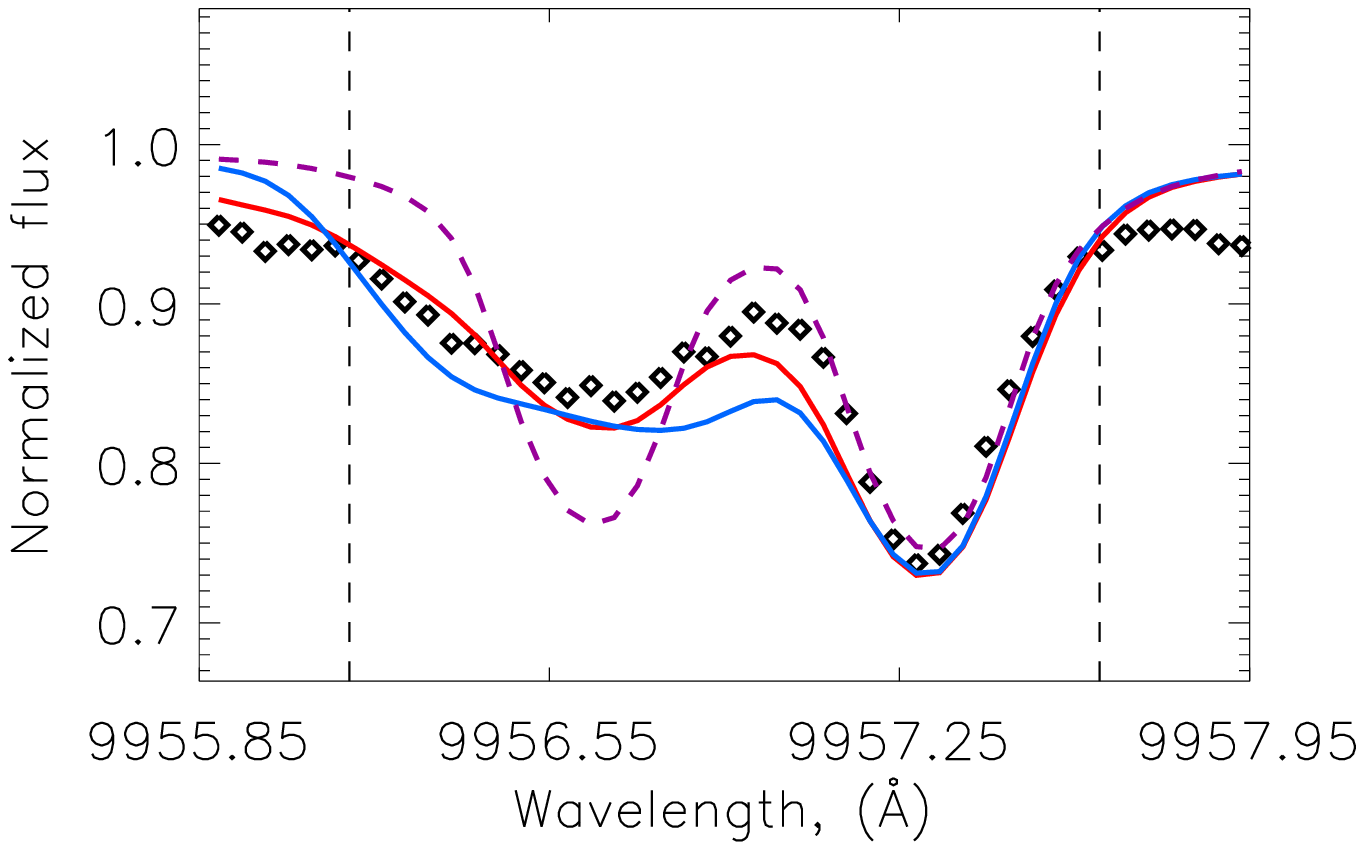}
\includegraphics[width=0.33\hsize]{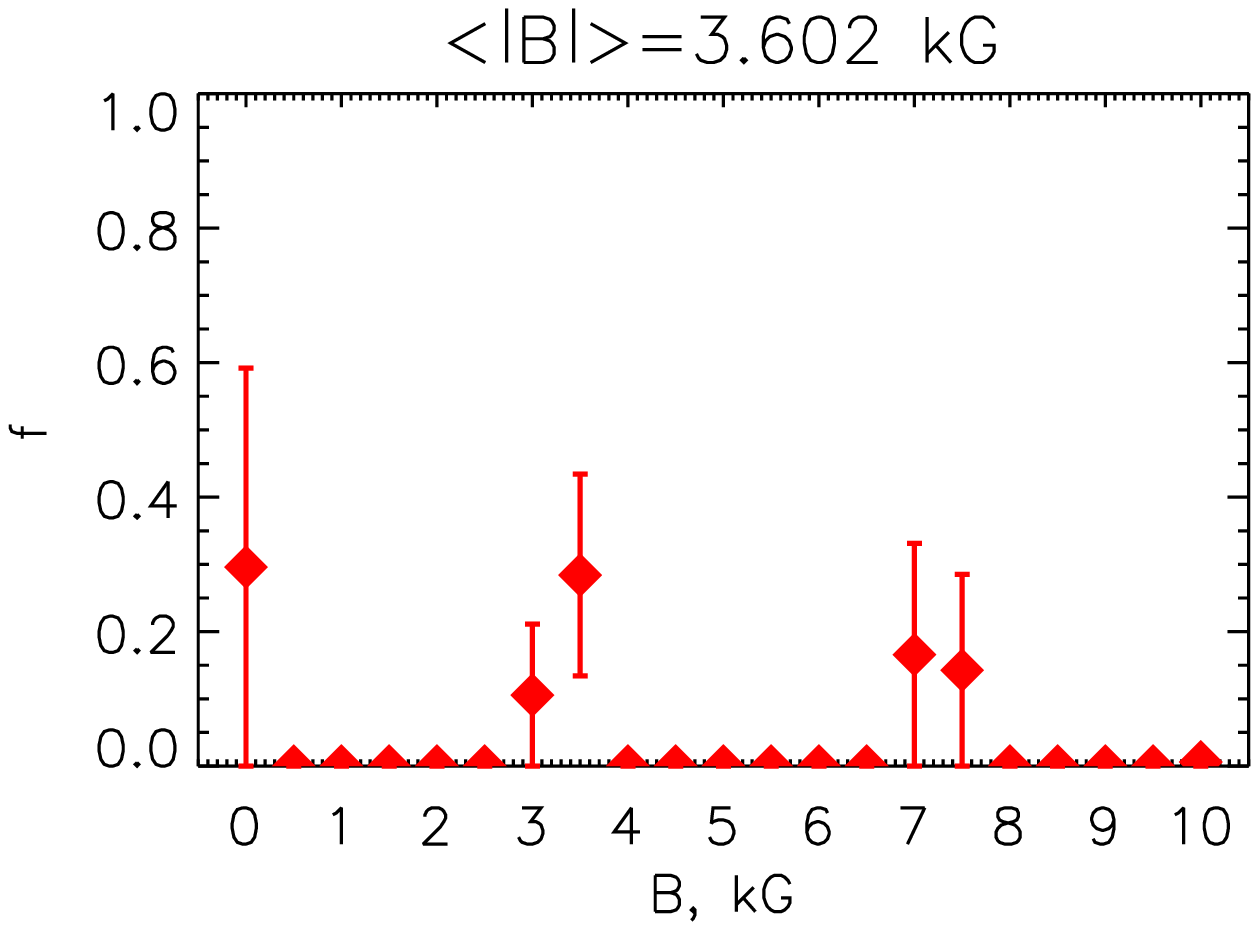}
}
\centerline{
\includegraphics[width=0.33\hsize]{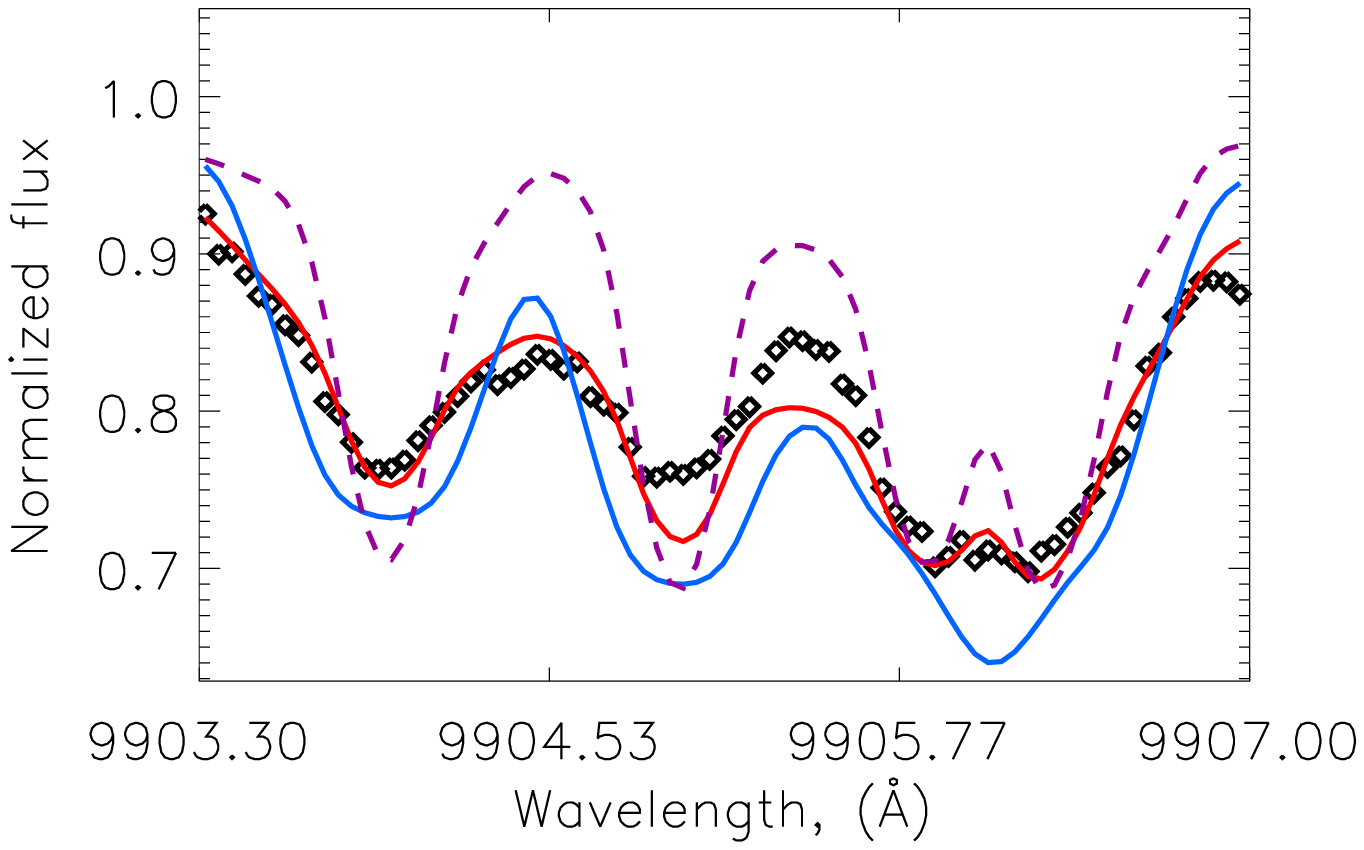}
\includegraphics[width=0.33\hsize]{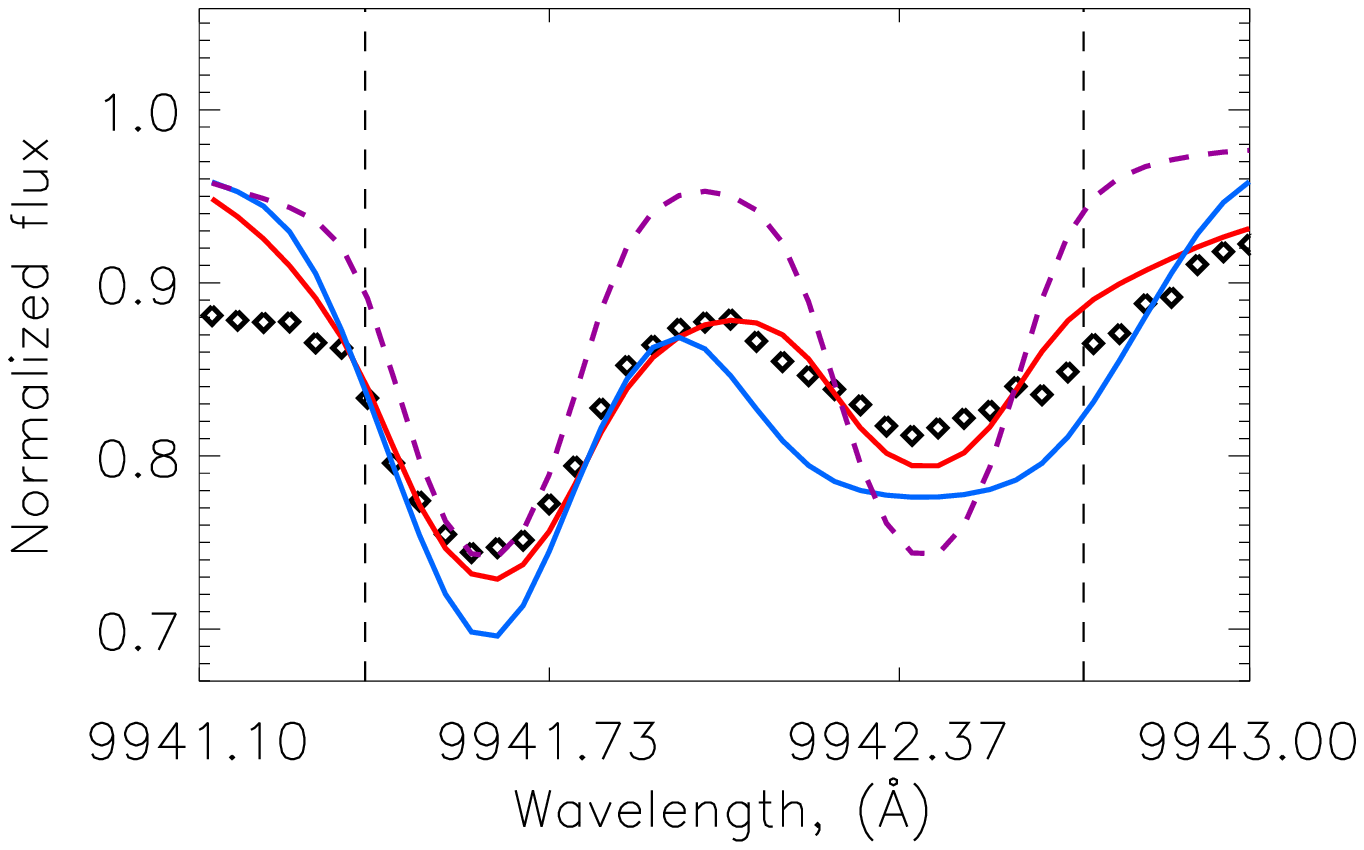}
\includegraphics[width=0.33\hsize]{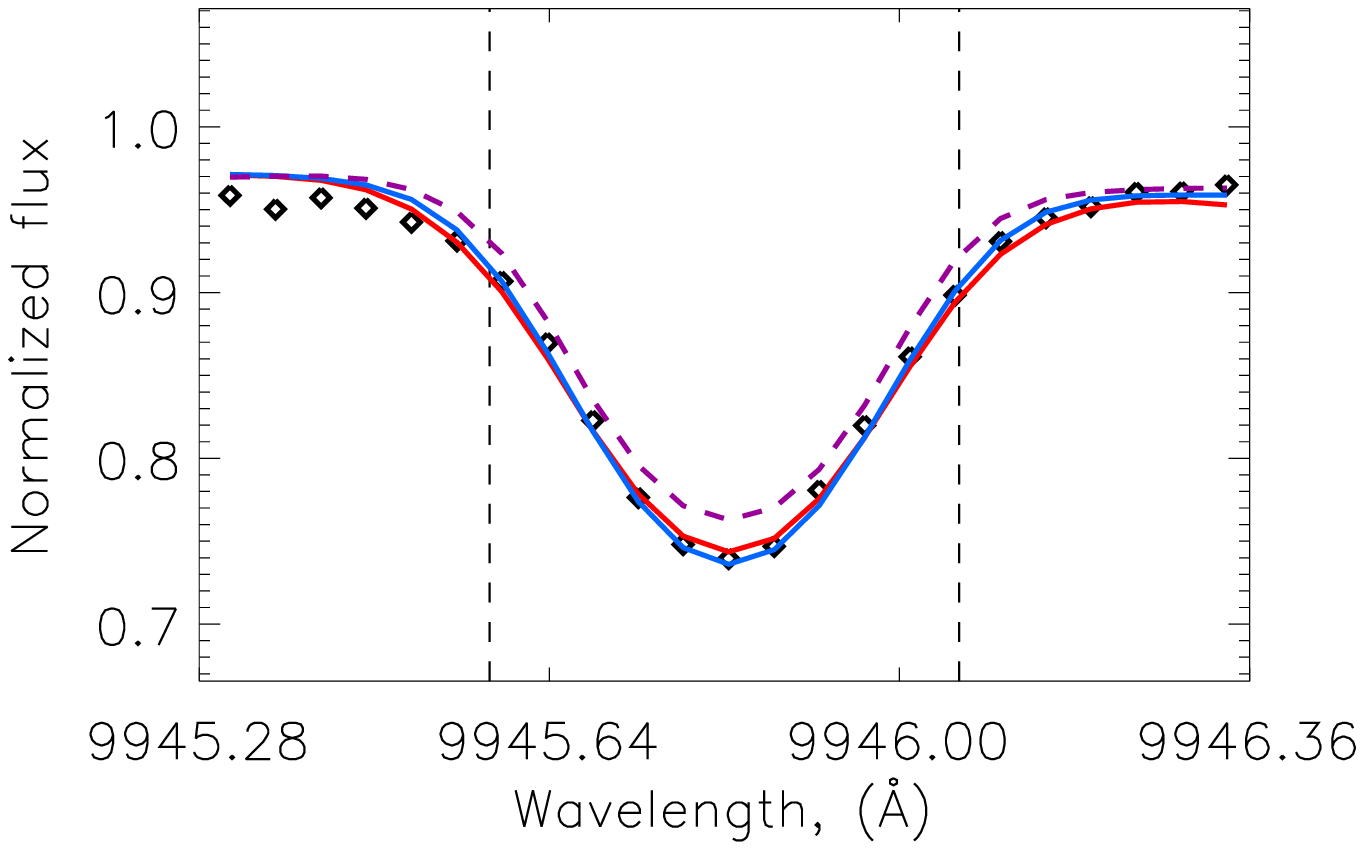}
}
\centerline{
\includegraphics[width=0.33\hsize]{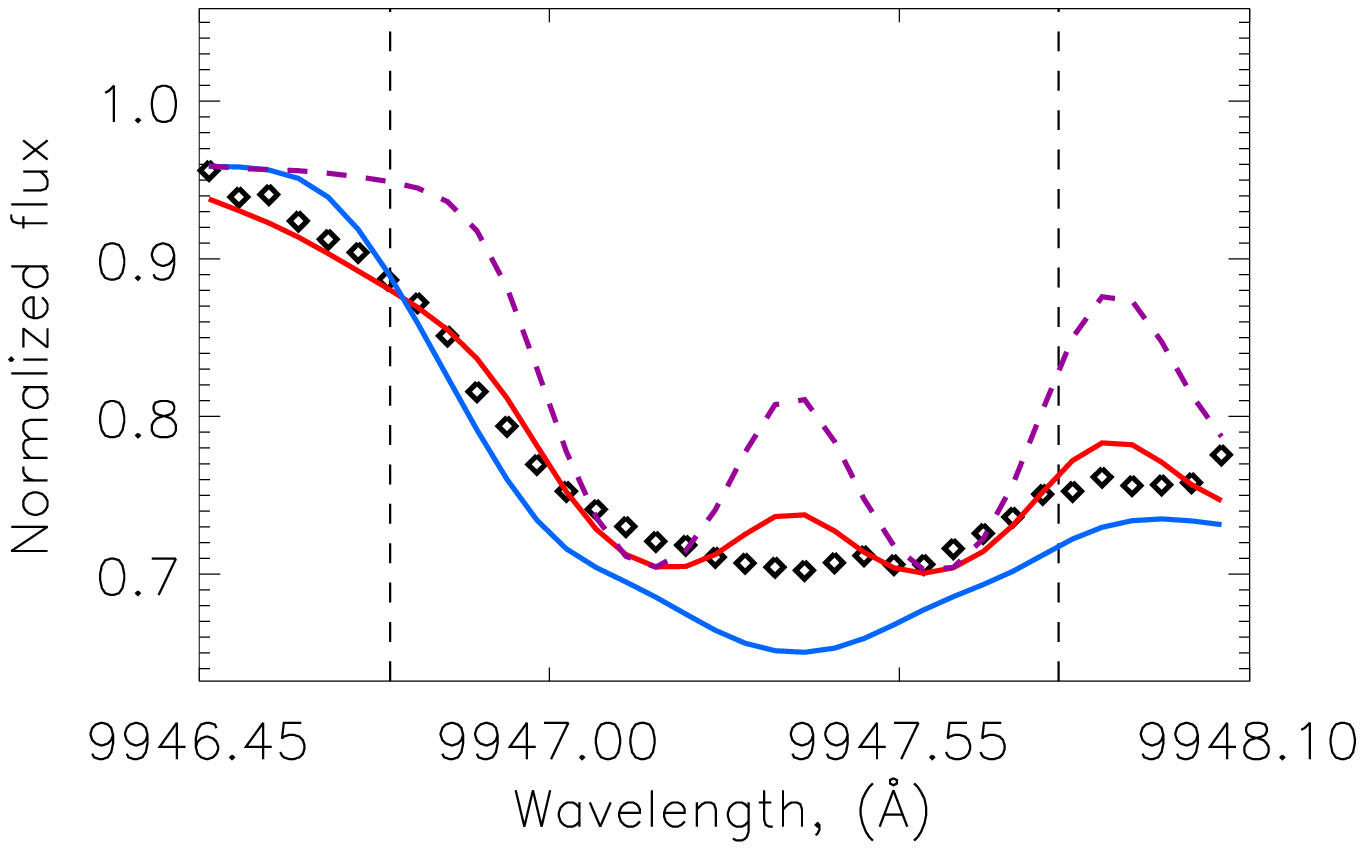}
\includegraphics[width=0.33\hsize]{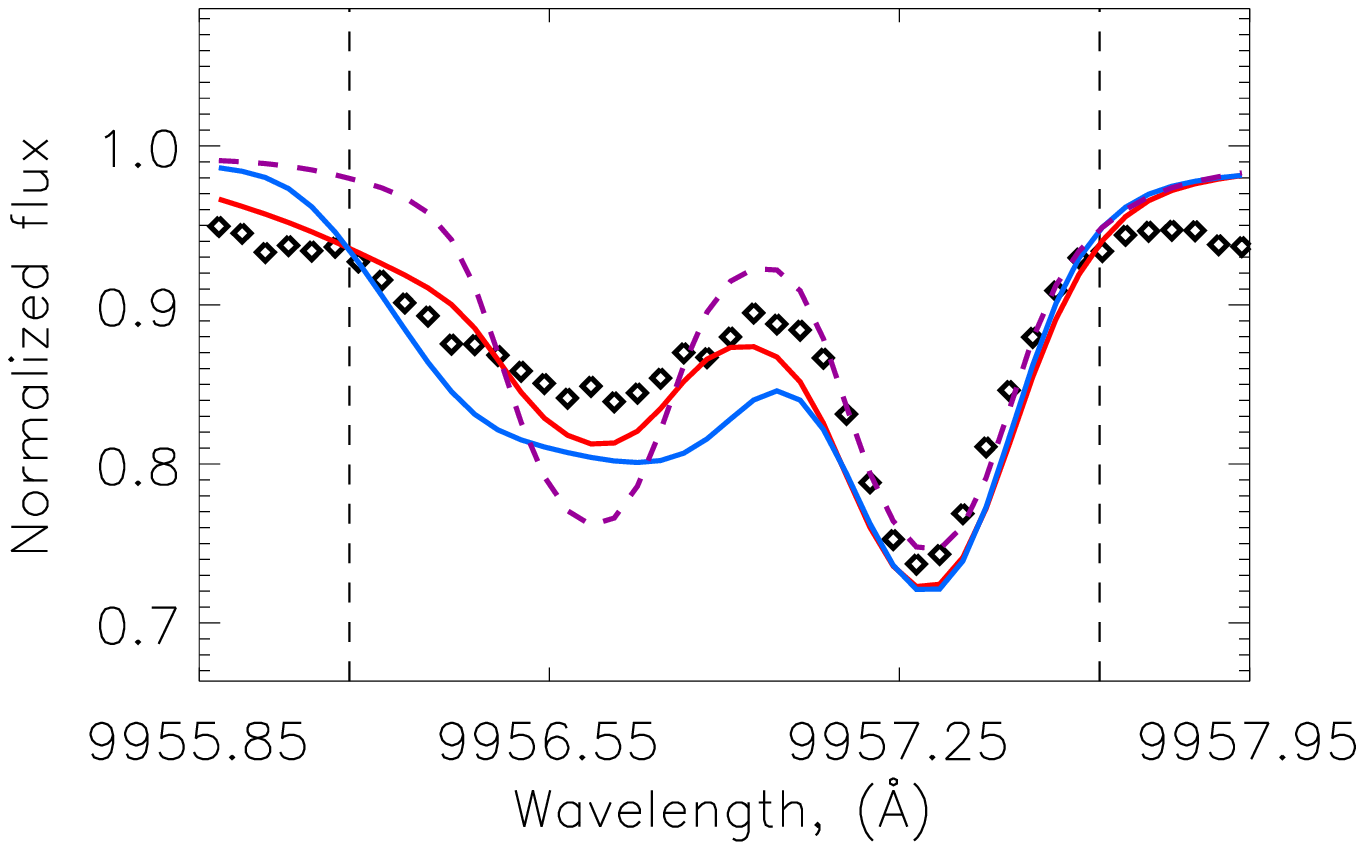}
\includegraphics[width=0.33\hsize]{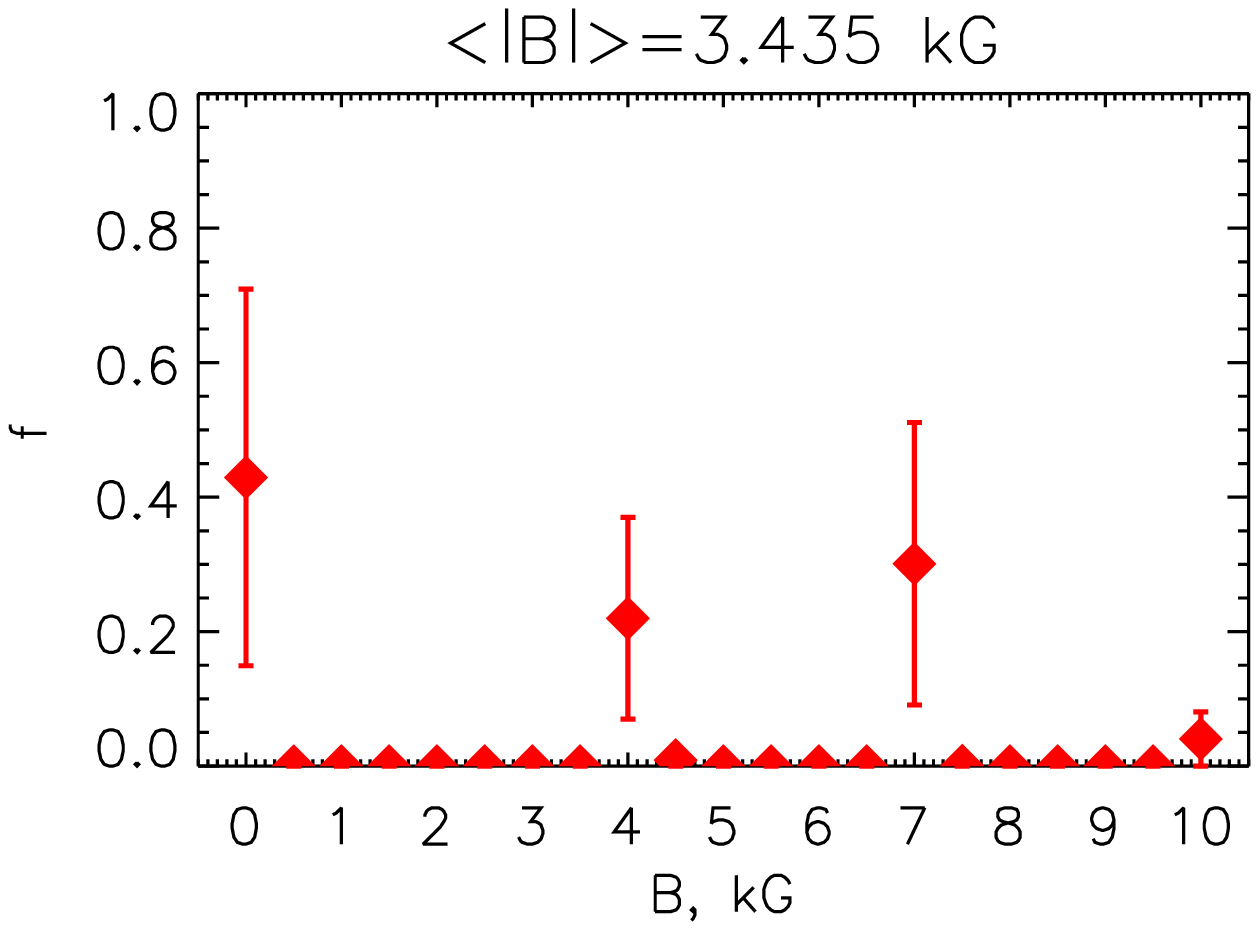}
}
\caption{Same as on Fig.~\ref{fig:gj388-fit} but for GJ~285.
Atmospheric parameters $\teff=3300$~K, $\abn{Fe}=-4.37$, $\vsini=6$~\kms.}
\label{fig:gj285-fit}
\end{figure*}

GJ~285 is another very active M$4.5$ dwarf with rotational velocity of about $\vsini\approx5$~\kms. RB07 derived
a lower limit for the mean surface magnetic field of $\approx3.9$~kG. We find a stable solution for filling factors
using an enhanced iron abundance and a faster rotation velocity of $\vsini=6$~\kms, as illustrated in Fig.~\ref{fig:gj285-fit}.
The RC model reveals five significant magnetic field components: one zero field, two components with
$3$~kG and $3.5$~kG field, and two with very strong $7$~kG and $7.5$~kG fields. They give in total a mean magnetic field
of $\bs=3.6$~kG, which is $0.3$~kG lower then reported in RB07. Employing  $\vsini=5$~\kms\ results  in $\bs=3.9$~kG,
but this increase in $\bs$ is then caused by the appearance of a non-zero $10$~kG field component which (as we {argued} for the case 
of GJ~1002) is likely spurious because it lowers $\chi^2$ by a better match to the far wings only.
Using $\vsini=6$~\kms\ provides the best overall fit between observed and predicted spectra.
The MC model gives a very similar distribution of filling factors with a zero-field, $4$~kG,
and $7$~kG components. But the resulting mean magnetic field is then $0.2$~kG lower compared to RC model, 
i.e. $\bs=3.4$~kG.
From Fig.~\ref{fig:gj285-fit} one can see that the fit to FeH lines is not always accurate. 
Lines like $\lambda9905$ are possibly affected by data reduction {issues}
because they are fit accurately in spectra of, e.g., GJ~388. Next, the observed line {at} $\lambda9947$ 
does not seem to show {a doubled
profile but the synthetic spectra predicts one}. This may tell us that the intensity of the magnetic field could be even 
lower than what we have found here. From the comparison of line profiles it is clear that the RC model
provide a better fit to observations. Lines like FeH $\lambda9942$, and $\lambda9956$ are obviously
too strong and the central feature of $\lambda9947$ is not reproduced when using the MC model.

\subsection{GJ~406 (CN~Leo)}

\begin{figure*}
\centerline{
\includegraphics[width=0.33\hsize]{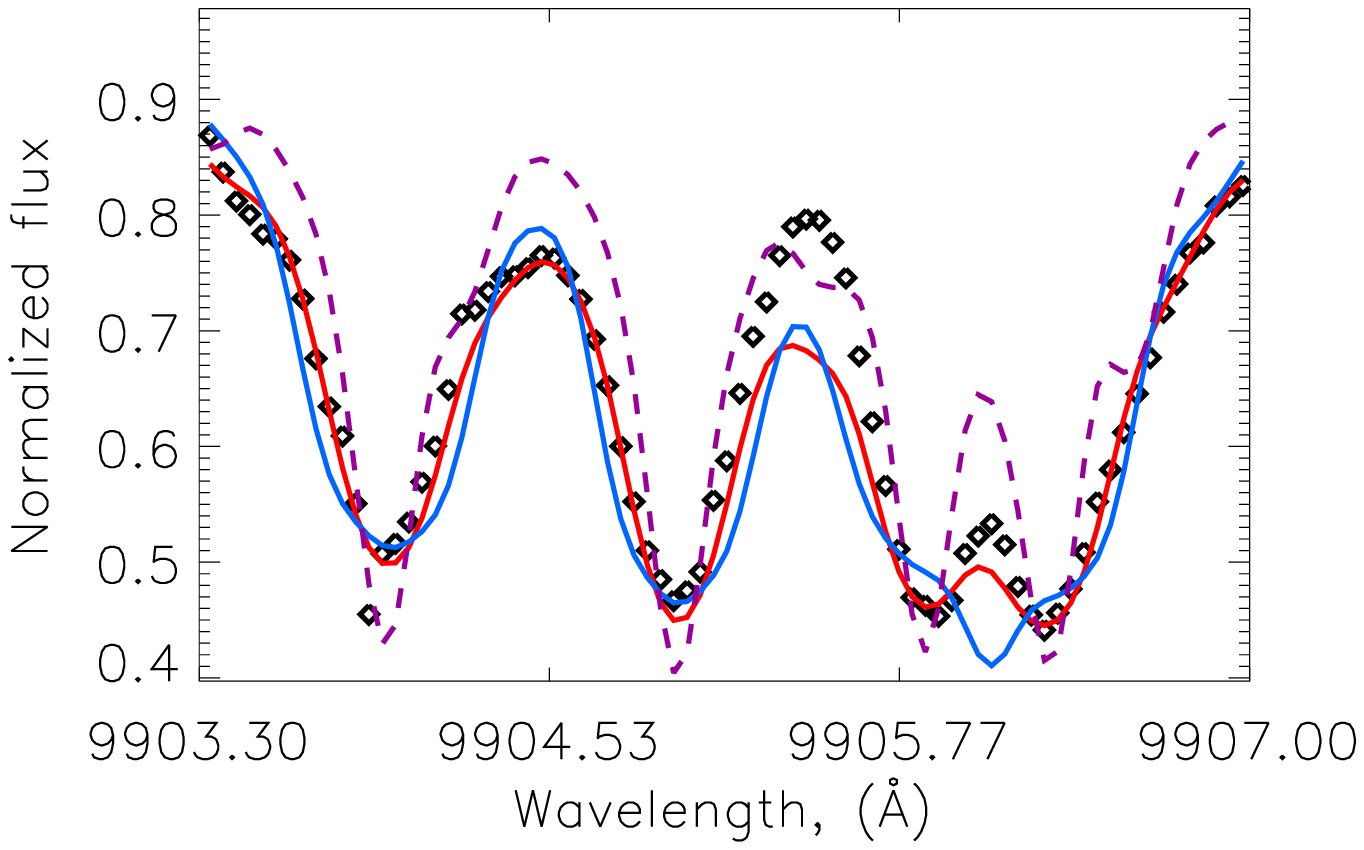}
\includegraphics[width=0.33\hsize]{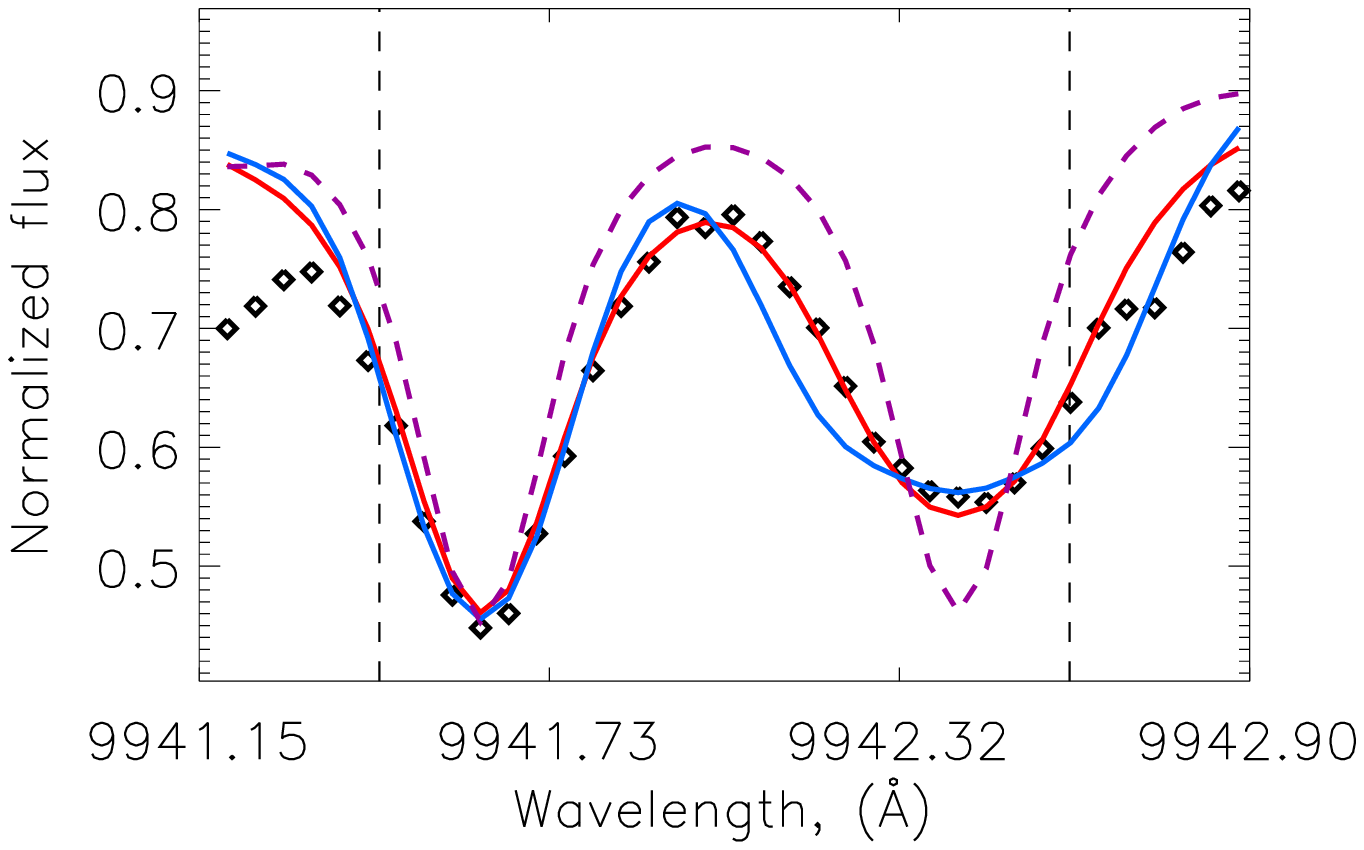}
\includegraphics[width=0.33\hsize]{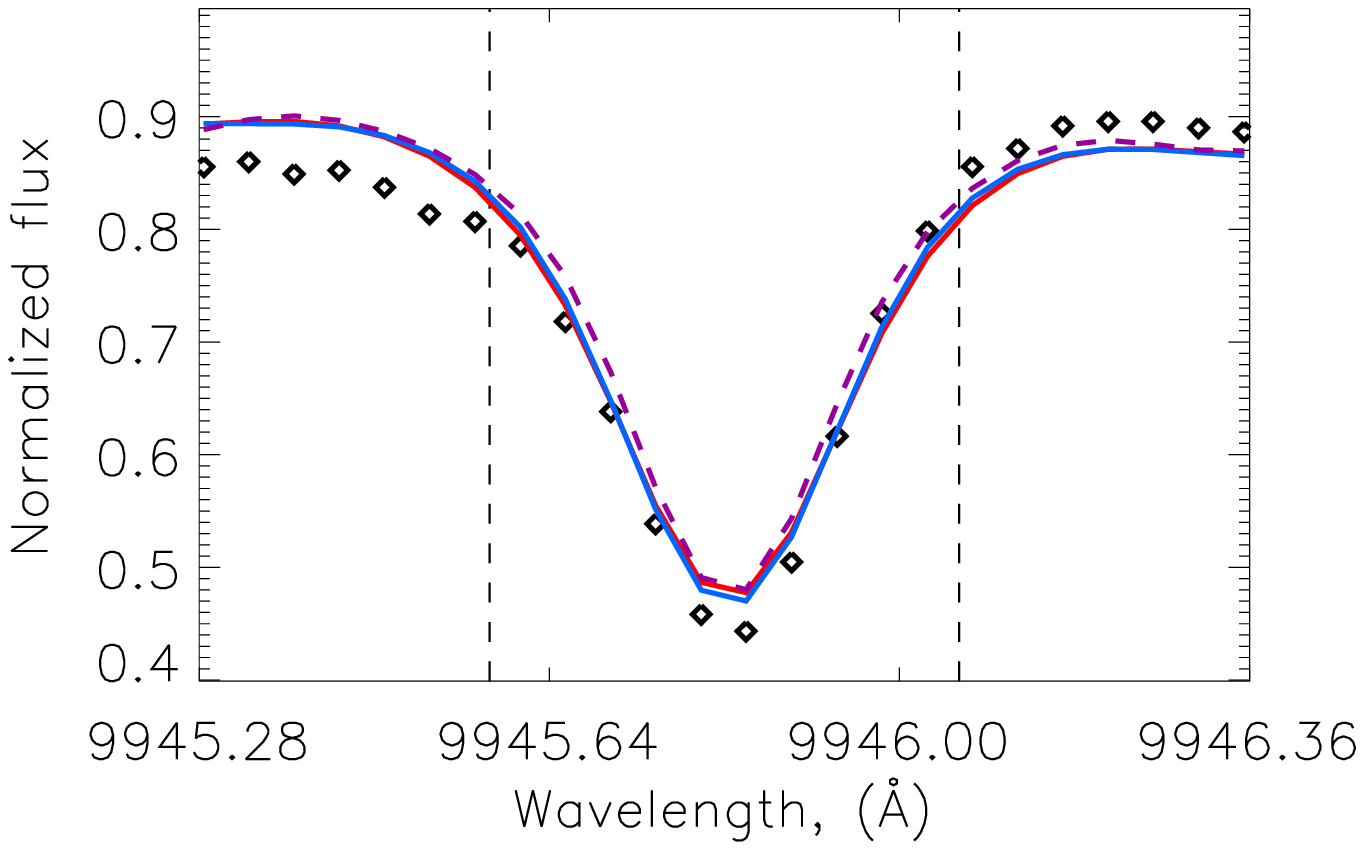}
}
\centerline{
\includegraphics[width=0.33\hsize]{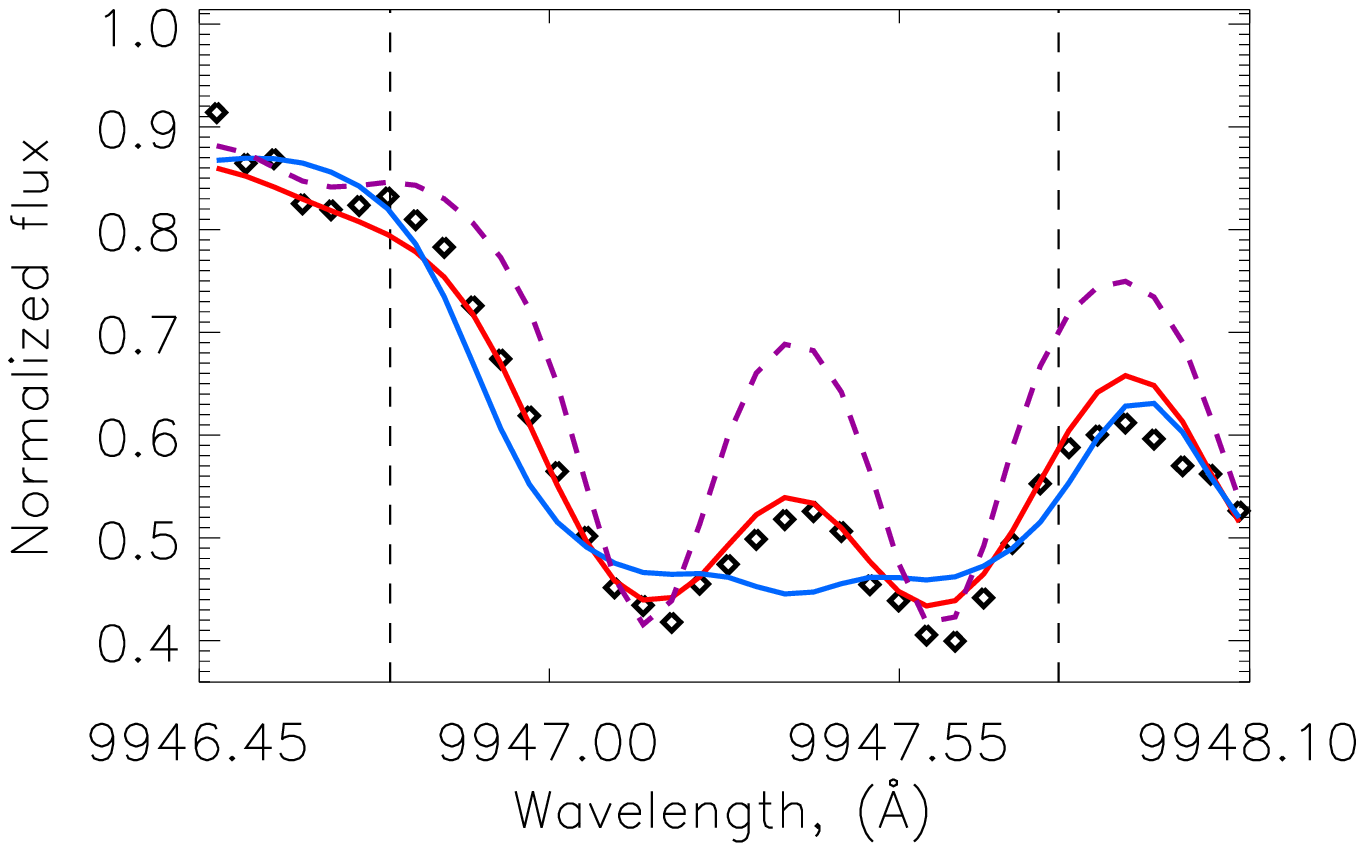}
\includegraphics[width=0.33\hsize]{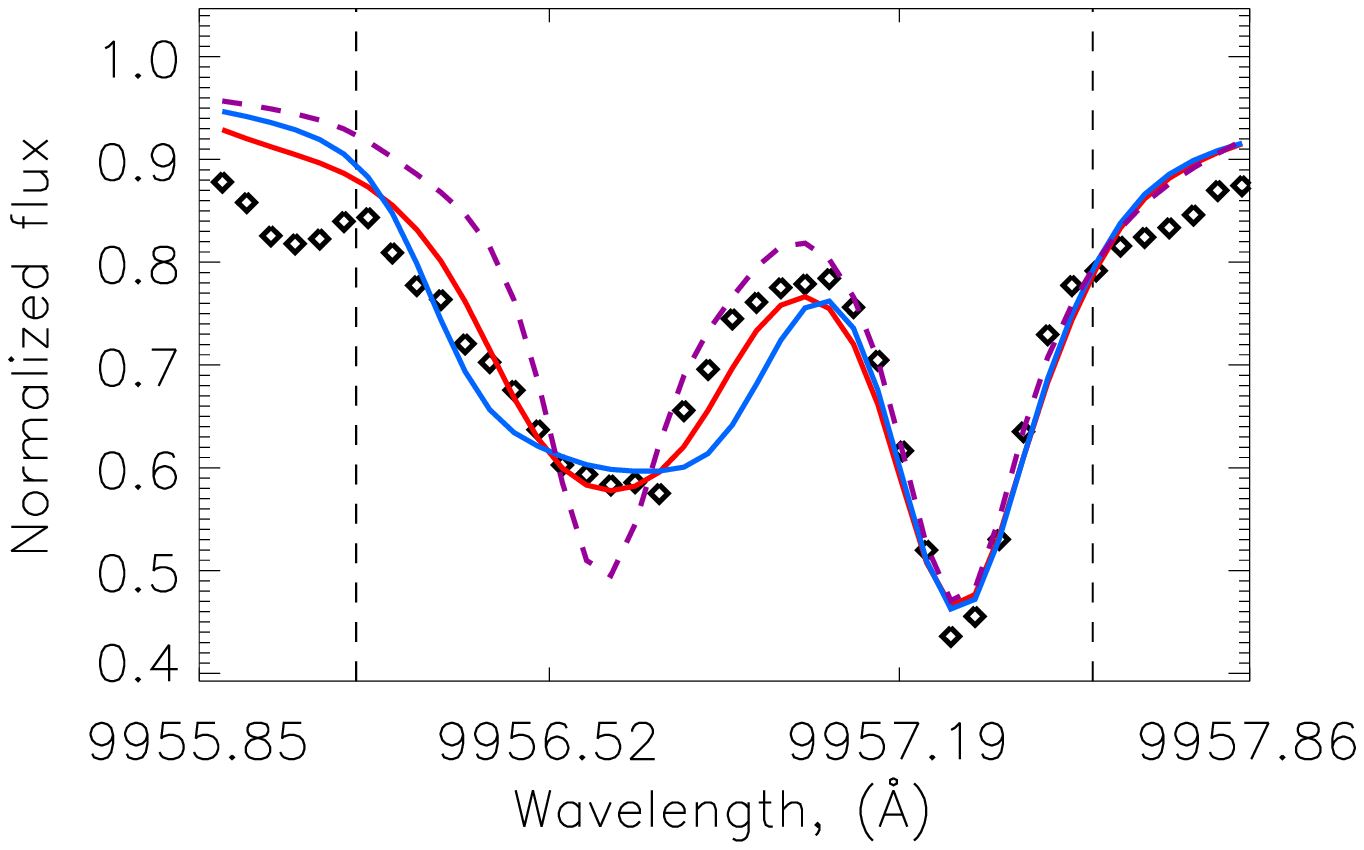}
\includegraphics[width=0.33\hsize]{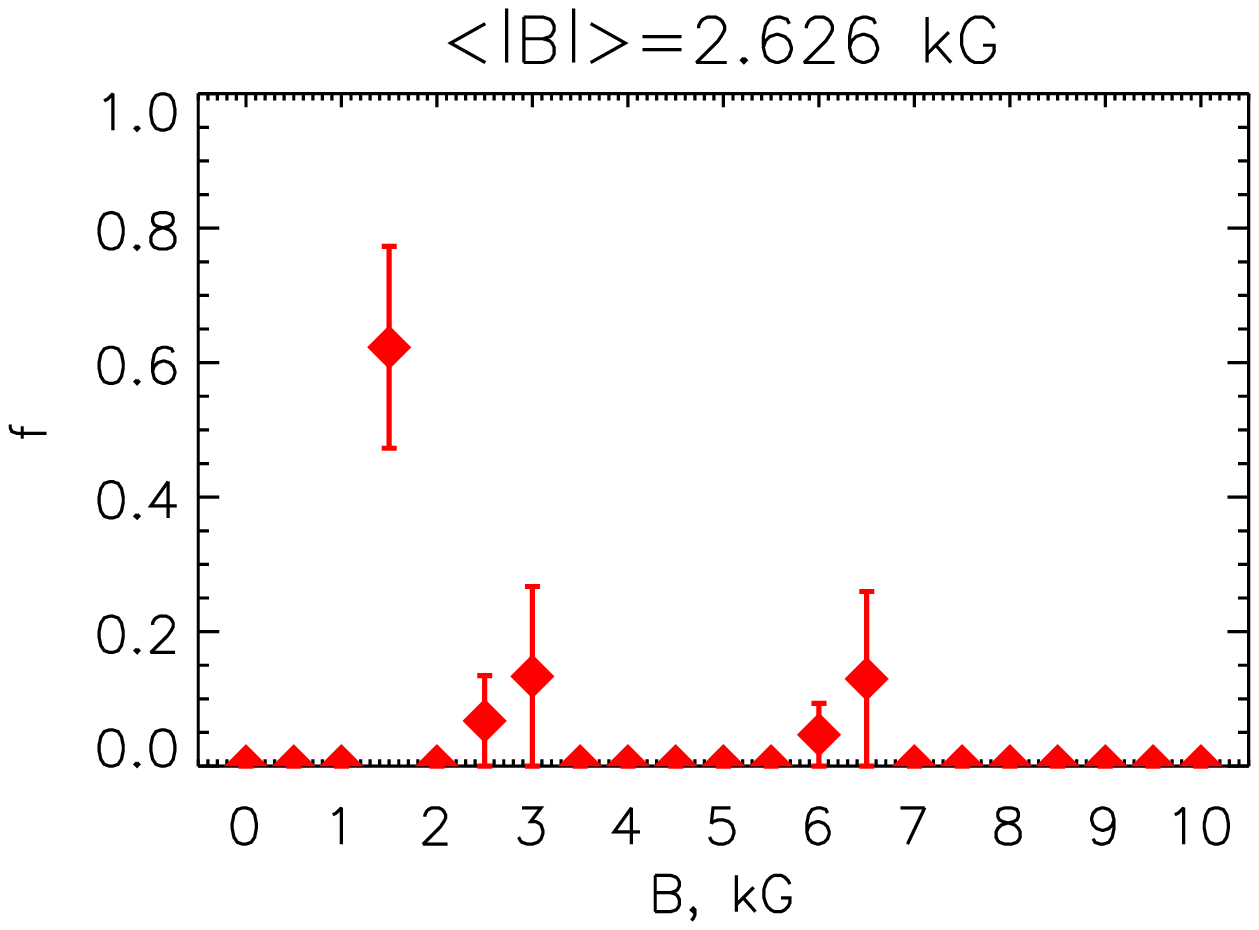}
}
\centerline{
\includegraphics[width=0.33\hsize]{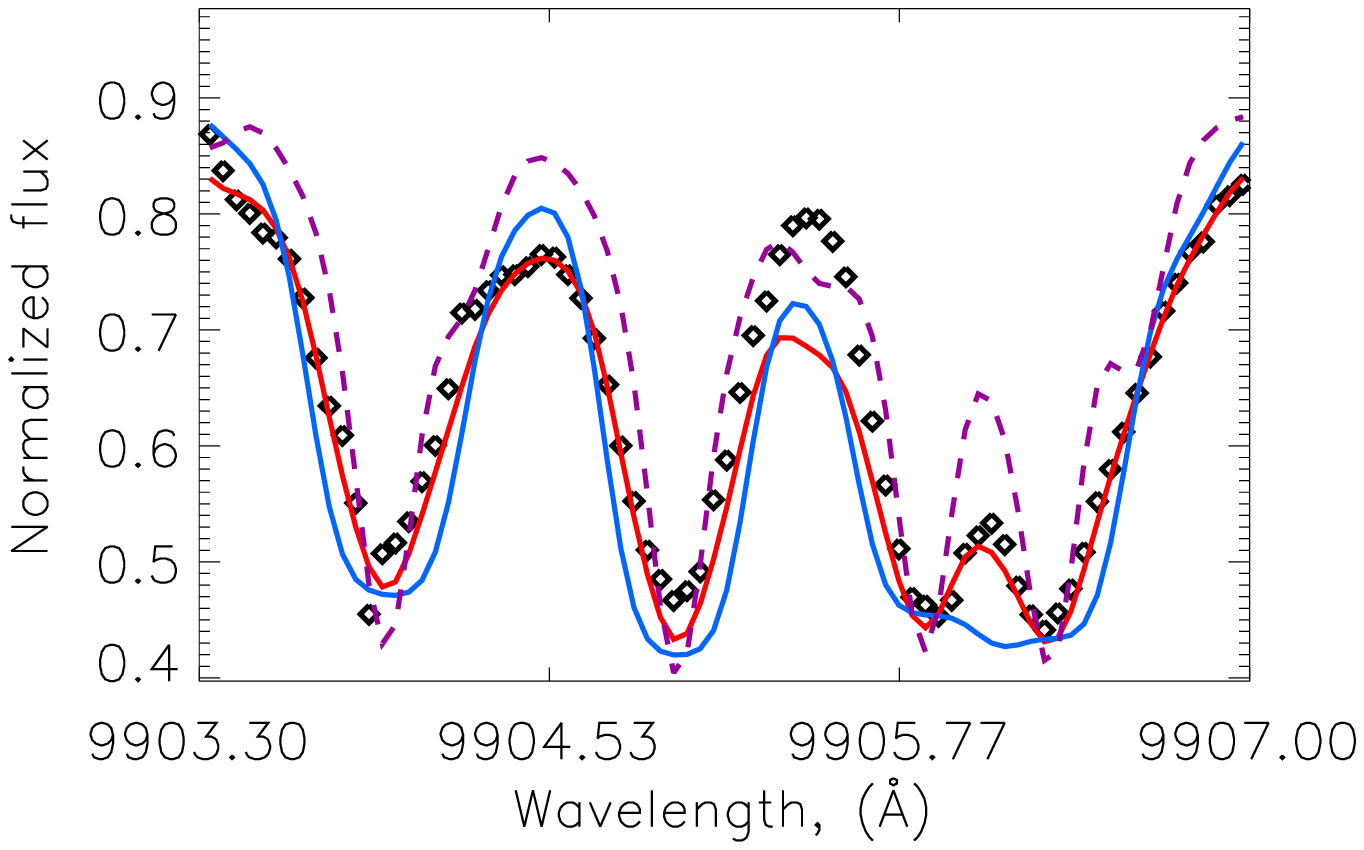}
\includegraphics[width=0.33\hsize]{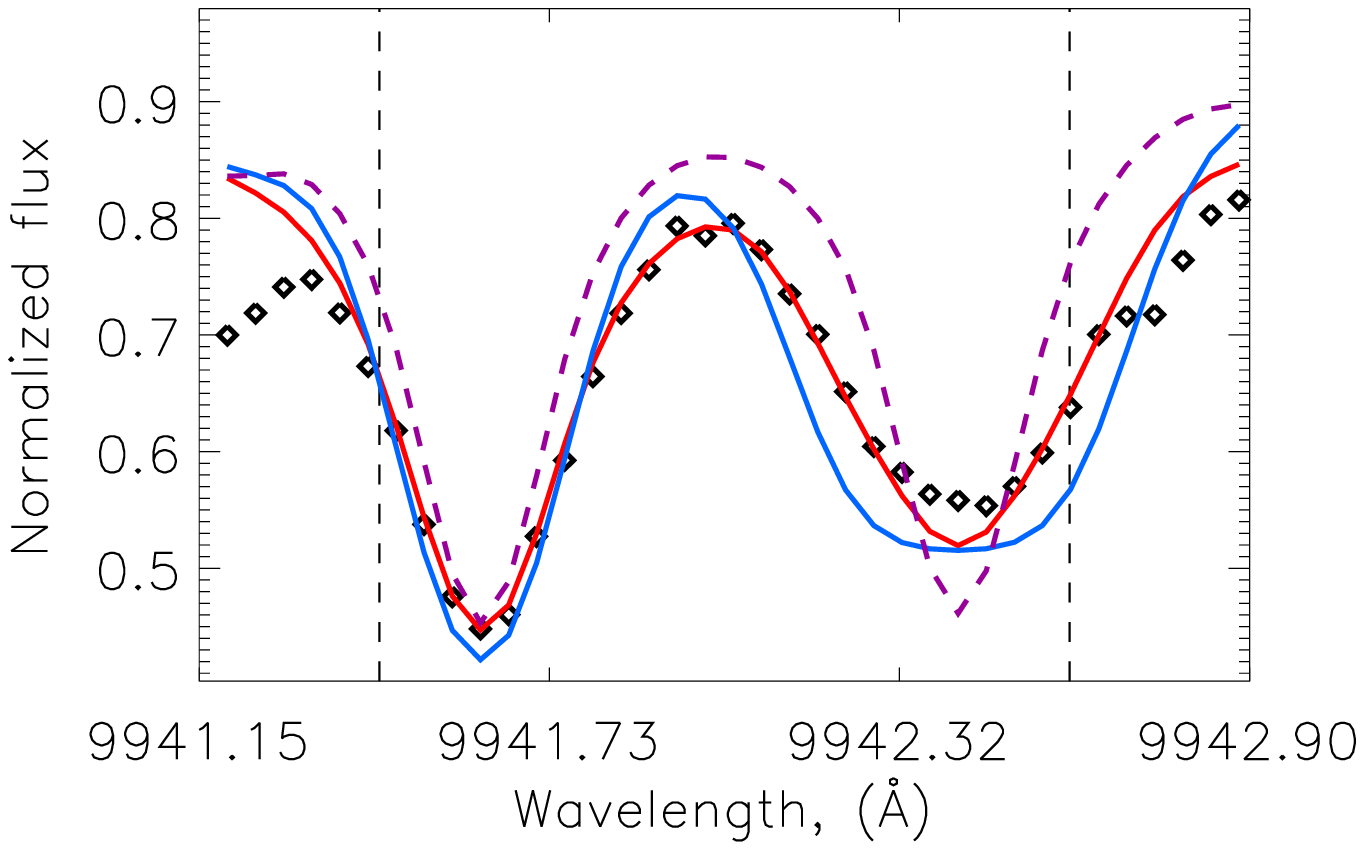}
\includegraphics[width=0.33\hsize]{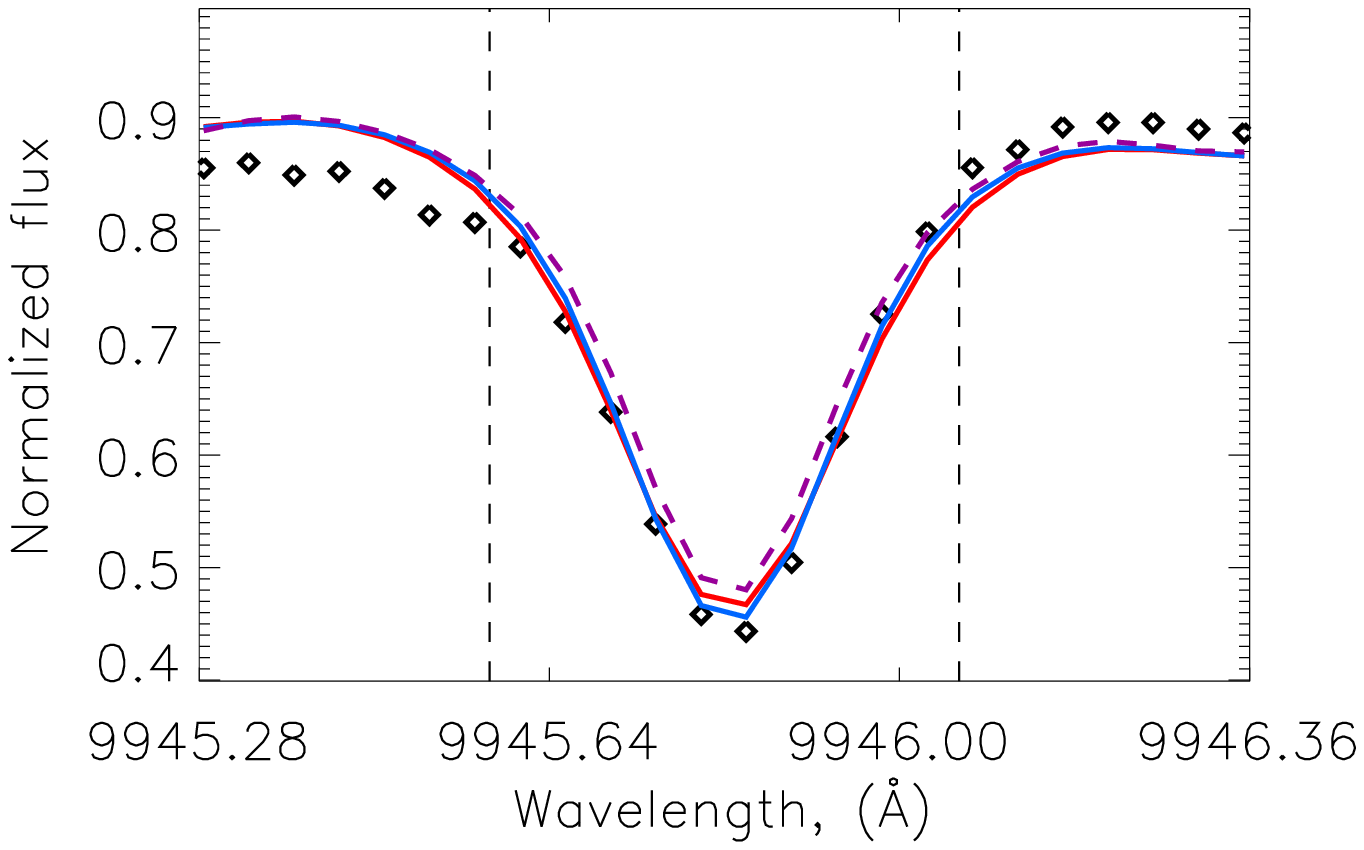}
}
\centerline{
\includegraphics[width=0.33\hsize]{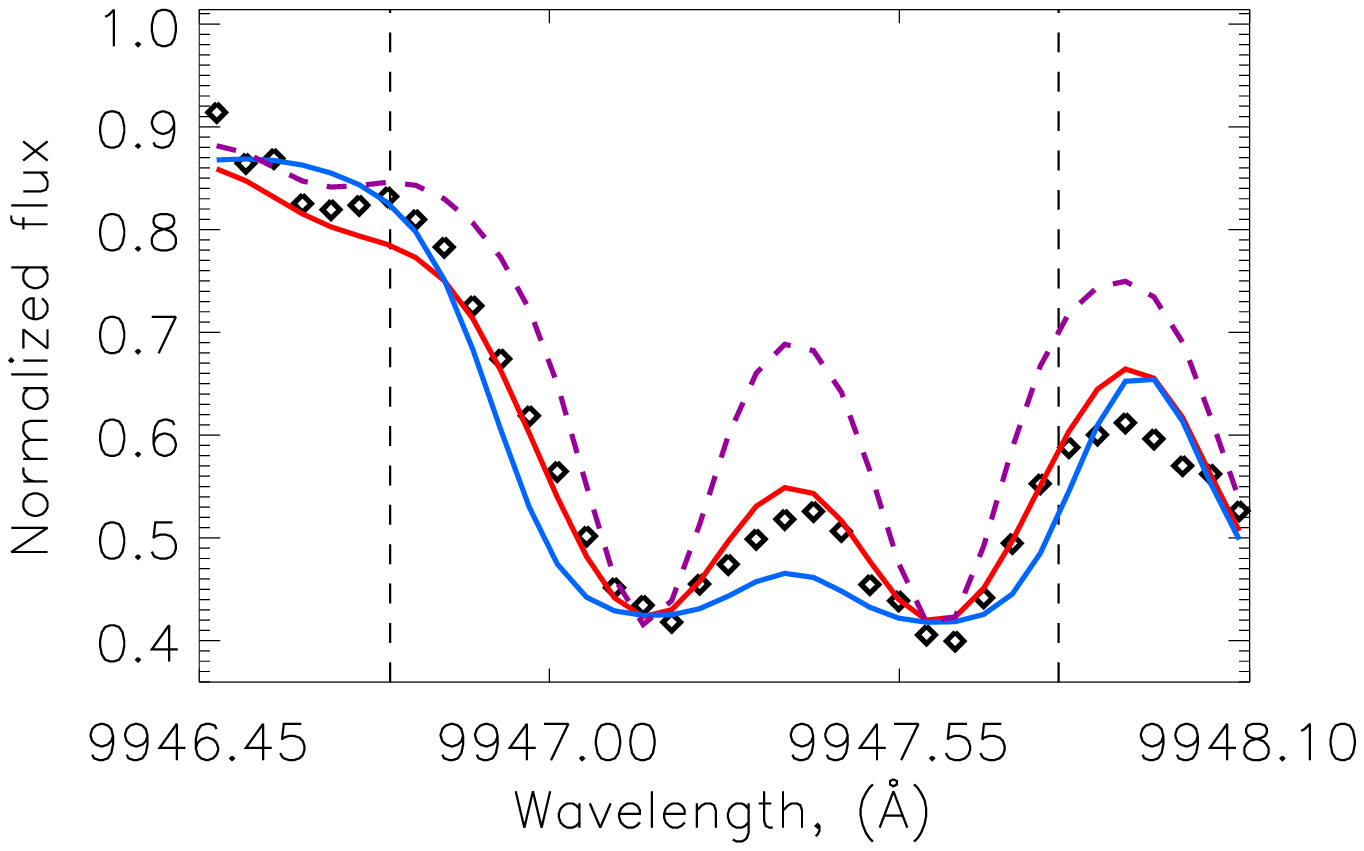}
\includegraphics[width=0.33\hsize]{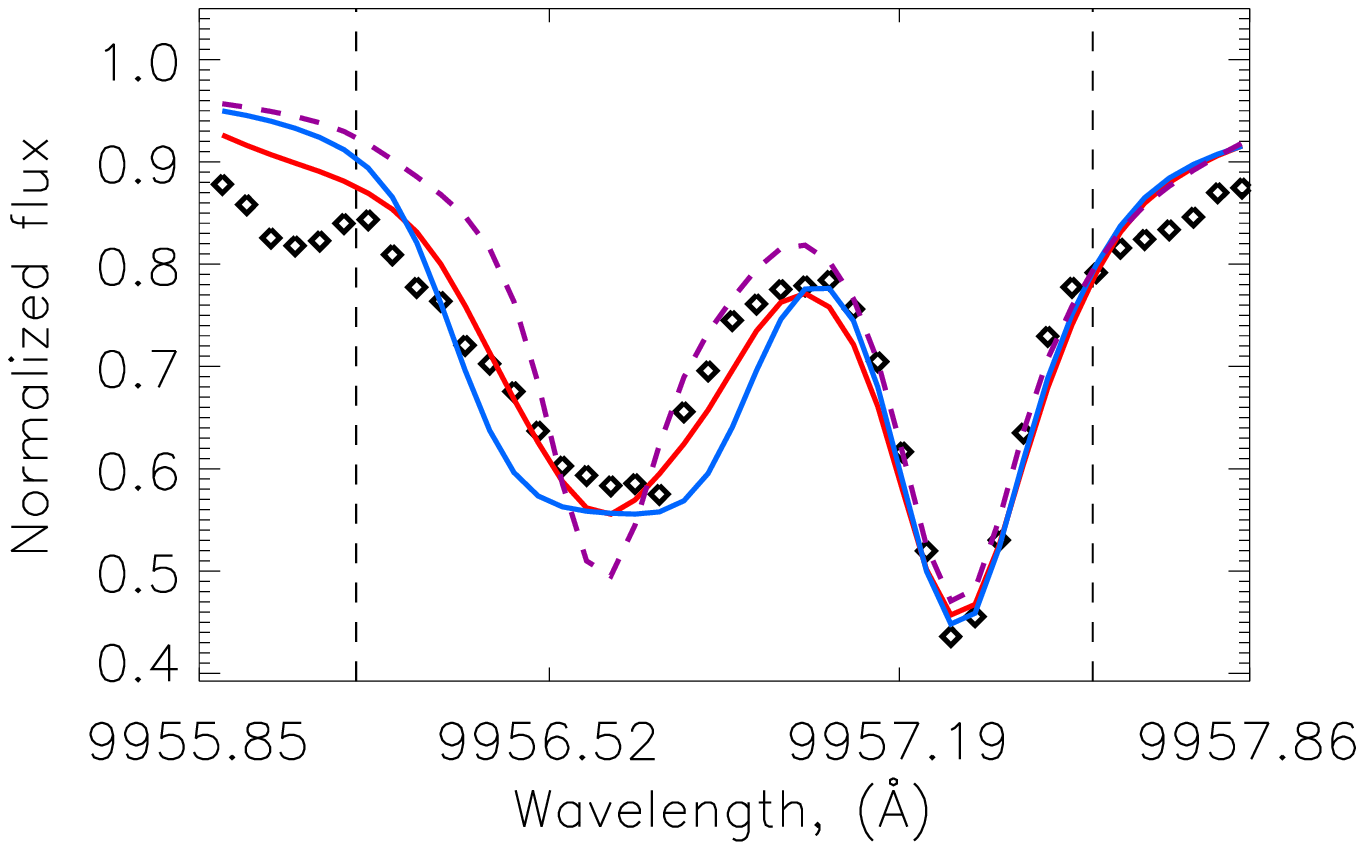}
\includegraphics[width=0.33\hsize]{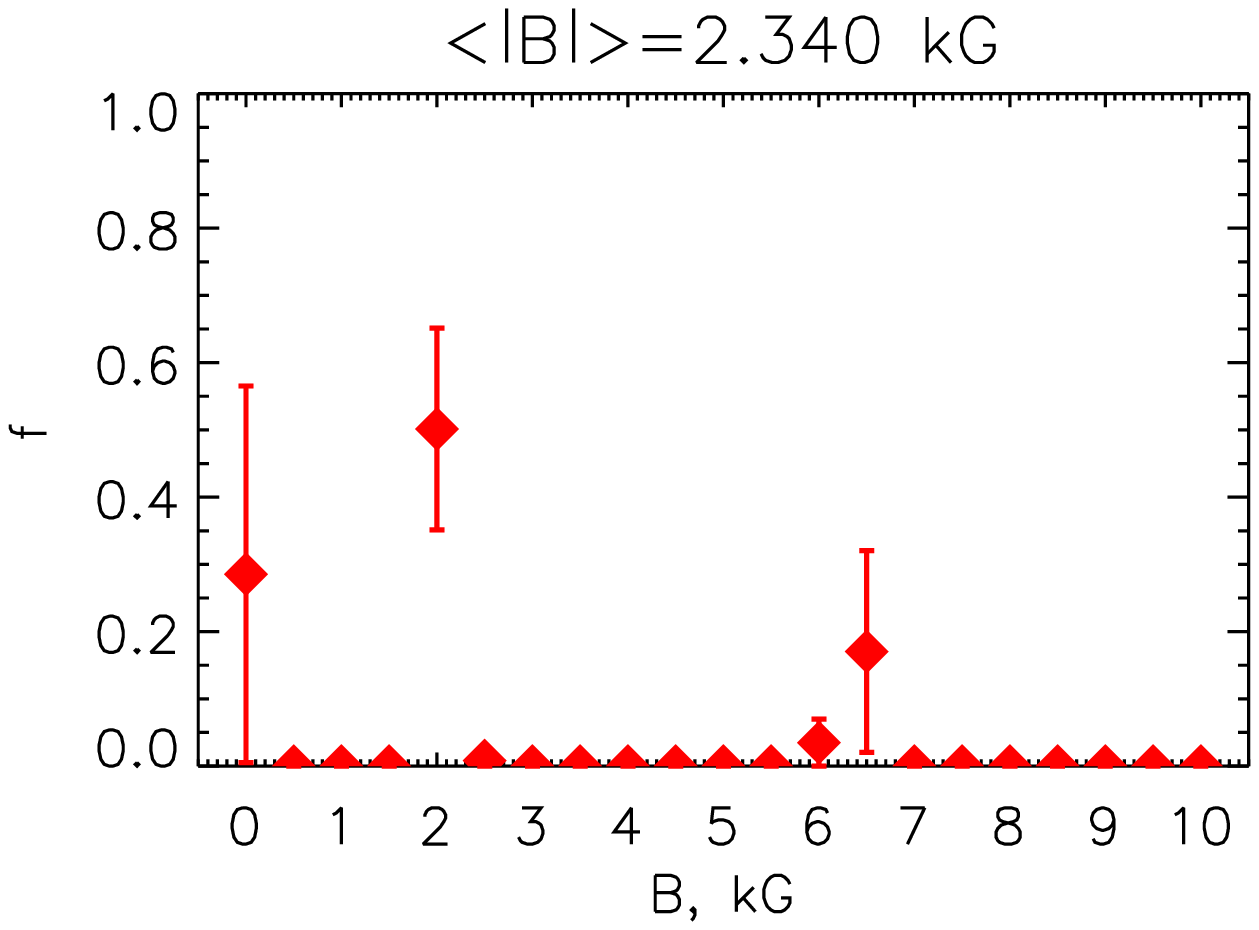}
}
\caption{Same as on Fig.~\ref{fig:gj388-fit} but for GJ~406.
Atmospheric parameters $\teff=3100$~K, $\abn{Fe}=-4.08$, $\vsini=3$~\kms.}
\label{fig:gj406-fit}
\end{figure*}

GJ~406 is a coolest target we analyze in this paper. It is of spectral type M$5.5$ and a slow rotator
with $\vsini=3$~\kms\ (RB07). We find the best fit atmospheric parameters $\teff=3100$~K, $\abn{Fe}=-4.08$, $\vsini=3$~\kms.
{Using a solar} iron abundance we were not able to
obtain an accurate fit {FeH lines such as} $\lambda9942$, $\lambda9950$, and $\lambda9956$ while matching profiles
of all other FeH lines. Moreover, {the} solar iron abundance systematically led to a much lower $\teff=2900$~K which
is inconsistent with the assigned spectral type of the star. Therefore we preferred a model with  $\teff=3100$~K thought
with {a} high iron abundance of $\abn{Fe}=-4.08$.
Figure~\ref{fig:gj406-fit} demonstrates the fit to {the FeH lines and the distribution} of filling factors. 
The latter reveals a strong $1.5$~kG component with $f=0.6$, two components with 
of $2.5$~kG and $3$~kG, and two components of $6$~kG and $6.5$~kG strength when employing
RC model. The {mean} magnetic field strength is $\bs=2.6$~kG. 
The MC model leads to a $f=0.3$ zero-field component and strong $f=0.55$ component of $2$~kG, as well as
weak $f=0.2$ component of $6.5$~kG. {This model gives a mean} $\bs=2.3$~kG, i.e. $\approx0.3$~kG lower compared 
RC case.

\begin{table*}
\caption{Magnetic field measurements in M-dwarfs.}
\label{tab:results}
\begin{center}
\begin{tabular}{lccccccccc}
\hline\hline

\multirow{3}{*}{Name}   & $\teff$         & $\abn{Fe}$          & $\vsini$          & \multicolumn{2}{c}{$\bs^a$}  & \multicolumn{2}{c}{$\sigma^b$}  & $Bf$ (RB07)$^c$                    & $B_{max}$$^d$ \\
                        & K               &  dex                & \kms              & \multicolumn{2}{c}{kG}       & \multicolumn{2}{c}{\%}          & kG                                 & kG            \\
                        &                 &                     &                   & RC  &  MC                    & RC  &  MC                       &                                    &               \\
\hline                                                                                   
                                                                                         
\multirow{3}{*}{GJ~388} & $3500$          & $-4.46$             & $2.9$             & $2.8$ & $2.9$                & $1.69$ & $1.92$                 & \multirow{3}{*}{$2.9\pm0.2$}       & $7.0$\\
                        & $3400$          & $-4.64$             & $3.0$             & $2.9$ & $2.9$                & $1.48$ & $1.53$                 &                                    & $6.5$\\
                        & $3300$          & $-4.73$             & $2.7$             & $2.7$ & $2.8$                & $1.64$ & $1.81$                 &                                    & $7.0$\\
\hline                                                                                   
                                                                                         
\multirow{3}{*}{GJ~729} & $3500$          & $-4.30$             & $4.4$             & $2.4$ & $2.3$                & $1.72$ & $1.68$                 & \multirow{3}{*}{$2.2\pm0.1$}       & $5.5$\\
                        & $3400$          & $-4.40$             & $4.0$             & $2.3$ & $2.1$                & $1.66$ & $1.74$                 &                                    & $6.0$\\
                        & $3300$          & $-4.54$             & $4.0$             & $2.2$ & $1.9$                & $1.67$ & $1.75$                 &                                    & $6.0$\\
\hline                                                                                   
                                                                                         
\multirow{3}{*}{GJ~285} & $3400$          & $-4.22$             & $6.2$             & $3.6$ & $3.4$                & $2.03$ & $2.11$                 & \multirow{3}{*}{$>3.9$}            & $7.5$\\
                        & $3300$          & $-4.37$             & $6.0$             & $3.6$ & $3.4$                & $2.00$ & $2.10$                 &                                    & $7.5$\\
                        & $3200$          & $-4.57$             & $6.2$             & $3.5$ & $3.6$                & $1.95$ & $1.97$                 &                                    & $7.0$\\
                                                                                         
\hline                                                                                   
                                                                                         
\multirow{3}{*}{GJ~406} & $3200$          & $-3.80$             & $3.0$             & $2.3$ & $2.0$                & $4.03$ & $4.36$                 & \multirow{3}{*}{$2.4\pm0.1$}       & $7.0$\\
                        & $3100$          & $-4.08$             & $3.0$             & $2.6$ & $2.3$                & $3.94$ & $3.87$                 &                                    & $6.5$\\
                        & $3000$          & $-4.26$             & $1.0$             & $2.2$ & $2.0$                & $4.10$ & $4.36$                 &                                    & $6.5$\\

\hline                                                                                   
                                                                                         
\multirow{2}{*}{GJ~1002}& $3100$          & $-4.37$             & $2.4$             & $<$0.1  &                    & $1.05$ &                        & \multirow{2}{*}{$0.0$}             & $1.0$\\
                        & $3100$          & $-4.47$             & $1.9$             & $<$0.02 &                    & $0.98$ &                        &                                    & $0.5$\\
\hline
\end{tabular}
\end{center}
$^a$~--~mean magnetic field derived assuming magnetic field models with dominating radial (RC) and meridional (MC) components\\
$^b$~--~deviation between observed and predicted profiles of FeH lines computed using Eqn.~(\ref{eqn:sigma})\\
$^c$~--~values are from \citet{2007ApJ...656.1121R}\\
$^d$~--~strength of the maximum field component detected
\end{table*}

\section{Discussion}

\subsection{Indicators of Zeeman splitting}

While most previous studies were based on only a few (or one) atomic lines at visual wavelengths,
in this work we attempted to extend the spectroscopic investigation of magnetic fields in M-dwarfs by
{using} numerous lines of the FeH molecule. We used CRIRES@VLT  to obtain high resolution $R=100\,000$ spectra of four well known
M-dwarfs in the FeH Wing-Ford band and \ion{Na}{i} lines in the K-band. Our goal was to measure the complexity of
the magnetic fields in these stars by using direct spectrum synthesis and up-to-date models of Zeeman splitting of FeH lines
developed and tested in \citet{2010A&A...523A..37S}. What we mean here by ``complexity'' of the field is the minimum number of
filling factors (each corresponding to a given magnetic field intensity) required to fit the observed line profiles 
for a given set of atmospheric parameters. The distribution and amplitude of filling factors  contain  information
about the possible structure of surface magnetic fields.

As shown by our forward simulations, the restored distribution of filling factors depends critically on the SNR of the observed spectrum.
If the SNR is of the order of $100$ the derived amplitude of filling factors may be biased if the real distribution is 
smooth, i.e., similar to those presented in the lower left plot of Fig.~\ref{fig:forward-fit}. On the other hand, the value
of the average surface magnetic field can be reproduced in all cases. In our sample, GJ~285 and GJ~388 have SNR~$>200$, while
GJ~729 and GJ~406 have SNR~$100-130$. Because the shape of the magnetic field distribution is less affected by SNR
we do not expect strong changes in the corresponding results if higher SNR observations were available, but {we} note that
comprehensive analysis is possible only with SNR$>300$.

We confirm that FeH lines are suitable indicators of magnetic fields as they contain both magnetically very
sensitive and magnetically insensitive lines in a $\approx$100~\AA\ window of the Wing-Ford band. 
Additionally, there are very few lines
of other species in that region except FeH, only some very weak atomic lines, i.e.  FeH lines do not suffer blends
{of other species}.
Lines at $\ll9903-9907$ and $\lambda9947$ are also excellent diagnostics of the complexity of the magnetic field
because of the characteristic behaviour of their line shapes as a function of the field distribution.

On the other hand, we failed to fit lines of \ion{Na}{i} even in a presumably non-magnetic
reference star GJ~1002. Since this spectral region is known to be crowded by numerous stellar water lines 
it is likely that we failed to perform an appropriate continuum normalization of the spectra. 
Unfortunately, CRIRES only provides data in very narrow wavelengths intervals, about 
$\Delta\lambda\approx100$~\AA\ in that region, which makes a reliable continuum determination very difficult.

As additional check for the consistency between atomic and FeH spectra, we used strong \ion{Ti}{i} lines at $1.1$\mum. 
Some of these lines are wide features with weak or no detectable blending. In {the} case where our method detects
no or an insignificant magnetic field from FeH lines, it resulted in a detection when only Ti lines were used. 
This is because of a mismatch between observed and {predicted profiles in the wings of the Ti lines}. In order to obtain
a solution that is consistent with a field-free situation
we had to decrease $\teff$ of the star by $50$~K and slightly increase $\vsini$. Obviously, we do not
expect FeH and Ti lines to form under such different physical conditions: the atmosphere of the star should be homogeneous, 
in particular in
absence of substantial magnetic activity. The difference may partly be explained by atomic and molecular lines being formed 
at different
atmospheric layers, but {it is} more likely a sign
of some missed physics, e.g., van der Waals broadening of Ti lines at such cool temperatures that differs from the
classical treatment that we use in our spectral synthesis.

\subsection{Comparison to earlier field measurements}


We summarize our magnetic field measurements in Table~\ref{tab:results}.
For each star, we {provide} the three best-fit solutions, one for our best estimate of $\teff$, and in addition, 
two solutions assuming a $\pm100$~K uncertainity on $\teff$, which corresponds to roughly $0.5$ stellar spectral types. 
With this assumption, $\abn{Fe}$ and $\vsini$ were determined by least-square fitting to the 
magnetically insensitive FeH features listed in Sect.~\ref{sect:methods} before the magnetic field distribution 
was calculated including all available spectral features. We note that in most cases the simultaneous least-square fit 
to all atmospheric parameters did not yield a statistically meaningful solution (the formal reduced $\chi^2$ was always 
much larger than $1$), which results in underestimation of uncertainties, in particular in the average magnetic field. 
Therefore, we do not provide formal uncertainties from $\chi^2$ intervals but calculate the effect 
of varying atmospheric parameters, which is the dominating source of error. The difference between the mean magnetic fields 
for the three solution in $\teff$ are shown in Table~\ref{tab:results}, {and} they can be used as a measure of the uncertainity 
of the magnetic field measurements. In addition, columns 5 and 6 of the Table~\ref{tab:results} contain values of the 
cumulative deviation between observed and predicted line profiles computed as following:

\begin{equation}
\sigma=\displaystyle\sqrt{\displaystyle\frac{1}{N}\sum_i^N \left(f_{i}^{\rm obs}-f_{i}^{\rm pred}\right)^2}.
\label{eqn:sigma}
\end{equation}

\noindent
The last two rows of Table~\ref{tab:results} list the two sets of atmospheric parameters of GJ~1002 discussed in Sect.~\ref{sect:methods}.

It is interesting to compare our results to measurements of magnetic fields in
the literature.  First, \citet{1985ApJ...299L..47S} used five Ti lines in the
K-band to measure the magnetic field of GJ~388. They found an average
$\b=3.8$~kG and $f=0.73$ so that $\btimesf=2.8$~kG. Later,
\citet{2000ASPC..198..371J} employed a method similar to the one we used 
to a few atomic lines finding $\btimesfi=3.3$~kG. This is very close
to $\btimesfi=3.2$~kG derived by \citet{2009AIPC.1094..124K}. Using the results from
\citet{1996ApJ...459L..95J}, RB07 report a value of $\bs=2.9$~kG, which is
identical to the value we derive here from FeH lines, independent upon the
magnetic field model used.

For GJ~729 \citet{2000ASPC..198..371J} found $\btimesfi=2.0$~kG, the corresponding
value from RB07 is $\bs=2.2$~kG, again in a good agreement with what we find:
$\bs=2.3$~kG and $\bs=2.1$~kG employing RC and MC models respectively (recall
that the average surface magnetic field $\bs=\btimesfi$, where $\bs_i$ runs from
$0$~kG to $10$~kG).

Next, for GJ~285 \citet{2000ASPC..198..371J} reported $\btimesfi=3.3$~kG, which is
much smaller then the lower limit of $\bs>3.9$~kG given by RB07. A recent
study by \citet{2009AIPC.1094..124K} provides $\btimesfi=4.5$~kG.  Our estimates are
$\bs=3.6$~kG and $\bs=3.4$~kG for the RC and MC models, respectively, which
fall between the earlier measurements.

For the coolest star in our sample, GJ~406, \citet{2007A&A...466L..13R} found
$\bs=2.2$~kG in a campaign spanning three observing nights. Intra-night
magnetic variability was found to be significant {at} the level of $100$~G. We
find $\bs=2.6$~kG and $\bs=2.3$~kG employing RC and MC models while RB07
derived a value of $\bs=2.4$~kG.

Thus, in general, measurements taken at different times are in good agreement
considering the substantial uncertainties reported by all authors and the
different spectral indicators used.  Nevertheless, it is interesting to note
that the highest variability (both absolute and relative) occurs in the most
active star GJ~285 {which has the strongest mean surface magnetic field 
and the highest level of x-ray luminosity among other stars in our sample \citep{2004A&A...417..651S}}. 
To what extent this reflects the variability of the
magnetic fields or differences between measurement techniques remains to be
identified.

\subsection{Field distributions}

For our sample stars, we have determined distributions of magnetic
fields rather than average values alone. A great advantage of our
method is that it is able to capture the full magnetic field
distribution on the surface of a star regardless of the field polarity
and geometry. All four stars show distinct groups of different
magnetic field strength. We did not find solutions with homogeneous
field distributions ($f = 1$ for the component that equals the average
field), and we did not find solutions in which different field
strengths are equally represented on the stellar surfaces ($f =
const.$ for all values of $B$ up to $B_{\rm max}$). This is an
interesting result because the \emph{average} fields on the four stars
are on the order of the \emph{maximum} field strength we observe in
sunspots. This implies that large surface regions contain
fields that are always on the order of those in small scale sunspots.
Magnetic field saturation
\citep[see][]{2009ApJ...692..538R} would then occur because the entire
stellar surface is populated with a field strength on the order of
2--3\,kG, while the local field strength does not change. Our results,
and earlier reports for example in \citet{2000ASPC..198..371J} clearly
show that this is not the case, but instead local magnetic flux
densities {occur that are much larger} than those found in sunspots, and
they co-exist with groups of much lower flux densities.

For our spectrum synthesis calculations we assumed two geometrically
different magnetic field models: one with a dominating radial
component (RC) and another one with a dominating meridional component
(MC). The two cases represent two extreme realizations of field
geometries and can provide useful information to estimate the
uncertainty in the field distribution introduced by our ignorance of
the field geometry. For the RC model, we find that three of the four
stars of our sample, GJ~388, GJ~729, and GJ~406 share very similar
field distributions. Their surfaces are entirely covered with regions
of significant magnetic fields, i.e., they show no zero-field
component ($f=0$ for $\bs=0$~kG). In them, the weakest field component
shows $1-1.5$~kG field strength covering the largest fraction of the
star ($f=0.35-0.55$). There is a lack of field strengths of $2$~kG
but a few smaller components with fields between $2.5-4$~kG exist
($f=0.1-0.3$). These components cover $10-30$~\% of the stellar
surface. This group again is distinct from the strongest component
with a magnetic field between $5.5$ and $6.5$~kG ($f=0.15-0.20$).  In
the RC case, the field distribution of the fourth star of our sample,
GJ~285, consists of three components as well, but this one is
different from the other three stars; it shows a significant
zero-field component ($f=0.3$). The second component has a field
strength on the order of $3.5$~kG and is of comparable size
($f\approx0.3$), and the third component has a very strong magnetic
field of $7.5$\,kG again covering about a third of the star.

Our second model of the magnetic field geometry, the MC model, leads
to very similar field distributions. In all stars, we detect three
distinct groups of field components similar to the previous case of
the RC model. Their relative strengths are slightly different while
the maximum field strength required for our MC model is the same as
for the RC model. In the MC case, a zero-field component is always
present (but it is not generally the dominating one).  GJ~388 shows an
additional fourth component of $4.5$~kG with a relatively small (yet
significant) filling factor of $f=0.15$.

It is remarkable that very similar magnetic field distributions are found in
all four sample stars and that the distributions are not much affected by the
choice of the magnetic field geometry. This suggests that field distributions
indeed follow some geometrical order ruled by the parameters of a star and its
dynamo. Furthermore, our target stars populate the region around the boundary
between partially and fully convective stars; GJ~388 and GJ~729 are on the
cool side of this threshold while GJ~406 and GJ~285 are probably fully
convective. Spectropolarimetric observations
of stars in this spectral range indicate
the existence of two distinct magnetic field geometries (with a few exceptions, see, e.g., 
\citet{2010MNRAS.407.2269M}): 
partly convective
objects seem to harbor non-axisymmetric, toroidal fields, while fully
convective objects prefer axisymmetric, poloidal fields. However, if such a
dichotomy of geometries exists, we would expect it {to affect} the distribution
of magnetic field components among our sample stars. For instance, large scale
axisymmetric poloidal fields would show something like one dominating
component while non-axisymmetric, toroidal fields would have a more uniform
distribution of field components. The pattern of the magnetic field
distributions we find {shows} no evidence for such a transition at the convection
boundary. We therefore conclude that our Stokes~I measurements of the entire
magnetic field cannot confirm a difference in magnetic field geometries
between partially and fully convective stars.

While the overall field distribution of the four sample stars is very
similar, we find evidence for more subtle differences between the
distributions of GJ~285 and the other three stars. In GJ~285, we
always detect a zero-field component independent of the magnetic field
model used. A zero-field component is also found in the MC models of
the other stars but not in their RC models. We cannot determine the
geometry of the field lines, but this difference is a first hint to a
subtle physical difference of the field geometries. Furthermore, the
MC model provides a rather poor fit in GJ~285 (see
Fig.~\ref{fig:gj285-fit}) so that in this case we believe the RC
solution is more probable. If the other three stars host a zero-field
component, their magnetic fields are better described by the MC
model. Although the differences are rather subtle, this implies that
the field configuration of GJ~285 is different from the other three:
either GJ~285 has a significant zero-field component while GJ~388,
GJ~729, and GJ~406 have not, or the latter three stars must be better
described by the MC geometry while for GJ~285 the RC geometry is
preferred. 

In what sense is GJ~285 different from the other three? We discussed above
that the status of convection is probably not the determining parameter
because at least one of the three other stars is very likely to share the same
convection properties as GJ~285. Another possible difference between GJ~285
and the other stars is the rotation rate; GJ~285 shows the highest value of
$v\,\sin{i}$ in our sample (see Table\,\ref{tab:results}). 
Recent measurements of \cite{2011ApJ...727...56I} provide {the} rotation period of GJ~285 $P=2.78$\,d 
which agrees well with the $P=2.7758$~d derived from Zeeman Doppler Imaging by \citet{2008MNRAS.390..567M}.
From this, \cite{2012AJ....143...93R} estimate an equatorial rotation velocity of $v_{\rm eq} =
5.8$\,km\,s$^{-1}$, which is consistent with our determination of $v\,sin{i}$
and $i=90^{\circ}$. There is no evidence for equally fast rotation of GJ~729
or GJ~406 \citep[see][]{2012AJ....143...93R}, but \cite{2009AIPC.1135..221E} report a rotation
rate of $P=2.23$\,d for GJ~388 ($P=2.2399$~d from \citet{2008MNRAS.390..567M}) 
meaning that this star may actually be rotating
faster than GJ~285. At this point, the situation remains inconclusive, and
more information from stars with measured rotation periods and magnetic field
distributions is required.

In our sample, the maximum field strength correlates with (projected) rotation
rate. While the average field strength saturates at Rossby numbers $Ro \sim
0.1$ and $B\approx4$\,kG \citep{2009ApJ...692..538R}, the local field strength
found in the present work in individual components does not saturate. From our small sample, we
cannot conclusively answer the question whether local field strength scales
with rotation rate, but we can conclude that it grows beyond the typical field
strengths we find in sunspots, and that there is a clear trend that 
$B_{\rm max}$ scales with rotation. Again, a larger sample is required to fully
address this question.

The detection of localized field components with the strength of up to
$7.5$~kG (GJ~285) provides important information about the surface {field} geometry of
active M-dwarfs.  If these structures are stable over long time intervals they
must be in equipartition with the surrounding plasma -- a situation similar to
sunspots. As can be computed using the available model atmospheres, 
the equipartition field at the level of the non-magnetized
photosphere of a mid-M dwarf is of the order of only $4-5$~kG. As in sunspots,
the equipartion field strength is higher deeper in the atmosphere. Therefore,
the high localized field strength can be in equipartition with the surrounding
plasma if we observe these magentic structures at deeper {atmospheric}
levels. This is possible if the regions of strong magnetic fields have cooler
temperatures compared to the rest of the atmosphere. This effect is known as
{the} Wilson depression: the suppression of convection by the magnetic field in the
sunspot leads to a temperature drop to $\approx4000$~K, lower gas pressure,
and allows one to observe deeper {layers} when looking through the spot.
Alternatively, if one assumes that the localized regions of strong fields are
not in the equipartition with the surrounding plasma then they must be
transient events. Consequently, the distribution and/or amplitude of filling
factors are expected to change in time. Unfortunately, no accurate estimate of
the temperature contrast is available for M-dwarfs because their surfaces
remain unresolved, and high quality time series of spectroscopic observations
in Stokes~I are not yet available.
{A line depths ratio method proposed by
\citet{2002A&A...394.1009C} to measure temperature contrast in cool stars
gives uncertain results for stars of types M \citep[see][]{2005LRSP....2....8B}.
Photometry seems to be a promissing method to test the spot contrast, but 
there are two basic effects that may (and do) noticeably modify the amplitude of the light curves:
stellar flares and spot distribution. The former can still be, to a certain degree, accounted for
and reasonably accurate light curves have resently been constructed at least for some stars
in \citep{2011ApJ...727...56I}. However, having an accurate flux variability curve
does not allow one to reconstruct the temperature contrast uniquely because the spot distribution is not known
at first place. In other words, the light curve amplitude will look similar in case of a single small spot with high
temperature contrast or a few small spots each having less strong temperature contrast.}
In any case, accurate phase-resolved
observations can provide the information required to probe the magnetic and
temperature structure of spots in low-mass stars.

\begin{acknowledgements}
We wish to thank Prof. Manfred Schuessler for his useful comments on the paper.
We acknowledge financial support from CRC~963~--~Astrophysical Flow Instabilities and Turbulence 
(project A16)
{and Deutsche Forschungsgemeinschaft (DFG) Research Grant RE1664/7-1} to DS and
funding through a Heisenberg Professorship, RE 1664/9-1 to AR.
We also acknowledge the use of electronic databases (VALD, SIMBAD, NASA's ADS) and 
cluster facilities at the computing centre of Georg August University G\"ottingen (GWDG).
This research has made use of the Molecular Zeeman Library (Leroy, 2004).
\end{acknowledgements}


\listofobjects


\begin{thebibliography}{}
\bibitem[Afram et al.(2008)]{2008A&A...482..387A} Afram, N., Berdyugina, S.~V., Fluri, D.~M., Solanki, S.~K., \& Lagg, A.\ 2008, \aap, 482, 387 
\bibitem[Asensio Ramos \& Trujillo Bueno(2006)]{2006ApJ...636..548A} Asensio Ramos, A., \& Trujillo Bueno, J.\ 2006, \apj, 636, 548 
\bibitem[Berdyugina(2005)]{2005LRSP....2....8B} {Berdyugina, S.~V.\ 2005, Living Reviews in Solar Physics, 2, 8 }
\bibitem[Berdyugina \& Solanki(2002)]{2002A&A...385..701B} Berdyugina, S.~V., \& Solanki, S.~K.\ 2002, \aap, 385, 701 
\bibitem[Catalano et al.(2002)]{2002A&A...394.1009C} {Catalano, S., Biazzo, K., Frasca, A., \& Marilli, E.\ 2002, \aap, 394, 1009 }
\bibitem[Donati et al.(2008)]{2008MNRAS.390..545D} Donati, J.-F., Morin, J., Petit, P., et al.\ 2008, \mnras, 390, 545 
\bibitem[Donati et al.(2006)]{2006Sci...311..633D} Donati, J.-F., Forveille, T., Collier Cameron, A., et al.\ 2006, Science, 311, 633 
\bibitem[Donati et al.(1997)]{1997MNRAS.291..658D} Donati, J.-F., Semel, M., Carter, B.~D., Rees, D.~E., \& Collier Cameron, A.\ 1997, \mnras, 291, 658 
\bibitem[Dulick et al.(2003)]{2003ApJ...594..651D} Dulick, M., Bauschlicher, C.~W., Jr., Burrows, A., et al.\ 2003, \apj, 594, 651 
\bibitem[Engle et al.(2009)]{2009AIPC.1135..221E} Engle, S.~G., Guinan, E.~F., \& Mizusawa, T.\ 2009, American Institute of Physics Conference Series, 1135, 221 \bibitem[Gray, 1992]{gray}Gray D.F., 1992, The Observation and Analysis of Stellar Photospheres, Cambridge University Press
\bibitem[Irwin et al.(2011)]{2011ApJ...727...56I} Irwin, J., Berta, Z.~K., Burke, C.~J., et al.\ 2011, \apj, 727, 56 
\bibitem[Gustafsson et al.(2008)]{2008A&A...486..951G} Gustafsson, B., Edvardsson, B., Eriksson, K., et al.\ 2008, \aap, 486, 951 
\bibitem[Johns-Krull(2007)]{2007ApJ...664..975J} Johns-Krull, C.~M.\ 2007, \apj, 664, 975
\bibitem[Johns-Krull \& Valenti(2000)]{2000ASPC..198..371J} Johns-Krull, C.~M., \& Valenti, J.~A.\ 2000, Stellar Clusters and Associations: Convection, Rotation, and Dynamos, 198, 371 
\bibitem[Johns-Krull \& Valenti(1996)]{1996ApJ...459L..95J} Johns-Krull, C.~M., \& Valenti, J.~A.\ 1996, \apjl, 459, L95 
\bibitem[Harrison \& Brown(2008)]{2008ApJ...686.1426H} Harrison, J.~J., \& Brown, J.~M.\ 2008, \apj, 686, 1426 
\bibitem[Leroy, 2004]{mzl}Leroy, B. 2004, Molecular Zeeman Library Reference Manual (avalaible on-line at http://bass2000.obspm.fr/mzl/download/mzl-ref.pdf)
\bibitem[Morin et al.(2010)]{2010MNRAS.407.2269M} Morin, J., Donati, J.-F., Petit, P., et al.\ 2010, \mnras, 407, 2269 
\bibitem[Morin et al.(2008)]{2008MNRAS.390..567M} Morin, J., Donati, J.-F., Petit, P., et al.\ 2008, \mnras, 390, 567 
\bibitem[Kaeufl et al.(2004)]{2004SPIE.5492.1218K} Kaeufl, H.-U., Ballester, P., Biereichel, P., et al.\ 2004, \procspie, 5492, 1218
\bibitem[Kochukhov et al.(2009)]{2009AIPC.1094..124K} Kochukhov, O., Heiter, U., Piskunov, N., et al.\ 2009, 15th Cambridge Workshop on Cool Stars, Stellar Systems, and the Sun, 1094, 124 
\bibitem[Kochukhov(2007)]{2007pms..conf..109K} Kochukhov, O.~P.\ 2007, Physics of Magnetic Stars, 109 
\bibitem[Kupka et al.(1999)]{1999A&AS..138..119K} Kupka, F., Piskunov, N., Ryabchikova, T.~A., Stempels, H.~C., \& Weiss, W.~W.\ 1999, \aaps, 138, 119 
\bibitem[Piskunov(1999)]{1999ASSL..243..515P} Piskunov, N.\ 1999, Polarization, 243, 515 
\bibitem[Piskunov et al.(1995)]{1995A&AS..112..525P} Piskunov, N.~E., Kupka, F., Ryabchikova, T.~A., Weiss, W.~W., \& Jeffery, C.~S.\ 1995, \aaps, 112, 525 
\bibitem[Reiners et al.(2012)]{2012AJ....143...93R} Reiners, A., Joshi, N., \& Goldman, B.\ 2012, \aj, 143, 93 
\bibitem[Reiners(2012)]{2012LRSP....9....1R} Reiners, A.\ 2012, Living Reviews in Solar Physics, 9, 1
\bibitem[Reiners et al.(2009)]{2009ApJ...692..538R} Reiners, A., Basri, G., \& Browning, M.\ 2009, \apj, 692, 538 
\bibitem[Reiners \& Basri(2007)]{2007ApJ...656.1121R} Reiners, A., \& Basri, G.\ 2007, \apj, 656, 1121 (RB07)
\bibitem[Reiners et al.(2007)]{2007A&A...466L..13R} Reiners, A., Schmitt, J.~H.~M.~M., \& Liefke, C.\ 2007, \aap, 466, L13 
\bibitem[Reiners \& Basri(2006)]{2006ApJ...644..497R} Reiners, A., \& Basri, G.\ 2006, \apj, 644, 497 
\bibitem[Saar \& Linsky(1985)]{1985ApJ...299L..47S} Saar, S.~H., \& Linsky, J.~L.\ 1985, \apjl, 299, L47 
\bibitem[Schmitt \& Liefke(2004)]{2004A&A...417..651S} {Schmitt, J.~H.~M.~M., \& Liefke, C.\ 2004, \aap, 417, 651 }
\bibitem[Shulyak et al.(2010)]{2010A&A...523A..37S} Shulyak, D., Reiners, A., Wende, S., et al.\ 2010, \aap, 523, A37 
\bibitem[Valenti et al.(2001)]{2001ASPC..223.1579V} Valenti, J.~A., Johns-Krull, C.~M., \& Piskunov, N.~E.\ 2001, 11th Cambridge Workshop on Cool Stars, Stellar Systems and the Sun, 223, 1579 
\bibitem[Wallace et al.(1999)]{1999asus.book.....W} Wallace, L., Livingston, W.~C., Bernath, P.~F., \& Ram, R.~S.\ 1999, NSO Technical Report \#99-001; Tucson: National Solar Observatory, 1999
\bibitem[Wende et al.(2010)]{2010A&A...523A..58W} Wende, S., Reiners, A., Seifahrt, A., \& Bernath, P.~F.\ 2010, \aap, 523, A58 
\bibitem[Wende et al.(2009)]{2009A&A...508.1429W} Wende, S., Reiners, A., \& Ludwig, H.-G.\ 2009, \aap, 508, 1429 
\end{thebibliography}
\end{document}